\documentclass[fleqn,usenatbib,twocolumn]{mnras}

\usepackage[T1]{fontenc}

\usepackage{graphicx}	
\usepackage{amsmath}	
\usepackage{amssymb}	

\usepackage{amsmath}
\usepackage{color}
\usepackage{appendix}
\usepackage{multirow}
\usepackage[flushleft]{threeparttable}
\usepackage{soul}
\newcommand{\astj}{AJ}

\newcommand{\mobse}{\texttt{MOBSE}}

\newcommand{\bpop}{\texttt{B-POP}}

\newcommand{\rem}{{\rm rem}}

\newcommand{\Ms}{{\rm M}_\odot}

\newcommand{\iso}{{\rm iso}}
\newcommand{\dyn}{{\rm dyn}}

\newcommand{\gc}{{\rm GC}}
\newcommand{\yc}{{\rm YC}}
\newcommand{\nc}{{\rm NC}}

\newcommand{\eff}{{\rm eff}}

\title[Isolated and dynamical black hole mergers with \bpop]{Isolated and dynamical black hole mergers with \bpop: the role of star formation and dynamics, star cluster evolution, natal kicks, mass and spins, and hierarchical mergers}

\author[M. Arca Sedda et al.]{
Manuel Arca Sedda$^{1}$\thanks{E-mail: m.arcasedda@gmail.com}
Michela Mapelli$^{2,3,4}$,
Matthew Benacquista$^{5}$,
Mario Spera$^{3,6,7,8}$
\\
$^{1}$Astronomisches Rechen-Institut, Zentrum f\"ur Astronomie der
  Universit\"at Heidelberg, M\"onchhoofstr. 12-14, D-69120 Heidelberg,
  Germany\\
$^{2}$Physics and Astronomy Department Galileo Galilei, University of Padova, Vicolo dell'Osservatorio 3, I--35122, Padova, Italy\\
$^{3}$INFN--Padova, Via Marzolo 8, I--35131 Padova, Italy\\
$^{4}$INAF--Osservatorio Astronomico di Padova, Vicolo dell'Osservatorio 5, I--35122, Padova, Italy\\
$^{5}$University of Texas Rio Grande Valley (Emeritus) Box 2044, Red Lodge, MT 59068, USA\\
$^{6}$Scuola Internazionale Superiore di Studi Avanzati (SISSA), Via Bonomea 265, I--34136 Trieste, Italy\\
$^{7}$INFN--Trieste, via Valerio 2, I--34127 Trieste,  Italy\\
$^{8}$IFPU--Institute for fundamental physics of the Universe, Via Beirut 2, I--34014 Trieste, Italy
}

\date{Accepted XXX. Received YYY; in original form ZZZ}

\pubyear{2021}

\begin{document}
\label{firstpage}
\pagerange{\pageref{firstpage}--\pageref{lastpage}}
\maketitle

\begin{abstract}
The current interpretation of LIGO--Virgo--KAGRA data suggests that the primary mass function of merging binary black holes (BBHs) at redshift $z\lesssim 1$ contains multiple structures, while spins are relatively low. Theoretical models of BBH formation in different environments can provide a key to interpreting the population of observed mergers, but they require the simultaneous treatment of stellar evolution and dynamics, galaxy evolution, and general relativity. We present \texttt{B-POP}, a population synthesis tool to model BBH mergers formed in the field or via dynamical interactions in young, globular, and nuclear clusters. Using \texttt{B-POP}, we explore how BH formation channels, star cluster evolution, hierarchical mergers, and natal BH properties affect the population of BBH mergers. We find that the primary mass distribution of BBH mergers extends beyond $M_1 \simeq 200\,{}$ M$_\odot$, and the effective spin parameter distribution hints at different natal spins for single and binary BHs. Observed BBHs can be interpreted as members of a mixed population comprised of $\sim 34\%  \,{}(66\%)$ isolated (dynamical) BBHs, with the latter likely dominating at redshift $z>1$. Hierarchical mergers constitute the $4.6-7.9\%$ of all mergers in the reference model, dominating the primary mass distribution beyond $M_1 > 65\,{}$ M$_\odot$. The inclusion of cluster mass-loss and expansion causes an abrupt decrease in the probability for mergers beyond the third generation to occur. Considering observational biases, we find that $2.7-7.5\%$ of mock mergers involve intermediate-mass black hole (IMBH) seeds formed via stellar collisions. Comparing this percentage to observed values will possibly help us to constrain IMBH formation mechanisms.     
\end{abstract}

\begin{keywords}
black holes -- gravitational waves -- stellar evolution -- star clusters -- globular clusters -- 
galactic nuclei
\end{keywords}

\section{Introduction}
The LIGO-Virgo-Kagra collaboration (LVK) has recently released an updated catalogue of gravitational wave (GW) events, named GWTC-3 \citep{gwtc3}. This catalogue contains the properties of 55 candidate black hole binary (BBH) mergers, featuring asymmetric mergers like GW190412 \citep{gw190412} and several peculiar systems, such as GW190814 \citep{gw190814}, whose companion falls in the so-called lower mass-gap and might be the lightest BH to date, and GW190521 \citep{gw190521a,gw190521b}, the first BBH merger that produced a remnant with a mass of $\sim 140\,{}\Ms$, i.e. in the mass range of the elusive intermediate-mass black holes (IMBHs).
Although still relatively low, the number of BBH mergers detected so far permitted to place some constraints on the BBH population at redshift $z<1-2$. In particular, \cite{gwtc2} suggest that the mass distribution of primary black holes (BHs) is characterised by a complex structure, likely described by a power-law with two peaks at $M_1 \sim 20\,{}\Ms$ and $M_1\sim 40\,{}\Ms$ and a sharp truncation at values $M_1>100\,{}\Ms$ \citep{gwtc2}. Such a complex distribution is likely the result of different BBH formation channels. 

A BBH can form through a variety of branches, but at the first order we can distinguish two broad ensembles: isolated binaries, which form from the evolution of stars paired together at birth, and dynamical binaries, whose formation is mediated by strong stellar encounters in young (YCs), globular (GCs), and nuclear clusters (NCs). According to the field triple channel, which is one of the possible sub-branches of the isolated binary scenario, three stars already bound at birth undergo a complex stellar and dynamical evolution that culminates in the formation of a merging BBH. Similarly, triples and higher order multiples can form in dense star clusters; in this case the three objects can become a bound system via dynamical interactions. 
With regards to galactic nuclei, we can distinguish three different sub-branches: i) BBH pairing in NCs without a supermassive BH (SMBH), ii) BBH pairing in NCs with a central SMBH, iii) BBH pairing in active galactic nuclei (AGN) discs. In the first case, BBH formation is regulated by dynamical encounters \citep{antonini16,antonini19,arca20b} and is directly linked to the formation history of the NC \citep{arca20b}. In the second case, the presence of an SMBH can efficiently affect the long-term evolution of the binary, potentially driving secular effects like Kozai-Lidov resonances \citep{kozai62,lidov62}, which can shorten the BBH lifetime and thus contribute to the formation of merging BBHs \citep{antonini12,hoang18,fragione19,arca20b}. In the latter case, the BH pairing is facilitated by the drag force of the AGN disc \citep[e.g.][]{mckernan18,tagawa21a}. While theoretical estimates derived for different channels and sub-channels suggest that they all might contribute to the cosmic population of BBHs, it is still unclear whether it is possible to discern fingerprints of different formation channels from observed mergers, mostly owing to the degeneracies that characterise different formation channels.
Several studies suggest that a mixed population of BBHs is the most likely scenario to explain LVK sources (e.g., \citealt{arca19b,arca20,bavera20,zevin21,bouffanais21}, but see \citealt{roulet21} and \citealt{rodriguez21} for a different interpretation).

Different formation channels can leave fingerprints on the distribution of remnant mass and spins, impact the properties of the merging BHs, and determine the merger rate per BH mass. BH spin magnitudes are still far from being understood. Several BHs in high-mass X-ray binaries seem to be nearly maximally spinning  \citep[see e.g.][]{qin19,reynolds21}, while LVK BBHs support evidence for relatively low spins ($\chi < 0.1-0.2$) \citep{gwtc2}. This might imply that different formation channels are characterised by different BH spin distributions.

In this work, we present the results of \bpop, a tool capable of creating large samples of BBH mergers formed either in isolation or in dynamical environments, which takes into account state-of-the-art stellar evolution recipes for single and binary stars, a semi-analytic treatment for the formation of dynamical mergers, a flexible treatment for BH natal spins, and implements numerical relativity fitting formulae to calculate remnant masses, spins, and GW recoil kicks. Moreover, \bpop exploits prescriptions for observational biases of second-generation ground-based GW detectors, and prescriptions for BH formation across cosmic time.

Varying the relative amount of BBH mergers forming in one channel or another, the BH natal spin distribution, and stellar evolution, we show that the observed GW sources can be interpreted as a part of a global population equally contributed by isolated and dynamical mergers. Such population exhibits a primary mass distribution characterised by a long tail extending beyond $M_1 > 100-200\,{}\Ms$ which might be particularly difficult to access with ground-based detectors due to observational selection effects.

The paper is organised as follows: in Section \ref{sec:model} we describe the main features and improvements of \bpop\ and the main properties of our models; Section \ref{sec:res} introduces the main results of our work in terms of primary and total BBH merger mass, spin, and effective spin parameter; Section \ref{sec:disc} discusses the implications of our results in terms of hierarchical mergers and massive BH seeds; whilst in Section \ref{sec:end}  
we summarize our conclusions.

\section{The B-POP code}
\label{sec:model}

In this work, we exploit an improved version of a semi-analytic tool that combines stellar evolution prescriptions for single and binary stars, a treatment for formation of dynamical and isolated BBH mergers, a treatment for relativistic kicks, and numerical relativity fitting formulae to estimate the properties of the remnant \citep{arca19b,arca20}. 
This method enables us to explore how different BBH formation channels can affect the properties of mergers observed from ground-based detectors like LIGO and Virgo. 

\begin{figure*}
\centering
\includegraphics[width=0.7\textwidth]{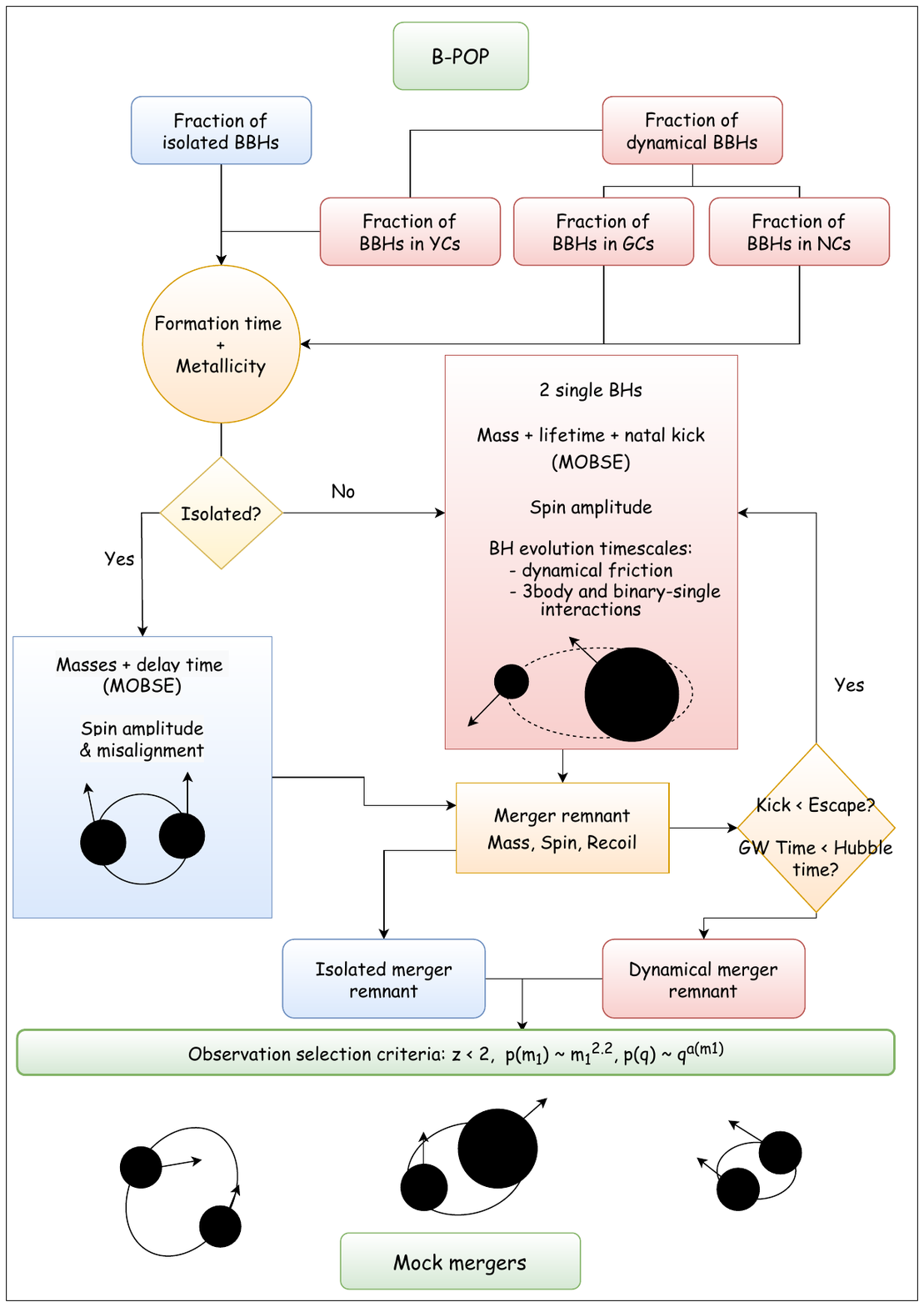}
\caption{\bpop\ workflow.}
\label{fig:flow}
\end{figure*}

Hereafter, we refer to the upgraded tool as \bpop\ (Black hole POPulation synthesis). \bpop\  implements a multi-stepped procedure that enables us to create populations of BBH mergers forming either via binary stellar evolution (isolated channel) or via gravitational encounters in star clusters (dynamical encounters). The \bpop\ workflow, which is shown in Figure \ref{fig:flow}, can be sketched as follows:
\begin{enumerate}
    \item Environment selection
    \begin{itemize}
    \item set the relative amount of BBH mergers forming via the isolated or the dynamical channel;
    \item for the dynamical channel, set the relative amount of mergers forming in YCs, GCs, and NCs;
    \item set the metallicity distribution of the host environment for dynamical and isolated channels;
	\item set the host cluster formation time for dynamical mergers and the binary formation time for isolated mergers;
    \end{itemize}
    
    \item Binary properties selection
    \begin{itemize}
    \item set the binary component masses according to a stellar initial mass function;
    \item for dynamical binaries, set the pairing criterion;
    \item use single/binary stellar evolution to calculate BH natal masses;
    \item assign to each BH a natal spin and set the level of spin alignment in the binary;
    \item assign to each merger a formation time inferred from the adopted star formation rate;
    \item assign to each merger a delay time, which is calculated accordingly to the formation channel as explained in the next sections;
    \end{itemize}
    
    \item Merger remnant
    \begin{itemize}
    \item calculate the remnant BH mass, spin, effective spin, and GW recoil using numerical relativity fitting formulae described in \cite{campanelli07}, \cite{lousto12}, and \cite{jimenez17};
    \item in the case of dynamical mergers, the remnant BH recoil is compared to the cluster escape velocity to model hierarchical mergers.
    \end{itemize}
\end{enumerate}

The mass and delay time for isolated BBHs and single BHs are obtained through the \mobse\ code \citep{mapelli2017,giacobbo18a,giacobbo18b, giacobbo20}. As detailed in Section \ref{dynBBH}, we exploit a semi-analytic method to assemble dynamical binaries.

The database of merging BBHs constructed this way is post-processed to take into account the intrinsic observational bias that might affect ground-based GW detectors. This procedure enables us to obtain two different populations: one ``raw'', that should reflect the population of mergers formed via dynamics or isolated stellar evolution, and the other ``weighted'' through the observation bias criteria adopted.

The aforementioned points summarize the main features of \bpop\footnote{We refer the reader to \cite{arca20} for further details about the previous version of our tool.}. Compared to the previous versions, \bpop\  implements several upgrades that are described in detail in the following.

\subsection{Black hole mass distribution}
\label{sec:bbhmass}

Several works that attempt to characterise the properties of dynamical mergers from a semi-analytic point of view generally assume that BHs participating in dynamical interactions have a mass distribution well described by the zero age main sequence (ZAMS) -- remnant mass relation of single stars \citep[see e.g.][]{OLeary09, OLeary16, antonini16b, gerosa17, fragione18, antonini20, doctor20}, i.e. a {\emph{simple single stellar BH mass spectrum}} (SSBH). 

Such a choice is reasonable as long as either the star cluster has little content in stellar binaries, or its stellar binary population is rapidly destroyed via strong interactions before stellar evolution plays a significant role, i.e. over a timescale $\ll 10$ Myr.

However, as pointed out in the recent literature, the actual mass spectrum of BHs participating in dynamical interactions can be much more complex, especially in clusters harboring a large population of primordial binaries \citep{dicarlo19,dicarlo21,gonzalez21,rizzuto21,rizzuto21b,rastello2021}. 

Some BHs can form out of collisions or mergers of massive stars initially paired in close binaries, while others can be former members of soft binaries that have been ionized via repeated interactions with other stars or compact objects \citep[see e.g.][]{spera19, dicarlo19, rizzuto21,rizzuto21b}. In the following, we will refer to this mass spectrum, associated with BHs formed from stellar collisions and former components of isolated binaries, as {\emph{ mixed single BH mass spectrum}} (MSBH).

In \bpop, the mass of BHs in star clusters can be drawn from both SSBH and MSBH, provided that the selected BH has a natal kick, which is provided by \mobse, smaller than the host cluster escape velocity. 
This implies that each cluster will be characterised by a peculiar BH mass spectrum that intrinsically depends on the cluster properties. 

\begin{figure*}
    \centering
    \includegraphics[width=0.8\columnwidth]{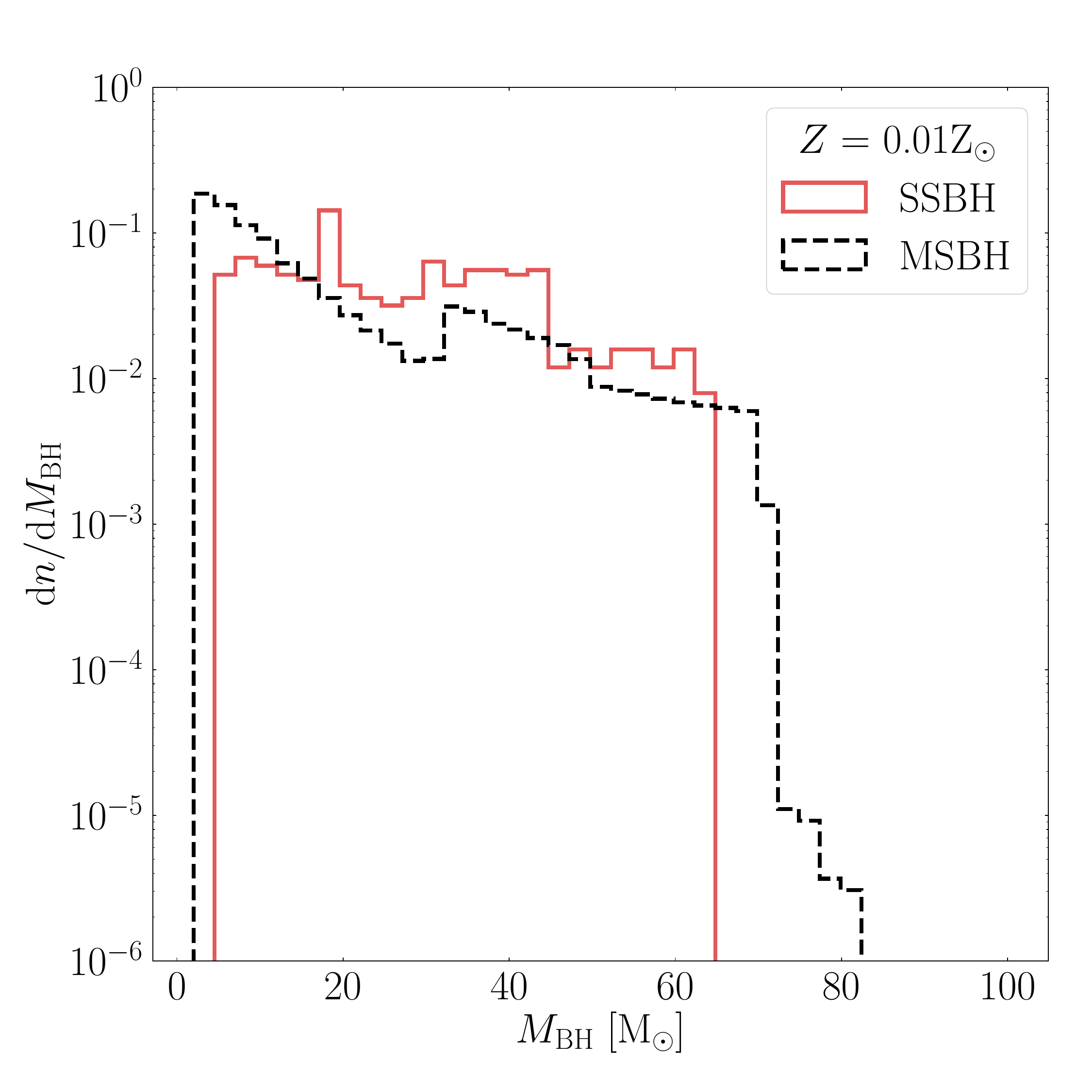}
    \includegraphics[width=0.8\columnwidth]{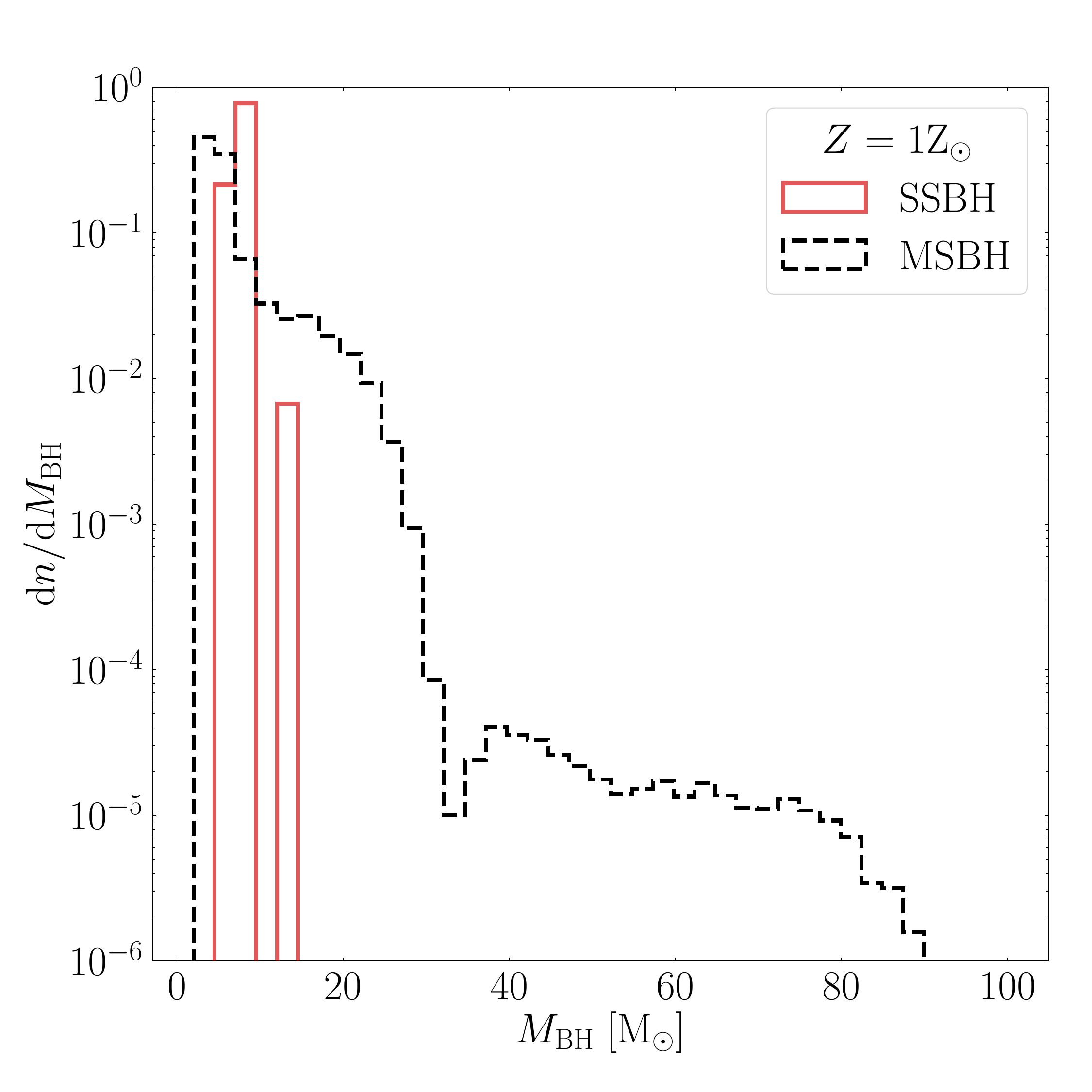}\\
    \includegraphics[width=0.8\columnwidth]{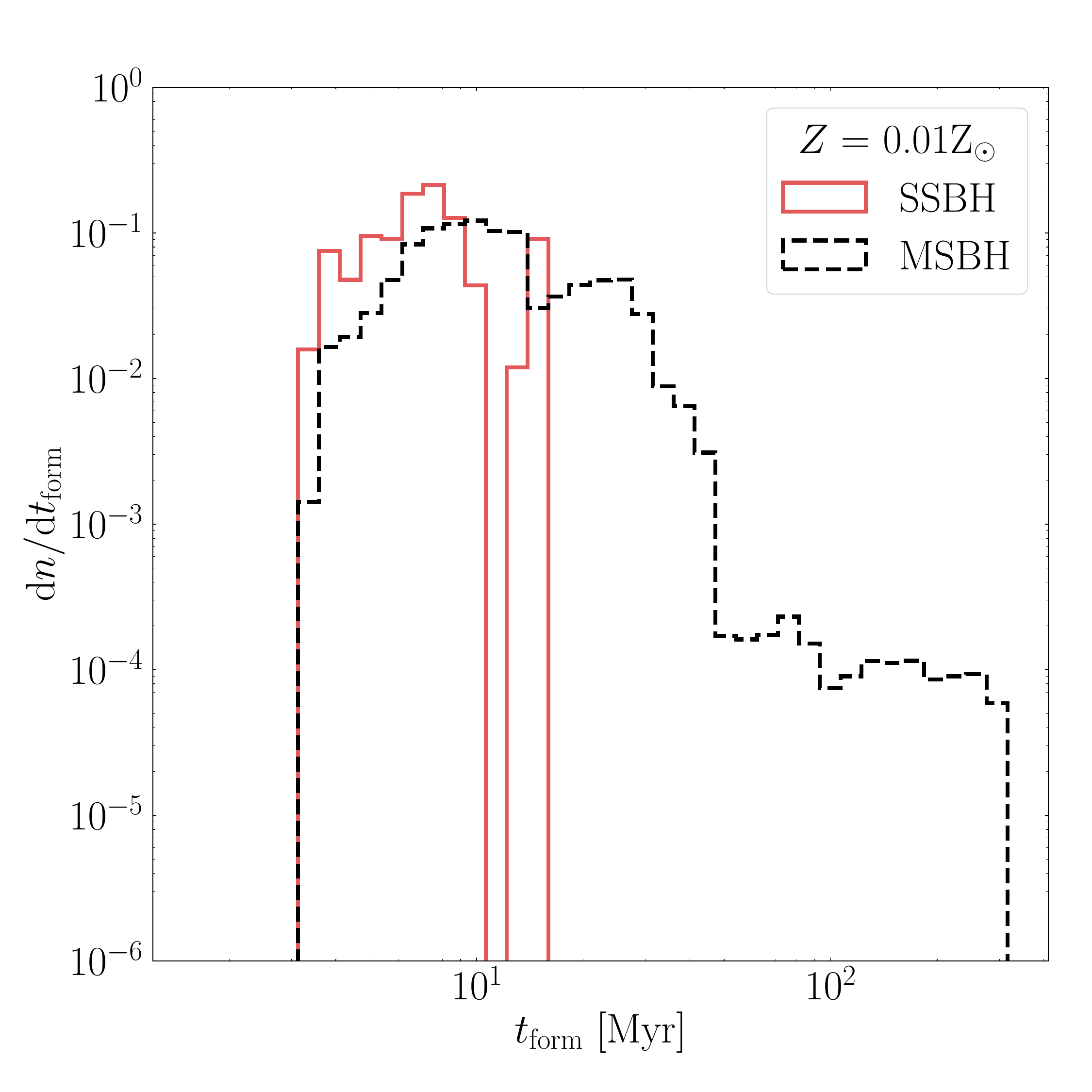}
    \includegraphics[width=0.8\columnwidth]{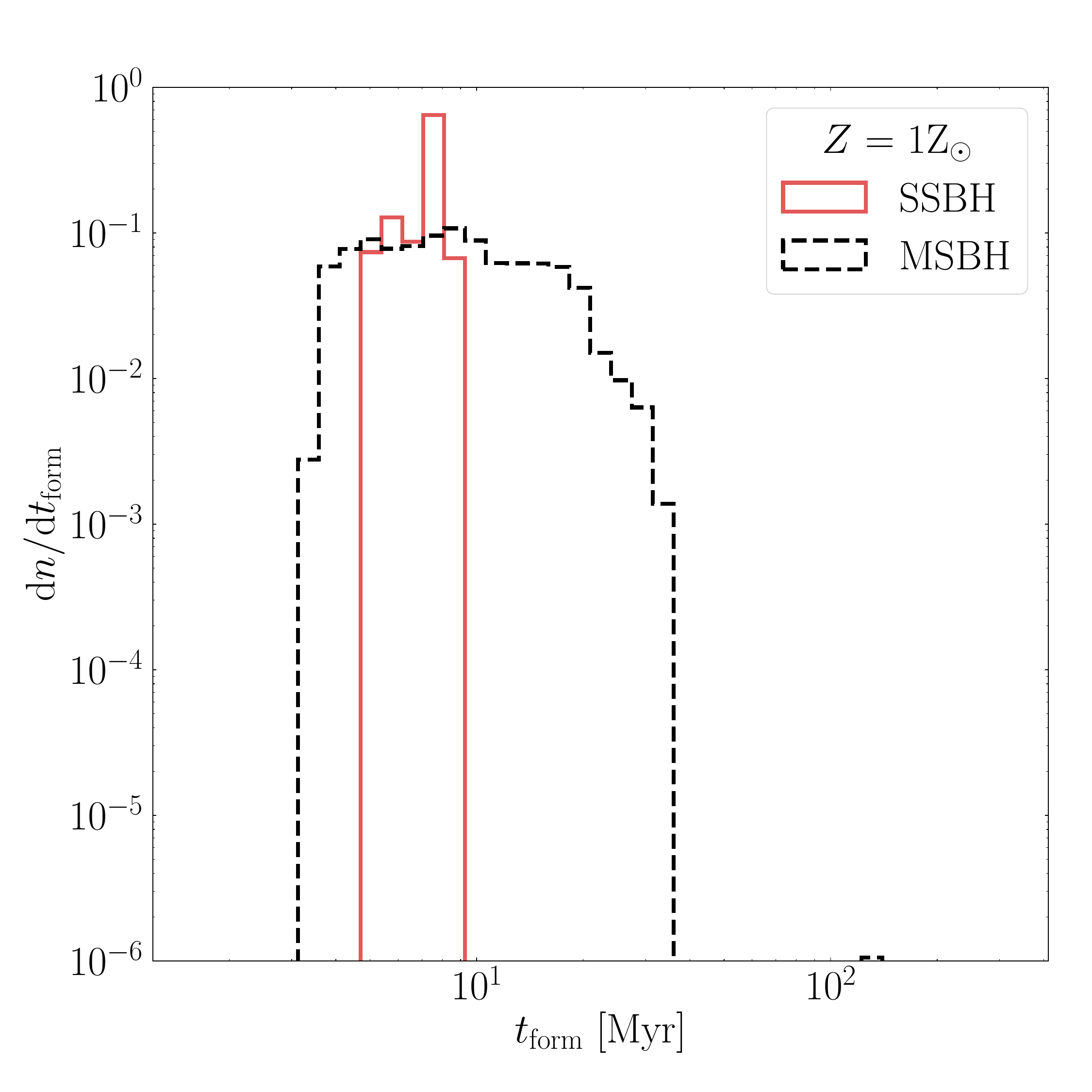}\\
    \includegraphics[width=0.8\columnwidth]{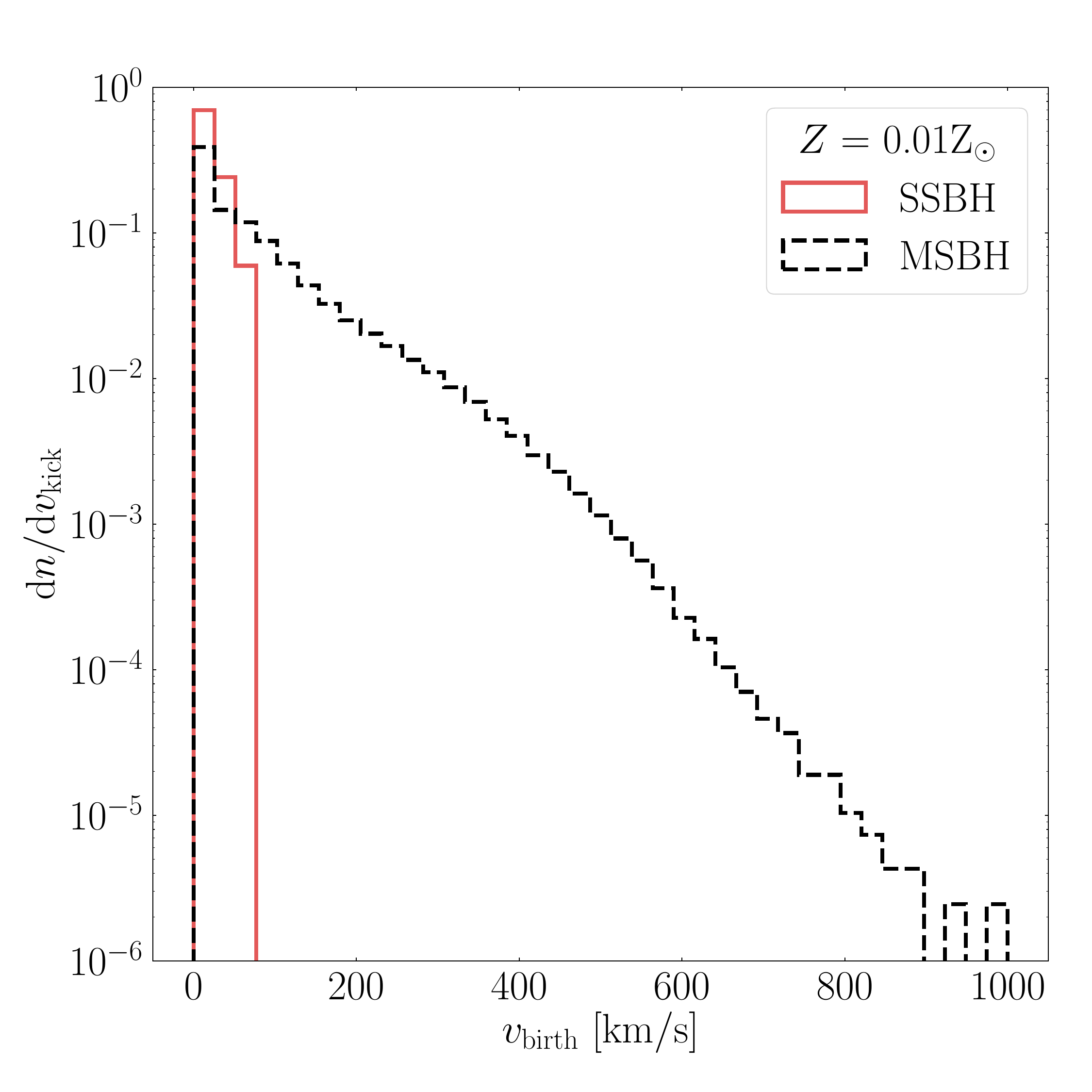}
    \includegraphics[width=0.8\columnwidth]{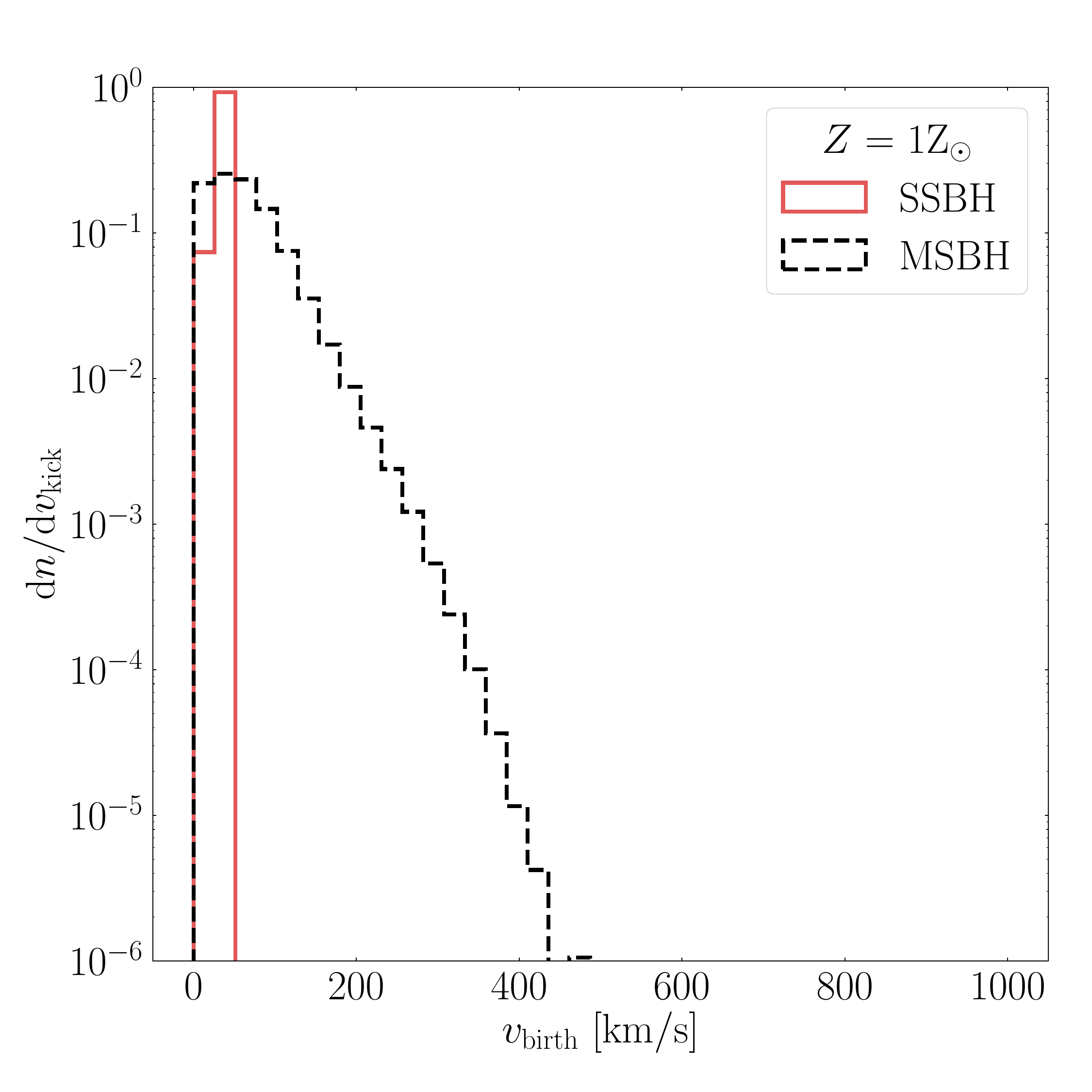}\\
    \caption{From top to bottom: mass, formation time, and natal kick distribution of BHs extracted from a single  (SSBH, red histograms) or mixed stellar BH population (MSBH, black histograms).}
    \label{fig:bhspect}
\end{figure*}

Figure~\ref{fig:bhspect} compares the distribution of BH mass, formation time, and natal kick amplitude for SSBH and MSBH mass spectra, assuming a metallicity $Z = (0.01-1)$ Z$_\odot$. In general, BHs from a MSBH spectrum exhibit a broader mass spectrum, with a clear peak at $M_{\rm BH} < 10\Ms$. At solar metallicity, MSBH produces a small fraction of BHs with masses in the $40-90\,{}\Ms$ mass range which cannot be covered with a simple SSBH.
Two additional features of the MSBH spectrum are a broader distribution of formation times, owing to the time needed for the stellar binary to coalesce and form the BH, and a broad distribution of natal kick velocities which extend up to $v_{\rm kick} \sim{500-800}$ km$/$s. 
The high-tail of the distribution owes to the adopted prescription for compact object natal kicks, which is described in \cite{giacobbo20}:
\begin{equation}
V_{\rm kick}=f_{\rm H05}\,{}\frac{\langle{}M_{\rm NS}\rangle{}}{M_{\rm rem}}\,{}\frac{M_{\rm ej}}{\langle{}M_{\rm ej}\rangle},
\end{equation}    
 where $\langle{}M_{\rm NS}\rangle{}$ and $\langle{}M_{\rm ej}\rangle$ are the average NS mass and ejecta mass from single stellar evolution, respectively, while $M_{\rm rem}$ and $M_{\rm ej}$ are the compact object mass and the ejecta mass \citep{giacobbo20}. The term $f_{\rm H05}$ is a random number drawn from a Maxwellian distribution with  one-dimensional root mean square $\sigma_\mathrm{kick}=265 \ \mathrm{km}\,{} \mathrm{s}^{-1}$.  According to this prescription, devised to match the proper motions of young Galactic pulsars \citep{hobbs} and to enforce low kicks for stripped and ultra-stripped supernovae \citep{tauris2017}, the magnitude of the kick scales with the ratio of the amount of ejected mass in the supernova event and the compact object final mass. The high kicks visible in Figure \ref{fig:bhspect} are generally associated with BHs lighter than $m_{\rm BH} \leq{} 5\,{}\Ms$, formed from progenitors that retained their H-rich envelope until the onset of core collapse.

\begin{figure}
    \centering
    \includegraphics[width=\columnwidth]{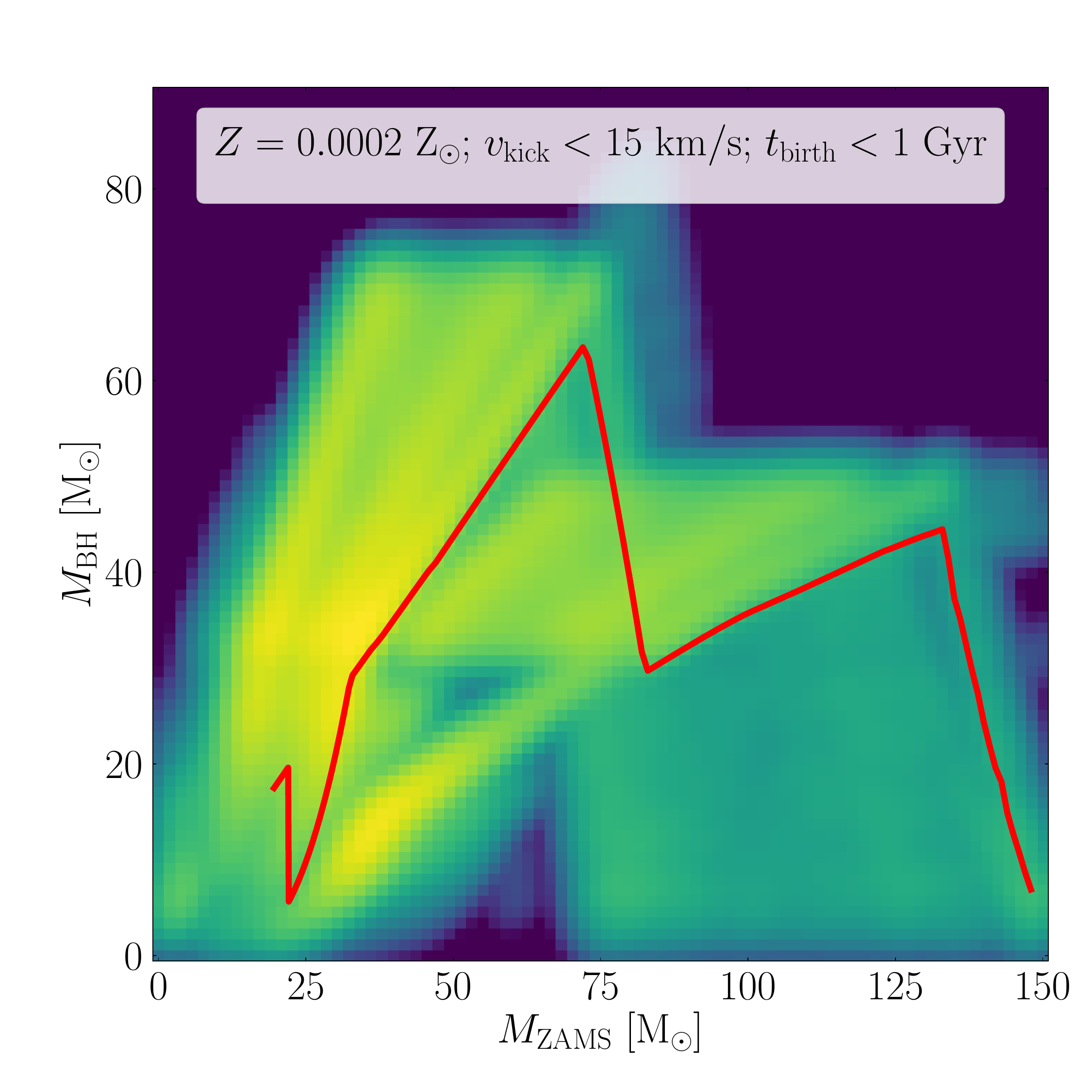}
    \caption{ZAMS--remnant mass distribution for MSBH (surface map) and SSBH (red line) spectra, assuming a metallicity $Z=0.0002$ and limiting the sample to BHs with a formation time $t_{\rm birth} <1$ Gyr and a natal kick $v_{\rm kick} < 15$ km$/$s.}
    \label{fig:msbh}
\end{figure}

Figure~\ref{fig:msbh} compares the ZAMS--remnant mass relation for SSBH and MSBH spectra assuming a metallicity $Z = 0.0002$ and limiting the sample to BHs with a natal kick smaller than $15$ km$/$s and a formation time $<1$ Gyr, i.e. typical values of GC velocity dispersion and half-mass relaxation time, respectively.
In the SSBH case, BHs with masses $<20\Ms$ can be produced only in a relatively narrow range of ZAMS masses around $25\,{}\Ms$ and around $150\,{}\Ms$, with the latter region disfavoured by the stellar initial mass function. Conversely, such light BHs can be produced in a wide range of ZAMS masses in the MSBH case, i.e. $M_{\rm ZAMS}=25-50\,{}\Ms$, thus highlighting how stellar binary mergers could impact the overall BH mass distribution. 

In this regards, SSBH and MSBH represent two limiting cases: the former describes clusters in which the evolution of primordial stellar binaries is irrelevant (either because binaries are not present or quickly destroyed by strong dynamical interactions), and the latter describes clusters containing a large population of primordial stellar binaries. Varying the mutual fraction of BHs coming from one channel or another can help us quantify the actual role of stellar binary mergers in sculpting the overall BH mass spectrum.

One additional pathway that can contribute to BH formation in star clusters is via stellar collisions and stellar accretion onto ``normal'' BHs, a process that can trigger the formation of IMBH seeds as massive as $100-500\Ms$ \citep{dicarlo19, arca20c, rizzuto21, gonzalez21}. Hereafter, we label the mass spectrum associated to IMBH seeds as {\emph{heavy BH mass spectrum}} (HSBH).

Although many efforts have been made toward a better comprehension of how BHs evolve in star clusters, it is still unclear whether typical BH populations in stellar systems are mostly dominated by a SSBH mass spectrum, or if the MSBH and HSBH spectra play a significant role in determining BH pairing and merger. Following a rather agnostic approach, we regulate the amount of BHs extracted from the SSBH, MSBH, or HSBH mass spectra via two parameters: the mixing fraction ($f_{\rm mix}$) and the seed formation probability ($f_{\rm seed}$). In practice, we assume that the whole population of BHs is composed of $(1-f_{\rm mix}-f_{\rm seed})$ BHs from the standard single BH mass spectrum, $f_{\rm mix}$ from processed primordial binaries, and $f_{\rm seed}$ BHs byproduct of repeated stellar collisions.

\subsection{Formation and delay times, metallicity distribution, and BH natal kicks}
\label{sec:form}

In \bpop, every BBH merger is characterised by at least two main time scales: 
the formation time $t_{\rm for}$, namely the time at which the two BH progenitors formed, and the delay time $t_{\rm del}$, namely the time elapsed from the binary formation to the BBH coalescence. 

Whilst the sum $t_{\rm for}+t_{\rm del}$ determines the cosmic time at which the merger takes place, the formation time alone represents a crucial quantity to determine the most likely value of the metallicity of the merger host environment. 

For isolated binaries and BHs forming in YCs, we extract a formation redshift from the cosmic star formation rate inferred by \cite{madau17}:
\begin{equation}
\psi(z)= \frac{0.01(1+z)^{2.6}}{1+[(1+z)/3.2]^{6.2}} ~\Ms~{\rm yr}^{-1}~{\rm Mpc}^{-3},
\end{equation}
whereas for BHs in GCs and NCs we extract the formation redshift from the GCs formation rate described in \cite{katz13}, which is nearly flat in the redshift range $z \simeq 2-6$. We convert the formation redshift into a cosmic time assuming the set of cosmological parameters provided by \cite{Planck15}, namely $H_0 = 67.74$ km~s$^{-1}$~Mpc$^{-1}$, $\Omega_{\rm M} = 0.3089$, $\Omega_\Lambda = 0.6911$.

The delay time is calculated either directly from \mobse\ for isolated binaries, or is inferred from the typical timescales of dynamical binary formation and mergers, as detailed in the next section. Following our previous paper \citep{arca20}, we assume that at redshift $z=2$ the metallicity of GCs and NCs can be described by a lognormal distribution limited between $Z = 0.0005 - 0.001$, i.e. the typical range of values observed in Galactic GCs, whilst for galaxies and YCs we adopt the metallicity distribution derived from the SDSS \citep{Gallazzi2006}, which is measured at redshift $z\simeq 0$. 

To take into account the fact that, on average, the larger is the redshift the lower the metallicity, we shift the metallicity distribution by a  redshift-dependent factor, $\delta Z$, defined as \citep[see e.g.][]{zevin21, bavera20}
\begin{equation}
{\rm Log}\left({\delta Z/{\rm Z}_\odot}\right) = -0.074z^{\alpha_Z},
\label{eq:zfac}
\end{equation}
with $\alpha_Z = 1.34$ in the case of isolated binaries \citep{bavera20} and $\alpha_Z = 1.2$ for star clusters. 

Moreover, it has been shown that the merger efficiency decreases with the environment metallicity in both the isolated and dynamical fields \citep[see e.g.][]{giacobbo18a,rastello2021, santoliquido21}, thus we weight the metallicity distribution with the probability that a merger occurs in an environment with given metallicity. Following our previous work \citep{arca20}, we model the weighting function with a power-law $p^\beta$ with slope $\beta = -1.5$. 

These assumptions imply that it is more likely for BBHs to develop at higher redshift, where the average metallicity is lower, and thus their merger probability is higher.

\subsection{Dynamical binaries and hierarchical mergers}
\label{dynBBH}

In \bpop, we extract cluster masses and half-mass radii from the observed distribution of GCs \citep{harris14} and NCs \citep{georgiev16}. For Galactic YCs, the dynamical mass is known for a handful clusters only, whilst the half-mass radius is easier to measure \citep{portegies2010}. For this reason, to select YC half-mass radii we adopt the observed distribution while for the mass we adopt the GC mass distribution lowered by 2.5 dex, as suggested by observed YCs in the MW and its satellites \citep{portegies2010, Gatto21}. 
Note that for YCs, these choices imply that the mass distribution peaks at around $10^4\Ms$, with the high-end tail extending up to $10^5\Ms$. The cumulative distribution of YCs obtained this way implies that the probability to have a YC with a mass $> 10^4\Ms$ is $50\%$. 
The global distribution of masses and half-mass radii for different cluster types is reconstructed using the Python built-in Gaussian kernel density estimator.

Cluster models are described by either a \cite{Plum} density profile or a power-law distribution with slope $\gamma$ \citep{Deh93}. This is crucial to determine the cluster central density and escape velocity. 

\begin{figure*}
    \centering
    \includegraphics[width=0.8\textwidth]{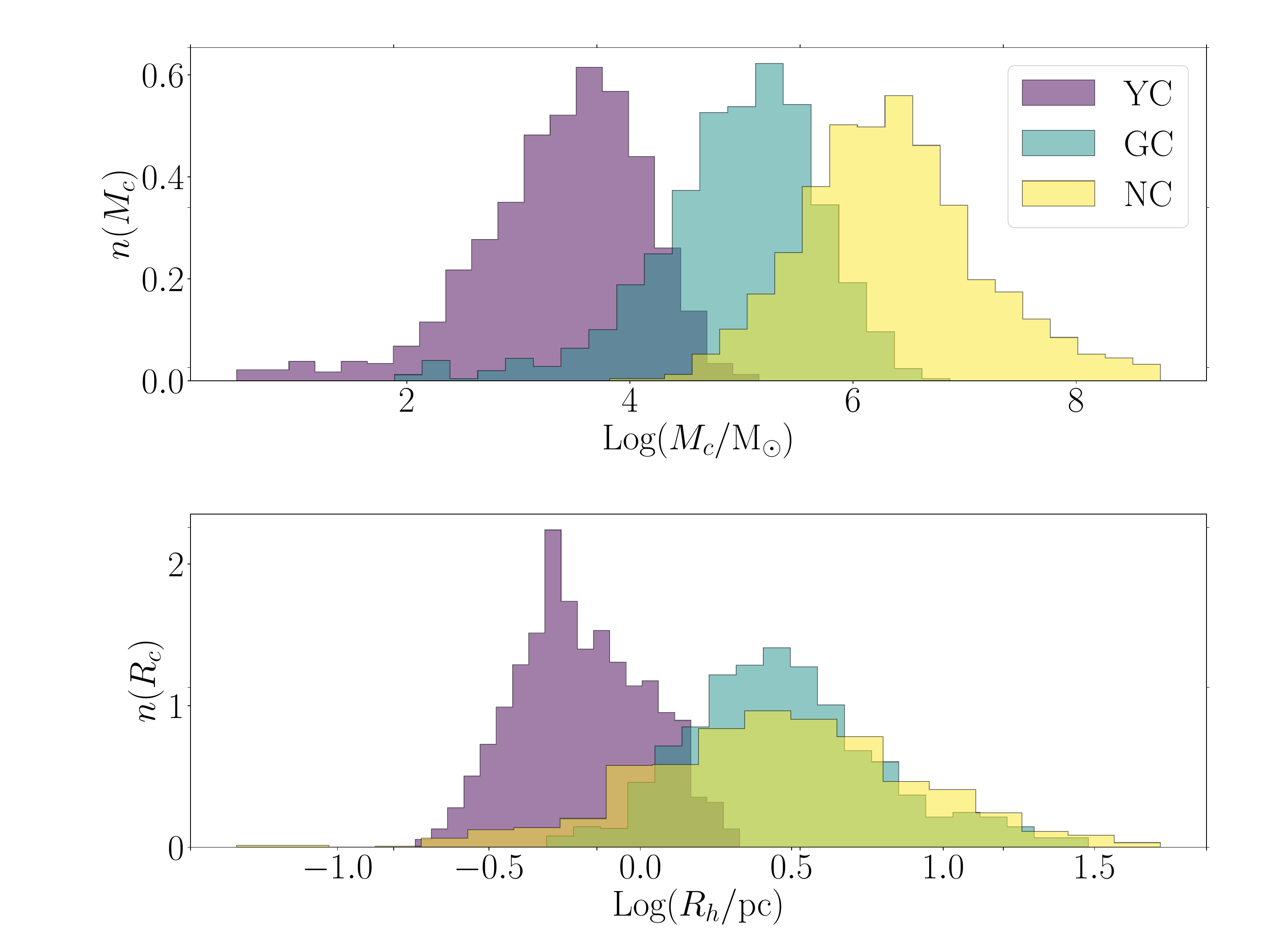}
    \caption{Distribution of mass (top panel) and half-mass radius (bottom panel) for YCs, GCs, and NCs. Data are taken from \citet{harris14} for GCs, \citet{georgiev16} for NCs, and \citet{portegies2010} for YCs. Given the scarcity of available data for YCs, we assume that the mass distribution of YCs is equal to the one for GCs but shifted by 2.5 dex \citet[e.g. see][]{}.}
    \label{fig:clprop}
\end{figure*}

For YCs and GCs, we assume that the cluster has a probability of $50\%$ to be described by a \cite{Deh93} sphere, and in such a case we adopt a value for the inner density slope of $\gamma = 1(1.5)$ for YCs(GCs). This choice enables us to explore how the matter distribution and the presence of a density cusp in the cluster innermost region could affect BBH formation. Otherwise, we adopt a \cite{Plum} sphere to model the cluster\footnote{It has been shown that \cite{Plum} closely resemble \cite{king62} models with an adimensional potential well $W_0\simeq 6$ \citep{aarseth2008}, which have been extensively used to fit the observed properties of Galactic GCs \citep{harris14}.}. For NCs, instead, we adopt a \cite{Deh93} model with $\gamma = 1.9$. 
According to this choice, the central escape velocity is thus determined as:
\begin{align}
    v_{\rm esc}^2  =\frac{2GM_c}{r_h}\times
    \begin{cases}
     1.3 & {\rm Plummer}, \\
     {\displaystyle [2^{1/(3-\gamma)}-1](2-\gamma)^{-1}} & {\rm Dehnen}.
    \end{cases}	
\end{align}
We note that in the case of YCs, the choice $\gamma=1$ 
leads to an escape velocity in \cite{Deh93} models $\sim 30\%$ smaller than in \cite{Plum} models -- at fixed value of $r_h$, while in the case of GCs, where $\gamma = 1.5$, $v_{\rm esc}$ is practically the same in both models. The choice of a flat (Plummer) or a cusp (Dehnen) density profile aims at capturing the different phases of cluster life, as early evolution, mass segregation, and binary formation can significantly affect cluster matter distribution.
If the remnant BH of a dynamical merger receives a kick $v_{\rm kick}<v_{\rm esc}$, we check whether the BH remnant can undergo one (or more) further mergers within a Hubble time. In case of multiple generation mergers we follow the evolution of the BH remnant until either: a) $v_{\rm kick}>v_{\rm esc}$, b) the number of BHs in the innermost region of the cluster is fully consumed in repeated mergers, c) the total delay time exceeds 14 Gyr. Figure \ref{fig:clprop} shows the mass and half-mass radius distribution for different types of cluster. 

We assume that the number of BHs participating in one or multiple mergers in star clusters is comparable to the number of BHs lurking inside the cluster scale radius, which is directly connected to the cluster half-mass radius:
\begin{equation}
    n_{\rm bhs} = f_{\rm bh} \,{}f_{\rm reten} \,{}f_{\rm encl}\,{} M_c,
\end{equation}
where $f_{\rm bh}=0.0008$ is the fraction of cluster mass in stellar BHs according to a \cite{kroupa01} initial mass function, $f_{\rm reten} = 0.5$ represents the fraction of BHs retained in the cluster
\citep[see e.g.][]{morscher15,AAG18a}, and $f_{\rm encl}$ is the fraction of cluster mass enclosed within the cluster typical radius.

As discussed in Section \ref{sec:bbhmass}, the BH masses are extracted from an SSBH, MSBH, or HSBH mass spectrum (depending on the user's choice), provided that the natal kick of the BH is smaller than the cluster escape velocity. As we will discuss in the following, once the BH masses are selected the code calculates the overall time needed for the BH to pair, harden, and merge, selecting only binaries with a total delay time shorter than the Hubble time.

In \bpop, the evolutionary timescales of dynamical BBHs are linked to the cluster evolution as follows.
For each BBH, we select a cluster formation time ($t_{\rm for}$) as explained in the previous section. We thus extract the formation time ($t_{\rm birth}$) from our \mobse\ catalogues and calculate the time over which the two BHs are expected to migrate into the cluster centre, i.e. the dynamical friction timescale $t_{\rm df}$ \citep[see e.g.][]{bt}
\begin{equation}
t_{\rm df} = 0.42 {\rm Gyr}  \left( \frac{10 m_*}{m_{\rm CO}} \right)  \left(\frac{r}{r_h}\right)^{1.76} \left(\frac{t_{\rm rel} }{4.2{\rm ~Gyr}}\right),	  
\end{equation}
where $m_*$ is the average mass of stars in the environment, $m_{\rm CO}$ is the mass of the heavy object (either single or binary), $r$ is its position, and $t_{\rm rel}$ is the relaxation time:
\begin{equation}
t_{\rm rel} = 4.2 {\rm ~ Gyr}  \left(\frac{15}{\log \Lambda}\right) \left(\frac{r_h}{4{\rm pc}}\right)^{3/2} \sqrt{\frac{M_c}{10^7\Ms}},		
\end{equation}
with $\log\Lambda$ the Couloumb logarithm.

Although generally $t_{\rm birth}$ is short compared to the dynamical friction timescale, some combination of  the stellar masses involved, the location of the stellar progenitors inside the cluster, and the cluster properties can lead to $t_{\rm df} < t_{\rm birth}$, especially in the case of young clusters or massive stars formed close to the innermost cluster regions. Thus, to calculate the total time over which a BH forms and segregates to the host cluster centre we add to the cluster formation time the maximum between BH birth time and the dynamical friction timescale.
Once they reach the core,  the BHs can pair up via different processes, the most likely being via three-body interactions, a process involving three unbound objects, which operates over a timescale
$t_{\rm 3bb}$ \citep{lee95} 
\begin{align}
t_{\rm 3bb} =& 4 {\rm ~Gyr} \left(\frac{10^6 {\rm \Ms~pc^{-3}}}{\rho_c}\right)^2 \left(\zeta^{-1}\frac{\sigma_c}{30{\rm ~km/s}}\right)^9 \times \nonumber \\
& \times \left(\frac{m_*}{m_{\rm CO}} 10\right)^{9/2} \left(\frac{10}{m_{\rm CO}}\right)^{-5}.
\label{t3bb}
\end{align}
Another possible BBH formation mechanism requires that BHs are captured in already existing stellar binaries, i.e. the so-called binary-single capture process, over a timescale
$t_{\rm bs}$ \citep{miller09,antonini16b}
\begin{align}
t_{\rm bs} =& 3{\rm ~Gyr} \left(\frac{0.01}{f_b}\right) \left(\frac{10^6{\rm~pc^{-3}}}{n_*}\right) \times \nonumber \\
& \times \left(\frac{\sigma_c}{30{\rm~km/s}}\right) \left(\frac{10\Ms}{a_h(M_1 + M_2 + m_p)}\right),
\end{align}
and the subsequent interaction of binaries containing at least one BH leads to the formation of a hard BBH over a timescale \citep{antonini16b}
\begin{equation}
    t_{\rm ex} = t_{\rm bs} \left(10\frac{m_*}{m_{\rm CO}}\right)^{1/2}.
\end{equation}
The capture process is a direct consequence of the fact that a binary with component mass $m_{1,2}$, semimajor axis $a$, and eccentricity $e$ that travels in an environment with density $n$ and typical velocity dispersion $\sigma$ will undergo strong binary-single scatterings at a rate
\begin{equation}
    \dot{R}_{\rm bs} = n\,{}\sigma\,{} \pi\,{} a^2\,{}(1-e)^2\,{}\left[1+\frac{G\,{}(m_1+m_2+m_p)}{a\,{}\sigma^2}\right].
\end{equation}

In the equations above, $\rho_c$($n_c$) and $\sigma_c$ represent the cluster matter(number) density and velocity dispersion, $\zeta\leq 1$ is a parameter representing the level of energy equipartition among the heavy and light population of stars -- we assume $\zeta = 1$ in our calculations, $f_b$ is the binary fraction, $m_p$ is the typical mass of stellar perturbers, and 
\begin{equation*}
a_h \simeq 59 {\rm AU} \left(\frac{M_1+M_2}{30\Ms}\right)\left(\frac{30{\rm km/s}}{\sigma_c}\right)^2
\end{equation*} 
is the hard binary separation. 

The actual value of each timescale has been selected from a Gaussian distribution peaking at the nominal value and assuming a dispersion of $10\%$. This choice takes into account the uncertainties in  cluster mass and half-mass radius, the cosmic star formation history, and the small scale physics regulating star formation in the galactic field.
We assume that a hard BBH is formed after a total time $t_{\rm for} + {\rm max}(t_{\rm birth}, t_{\rm df}) + {\rm min}(t_{3bb},t_{\rm bs}) + t_{\rm ex}$.

The further evolution of the binary is regulated through the binary-single interaction timescale $t_{1-2}$ \citep{gultekin04,antonini16}
\begin{align}
t_{1-2} =& \frac{0.02{\rm Gyr}}{\zeta}  \left(\frac{10^6 {\rm~pc^{-3}}}{n_c}\right) \left(\frac{\sigma_c}{30}\right) \times \nonumber \\
& \times \sqrt{\frac{10 m_* }{ M_1+M_2}}  \left(\frac{0.05{\rm AU} }{ a_h}\right) \left(\frac{20}{M_1+M_2}\right),
\label{t12}
\end{align}
which is the typical timescale in which the binary spends most of its lifetime. 

At this stage, we calculate two critical values of the BBH semi-major axis ($a$), namely the maximum value below which the binary gets ejected via further interactions ($a_{\rm ej}$) and the maximum value below which GW emission dominates the evolution ($a_{\rm gw}$) \citep{antonini16}:
\begin{align}
\label{eq:ejorin}
a_{\rm ej} =& 0.07 {\rm AU} \frac{\mu\,{} m_p^2 }{ (M_1 + M_2 + m_p)(M_1+M_2)} \left(\frac{v_{\rm esc}}{50{\rm km/s}}\right)^{-2}, \\
\label{eq:ejorin2}
a_{\rm gw} =& 0.05 {\rm AU} \left(\frac{M_1+M_2}{20\Ms}\right)^{3/5} \frac{ (M_2/M_1)^{1/5} }{ \left(1+M_2/M_1\right)^{2/5} }\times \nonumber \\
&\times \left(\frac{\sigma_c}{30{\rm km/s}}\right)^{1/5} 
\left(\frac{10^6{\rm \Ms pc^{-3}} }{\rho_c}\right)^{1/5},
\end{align}
where $\mu=M_1\,{}M_2/(M_1+M_2)$ is the reduced mass of the binary system.
If the BBH has $a_{\rm ej} > a_{\rm gw}$, the binary will be 
ejected and merge outside the cluster over a GW timescale ($t_{\rm GW}$), calculated according to \cite{peters64}. Otherwise, the BBH merger will be mediated by three-body encounters over a timescale $t_{\rm GW3} \simeq 5(M_1 + M_2)/m_p\,{} t_{1-2}$ \citep{miller02,antonini16b}.

If the sum of all timescales above is larger than a Hubble time ($t_{\rm H}$) we extract another BBH and recalculate all the relevant times until the total delay time is shorter than $t_{\rm H}$. If we do not find a suitable BBH in 10,000 tries, we pass to the next cluster model and label the cluster as {\emph{ merger-free}}. In this way, when the threshold is hit, we can place an upper limit on the probability ($<1/10,000$) to form a 1st generation merger in a cluster with a given mass, radius, and formation time and, at the same time, maintain a reasonable computational cost. Note that the extraction of 100,000 BBHs requires between 1 and 4 minutes on a single CPU, depending on the adopted parameters. 
In dynamical-only models (i.e. ID2, 3, 4) we find $\sim 25$ {\it merger-free} clusters out of 100,000, which generally are relatively light, $M_{\rm cl}<10^4\Ms$, forming at low redshift.

If the sum of all timescales above is shorter than a Hubble time, the BBH is labelled as a merger and the associated GW recoil kick is calculated. If the BBH merges inside the cluster and the GW kick is larger than $v_{\rm esc}$, the remnant is ejected from the cluster and the merger chain is halted, otherwise the remnant is displaced from the centre to a maximum distance $r_{\rm d} = r_h \sqrt{v_{\rm esc}^4 / (v_{\rm esc}^2-v_{\rm gw}^2)^2 - 1}$ \citep[see e.g.][]{antonini19,fragione19}. In the latter case, the remnant is assumed to come back to the cluster centre over a dynamical friction time $t_{\rm df}$, and to form a hard binary over a $t_{\rm bs}$ timescale. The whole procedure is repeated until either the reservoir of stellar BHs in the cluster centre is emptied, or the remnant is ejected from the cluster, via dynamical interactions if $a_{\rm ej} > a_{\rm gw}$ or GW kick if $v_{\rm kick} > v_{\rm esc}$.  

Whilst binaries form and evolve, their parent clusters evolve as well. The cluster evolution can be driven by internal (stellar evolution, mass-segregation, relaxation) and external (galactic tidal field, collision with giant molecular clouds) processes that progressively lead to the cluster evaporation and expansion. Mass loss and core expansion cause a dramatic decrease of the cluster density and central velocity dispersion, thus affecting the rate of three-body and binary-single interactions and, overall, the possible formation of first- and multiple-generation mergers. 

We model the cluster evolution following two different prescriptions. The first one is based on recent $N$-body simulations of young massive clusters with masses in the $(0.6-3.5\times10^5)\Ms$ range (Arca Sedda et al, in prep.):
\begin{eqnarray}
M_{\rm cl}(t) &=& M_{\rm cl,0}[1 + t/(0.1t_{\rm rel} (R_{\rm cl,0}/1{\rm pc})^{-3/2}]^{0.1},\\
R_{\rm cl}(t) &=& R_{\rm cl,0}[1+t/(0.45t_{\rm rel})]^{0.4},
\end{eqnarray}
which are tailored on metal-poor, massive and dense clusters with a high binary fraction ($\sim 30\%$). The second is based on well established semi-analytic and theoretical prescriptions that are based on the effect of two-body relaxation process in simple cluster models with a monochromatic mass spectrum \citep{cohn80, goodman84, gnedin99, bt}:
\begin{eqnarray}
M_{\rm cl}(t) &=& M_{\rm cl,0}\,{}\exp(t/(\xi t_{\rm rel})),\\
R_{\rm cl}(t) &=& R_{\rm cl,0}\,{}[1+t/(\xi t_{\rm rel})]^{2/3}.
\end{eqnarray}

The $N$-body recipe causes a faster mass-loss, reducing the cluster mass by $80\%$ in 10 relaxation times. Conversely, standard theoretical prescriptions predict a more gentle decline in mass, leading a mass-loss of around $20\%$ in 10 relaxation times.

We adopt a recursive method to account for  cluster evolution, dividing the binary formation, evolution, and merger in four main phases:
\begin{itemize}
\item BH formation and segregation time, i.e. $t_1 = {\rm max}(t_{\rm df},t_{\rm birth})$;
\item hard binary formation time, i.e. $t_2 = {\rm min}(t_{\rm 3bb}, t_{\rm bs})$;
\item binary hardening and merger, i.e. $t_3 = t_{1-2} + t_{\rm GW}$.
\end{itemize}
At the end of each phase, we update the cluster mass and radius and calculate the timescales during the next phase using the updated values. 

The procedure is repeated until either the merger remnant is ejected from the cluster or the cluster mass falls below $10\,{}\Ms$. 
While the evolution of the cluster can be quite relevant for YCs and GCs, it may be less effective for NCs since they are well embodied in the host galaxy potential well and thus their evolution should be self-regulated by the interactions with the galactic environment. 

To explore the effects of the cluster evolution onto the BBH population we run two additional model sets (ID 14, 15, and 16) entirely focused on GCs adopting the cluster evolution scheme described above. The results of this additional models are described in Section \ref{sec:GCevo}.

\subsection{Black hole natal spins}

One of the most debated aspects of stellar BH formation and pairing is the actual distribution of natal spins. Hereafter, we refer to the dimensionless, or Kerr, spin parameter, $\chi_{i} = c\,{}J_i/G\,{}M_i^2$, where $J_i$ represents the amplitude of the BH angular momentum, and $i=1,2$ for the primary and secondary BH, respectively.
Observations of BHs in low- and high-mass X-ray binaries suggest that BHs have large natal spin, up to 0.9 \citep[see e.g.][]{qin19}, whilst \cite{fuller19} suggest that efficient angular momentum transport triggered by the Tayler-Spruit dynamo \citep{spruit02} can lead to spins as low as 0.01 in BHs born from single stars. In this framework, the population of detected BBH mergers hints at a spin distribution for merging BHs attaining relatively low values, $\chi \simeq 0.02$ \citep{gwtc2}.

Given these uncertainties, in \bpop\ we allow for different choices. Throughout the paper, we extract spins in the range $0-1$ and explore four different cases:
\begin{itemize}
\item spins are drawn from a Gaussian distribution centered on $\chi = 0.5$ with dispersion $0.1$ (high spin model, denoted with GSS and letter H);
\item spins are drawn from a Gaussian distribution centered on $\chi = 0.2$ with dispersion $0.1$ (low spin model, denoted with GSS and letter L);
\item spins are drawn from a Maxwellian with dispersion 0.2 (model denoted with MXL);
\item spins are set to 0.01 according to \cite{fuller19} (model denoted with FM19);
\item spins are set to 0.01 for dynamical BHs whilst extracted from a Maxwellian, with dispersion 0.2, or a Gaussian centered on $\chi=0.5$ distribution for  BHs in isolated mergers (model denoted with FM19+MXL and FM19+GSS).
\end{itemize}

\subsection{Observational biases}

Several parameters can affect the probability to detect BBH mergers. Among others, the distance at which the merger takes place, the direction of the GW that hits the detector, and the binary orbital parameters. For ground-based detectors like LIGO and Virgo, the accessible cosmological volume $VT$ depends on the primary mass via a power-law $\propto M_1^{2.2}$, at least in the $10<M_1/\Ms<100$ mass range, and increases for increasing binary mass ratio \citep{fishbach17}. The volume-mass ratio dependence can also be described by a power-law in the form $\propto M_1^\beta$, with $\beta = 0.47-0.72$ depending on the primary mass \citep[see Figure 7 in][]{arca21}.

In \bpop, we first create a sample of BBH mergers following the method described in the previous sections, and then we sample ``mock'' observations of BBHs exploiting the $VT-M_1$ and $VT-q$ relations as selection criteria \citep[see also][]{arca20b}.   

Additionally, we require that mock BBHs happen at a redshift $z<2$, i.e. close to the maximum distance reachable with LIGO at design sensitivity \citep{abbott16detector, abbott20}.

Although rather crude, this approach enables us to study both the overall population of mergers forming in isolation or dynamically, and the sub-population of mergers that might be accessible with second-generation GW detectors.

\subsection{The reference model}
Our reference model has the following features: 
\begin{itemize}
\item Fraction of dynamical mergers (compared to the total): $f_{\rm dyn} = 0.5$;
\item Fraction of mergers coming from YCs, GCs, NCs: $f_{\rm YC, GC, NC} = 1/3$;
\item BBH merger formation time selected according to:
    \begin{itemize}
        \item the cosmic SFR from \cite{madau17} for isolated binaries and dynamical binaries in YCs, 
        \item the cluster formation rate from \cite{katz13} for GCs and NCs;
    \end{itemize}
\item Galaxy and YC metallicity distribution adapted from \cite{Gallazzi2006};
\item GC and NC metallicity distribution is assumed flat in logarithmic values as discussed in our previous work \citep{arca20} and similar approaches \citep[see e.g.][]{bavera20,zevin21};
\item As detailed in Section \ref{sec:form} above, the metallicity distribution is conveniently rescaled via the redshift-dependent factor shown in Equation \ref{eq:zfac};
\item We weight the metallicity distribution with the probability for a BBH to merge in an environment with a given metallicity, assuming a power-law with slope $-1.5$\footnote{The choice of a power-law with slope $-1.5$ returns results consistent with results from isolated binaries \citep[e.g.][]{giacobbo18a} and star cluster simulations \citep[e.g.][]{askar17}, as shown in our previous paper \citep{arca20}};
\item Dynamical BBH masses are extracted from SSBH and MSBH mass spectra assuming a mixing fraction of $f_{\rm mix} = 0.5$, whilst we neglect the contribution of IMBH seeds (thus $f_{\rm seed} = 0$);
\item BH spins are extracted from a Gaussian distribution centred on $\chi = 0.2$ with dispersion 0.1 truncated between 0 and 1; 
\item The polar angle $\theta$ between BH spins and the BBH angular momentum is extracted from:
\begin{itemize}
    \item a uniform distribution in $\cos\theta$ 
in the case of dynamical binaries, 
    \item the cumulative distribution $P_\theta = [(\cos \theta + 1)/2]^{n_\theta+1}$ \citep{arca19b} for isolated binaries. In this case we adopt $n_\theta = 8$, which implies $20(55)\%$ of binaries having $\theta_{1,2}$ values that differ by less than $5(20)\%$; 
\end{itemize}
\item The angle $\phi$ between the BH spin vectors is assigned assuming a uniform distribution in $\cos\phi$.
\end{itemize}

To explore the parameter space, we create different models varying the BH spin distribution, the fraction of dynamical mergers, the impact of the single BH mass spectrum adopted, and the role of IMBH seeds in determining the observed BH mass spectrum, as detailed in the next sections. For each model, we create a database of $100,000$ BBH mergers from which we select mergers happening at a redshift $z<2$ according to the adopted observational selection criteria. The selection, based on the acceptance-rejectance method, returns a sub-sample of around $7,000$ {\it mock} mergers per model.

\section{Results}
\label{sec:res}

In this section we present the main features of BBH mergers from the reference model and discuss how they compare with LVK data in terms of global properties, primary mass distribution, and effective spin parameters. 

\begin{table*}
\centering
    \begin{tabular}{c|cc|ccc|cc|cc|cc|ccc}
         \hline
         \hline
         {\bf ID} & \multicolumn{2}{c|}{{\bf Channel}} & \multicolumn{3}{c|}{{\bf Dynamics}} & \multicolumn{2}{c|}{{\bf Metallicity}} & \multicolumn{2}{c}{\bf OBS} &\multicolumn{2}{c}{{\bf Spins}} &
         \multicolumn{3}{|c}{\bf BHMF}\\
         \hline
         & $f_\iso$ & $f_\dyn$ & $f_\gc$ & $f_\nc$ & $f_\yc$ & iso+YC & GC+NC & $\alpha_{M_1}$ & $\alpha_{q}$ & $P(a_1)$ & $n_{\theta}$ &
         SSBH & MSBH & HSBH  \\
        \hline
        \multicolumn{15}{c}{\bf Reference model}\\
        \hline
        0(H/L) & 0.5 & 0.5 & 0.33 & 0.33 & 0.33 & SDSS & LOG & 2.2 & 0.4-0.7 & GSS(H/L) & 8 & 0.5& 0.5& 0 \\
        \hline
        \multicolumn{15}{c}{\bf Isolated channel}\\
        \hline
        1(H/L) & 1 & 0 & - & - & - & SDSS & - & 2.2 & 0.4-0.7 & GSS(H/L) & 8 & -& -& - \\
        \hline
        \multicolumn{15}{c}{\bf Dynamical channel}\\
        \hline
        2(H/L) & 0 & 1 & 0.33 & 0.33 & 0.33 & - & LOG & 2.2 & 0.4-0.7 & GSS(H/L) & - & 1& 0& 0 \\
        3(H/L) & 0 & 1 & 0.33 & 0.33 & 0.33 & - & LOG & 2.2 & 0.4-0.7 & GSS(H/L) & - & 0.5& 0.5& 0 \\
        4(H/L) & 0 & 1 & 0.33 & 0.33 & 0.33 & - & LOG & 2.2 & 0.4-0.7 & GSS(H/L) & - & 0& 1& 0\\
        \hline
        \multicolumn{15}{c}{\bf Spins amplitude and alignment}\\
        \hline
        5 & 0.5 & 0.5 & 0.33 & 0.33 & 0.33 & SDSS & LOG & 2.2 & 0.4-0.7 & MXL & 8 & 0.5& 0.5& 0 \\
        6 & 1 & 0 & - & - & - & SDSS & - & 2.2 & 0.4-0.7 & MXL & 0 & -& -& - \\
        7(H/L) & 1 & 0 & - & - & - & SDSS & - & 2.2 & 0.4-0.7 & GSS(H/L) & 0 & -& -& - \\
        8(H/L) & 0.5 & 0.5 & 0.33 & 0.33 & 0.33 & SDSS & LOG & 2.2 & 0.4-0.7 & GSS(H/L) & 4 & 0.5& 0.5& 0 \\
        9(H/L) & 0.5 & 0.5 & 0.33 & 0.33 & 0.33 & SDSS & LOG & 2.2 & 0.4-0.7 & GSS(H/L) & 2 & 0.5& 0.5& 0 \\
        10 & 0.5 & 0.5 & 0.33 & 0.33 & 0.33 & SDSS & LOG & 2.2 & 0.4-0.7 & FM19 & 8 & 0.5& 0.5& 0 \\
       \hline
        \multicolumn{15}{c}{\bf IMBH seeds}\\
        \hline
        11(H/L)& 0.5 & 0.5 & 0.33 & 0.33 & 0.33 & SDSS & LOG & 2.2 & 0.4-0.7 & GSS(H/L) & 8 & 0.4& 0.4& 0.2\\
        12(H/L)& 0.5 & 0.5 & 0.33 & 0.33 & 0.33 & SDSS & LOG & 2.2 & 0.4-0.7 & GSS(H/L) & 8 & 0.85& 0.05& 0.1\\        
    \hline  
        \multicolumn{15}{c}{\bf Mixed spin distribution}\\
        \hline
        15& 0.5 & 0.5 & 0.33 & 0.33 & 0.33 & SDSS & LOG & 2.2 & 0.4-0.7 & FM19+MXL & 8 & 0.5& 0.5& 0\\
        16& 0.5 & 0.5 & 0.33 & 0.33 & 0.33 & SDSS & LOG & 2.2 & 0.4-0.7 & FM19+GSS & 8 & 0.5& 0.5& 0\\        
    \hline    
    \hline  
        \multicolumn{15}{c}{\bf Cluster evolution}\\
        \hline
        14(noEvo)& 0 & 1 & 1 & 0 & 0 & - & LOG & 2.2 & 0.4-0.7 & GSS L & - & 1& 0& 0\\
        15($N$-body)& 0 & 1 & 1 & 0 & 0 & - & LOG & 2.2 & 0.4-0.7 & GSS L & - & 1& 0& 0\\
        16(Theory)& 0 & 1 & 1 & 0 & 0 & - & LOG & 2.2 & 0.4-0.7 & GSS L & - & 1& 0& 0\\
    \hline
    \end{tabular}
    \caption{Main properties of the simulated models. Col 1: model ID. Col 2-3: fraction of isolated and dynamical BBHs. Col 4-6: fraction of BBHs in YCs, GCs, and NCs. Col 7-8: metallicity distribution adopted for isolated and dynamical BBHs. Col 9-10: observational selection functions for primary mass and mass-ratio. Col 11: spin distribution adopted, H/L denote high/low spin models. Col 12: slope of the polar angle distribution. Col 13-15: fraction of BHs with masses extracted from the simple single mass function (SSBH), mixed single mass function (MSBH), and the heavy seed mass function (HSBH). Models 14-16 correspond to the same set of initial conditions, but assuming only GCs that do not evolve (14), follow the $N$-body evolution recipes (15), or those predicted by the cluster evolution {\it classical} theory (16).}    
    \label{tab:my_label}
\end{table*}

The adopted selection criteria and the requirement that mergers must occur at $z<2$ clearly impact the actual fraction of mergers coming from one formation channel or another. Compared to the initial assumptions, i.e. $f_{\rm iso} = f_{\rm dyn}$, we find that the actual fraction of mock BBHs coming from the isolated (dynamical) channel is $f_{\iso\,{}(\dyn), ~ mock}\simeq 34\,{}(66)\%$. Similarly, the selection affects the amount of BBHs forming in YCs, GCs, and NCs, leading to $f_{\yc,\gc,\nc, ~mock} \sim (29,~53,~18)\%$, with a difference of $< 1-2\%$ from one models set to another. Note that this owes entirely to the selection criteria. 

\subsection{Component masses, effective spin parameter, and mass ratios of merging BBHs}

\begin{figure*}
\centering
\includegraphics[width=0.45\textwidth]{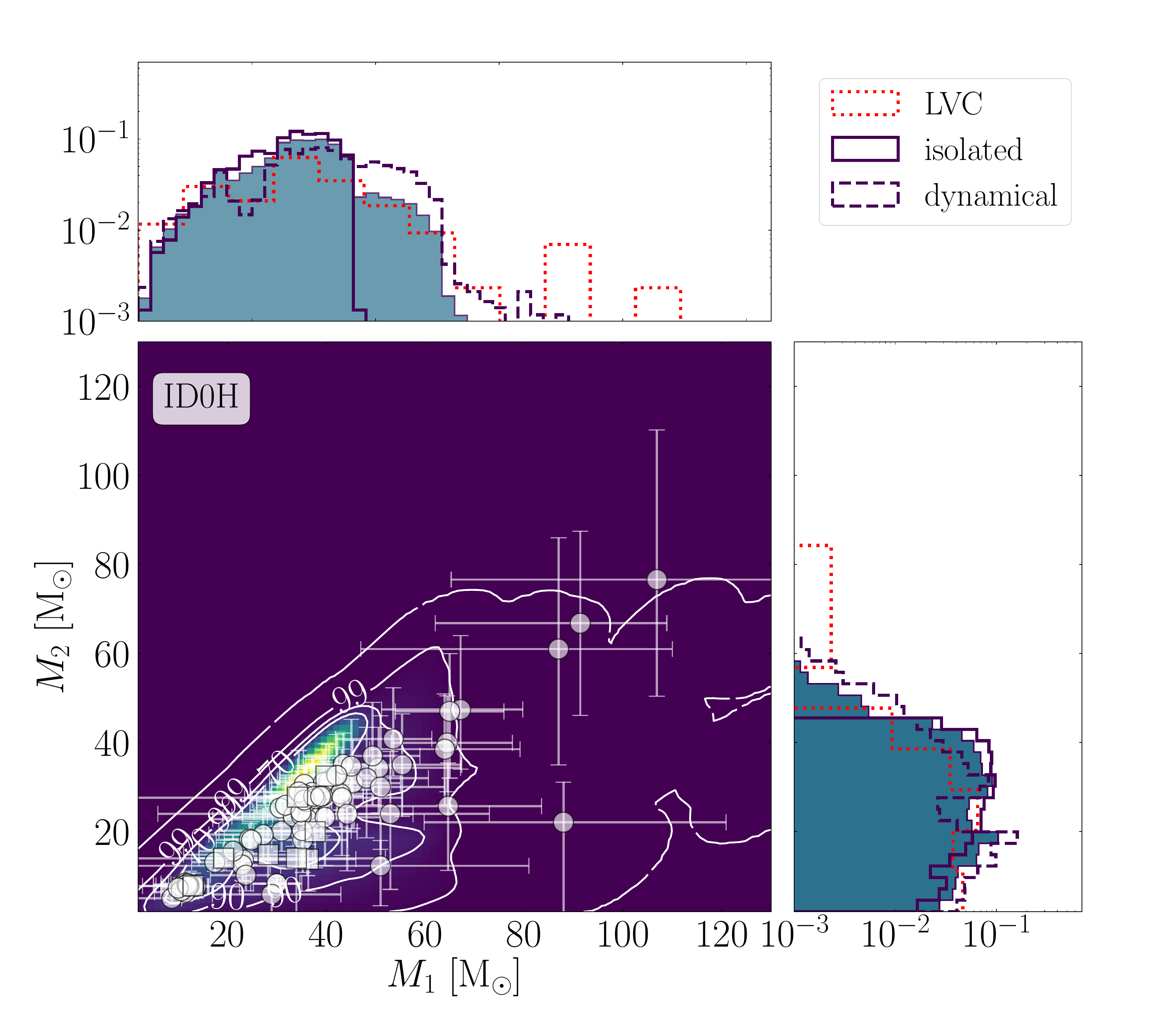}
\includegraphics[width=0.45\textwidth]{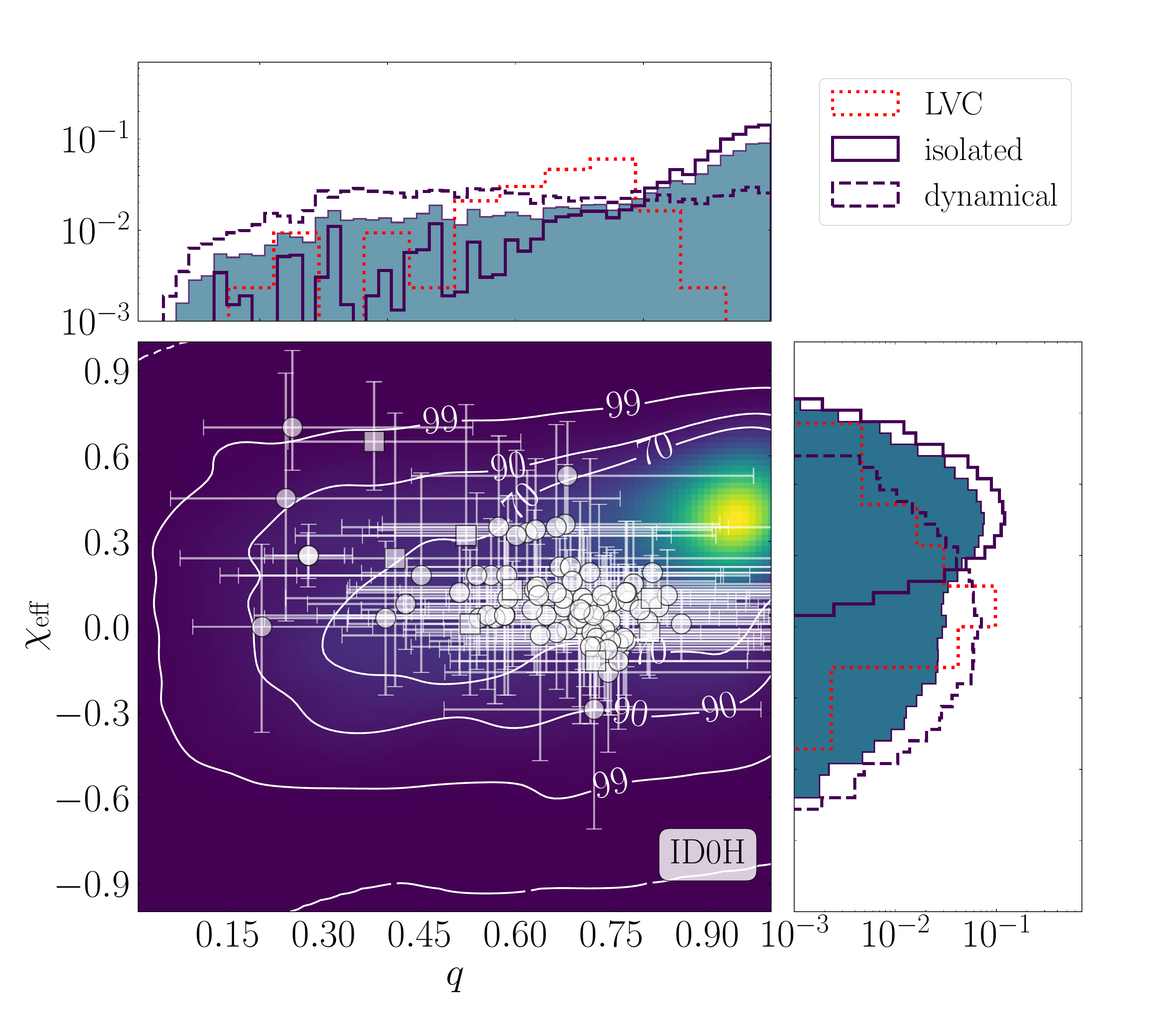}\\
\includegraphics[width=0.45\textwidth]{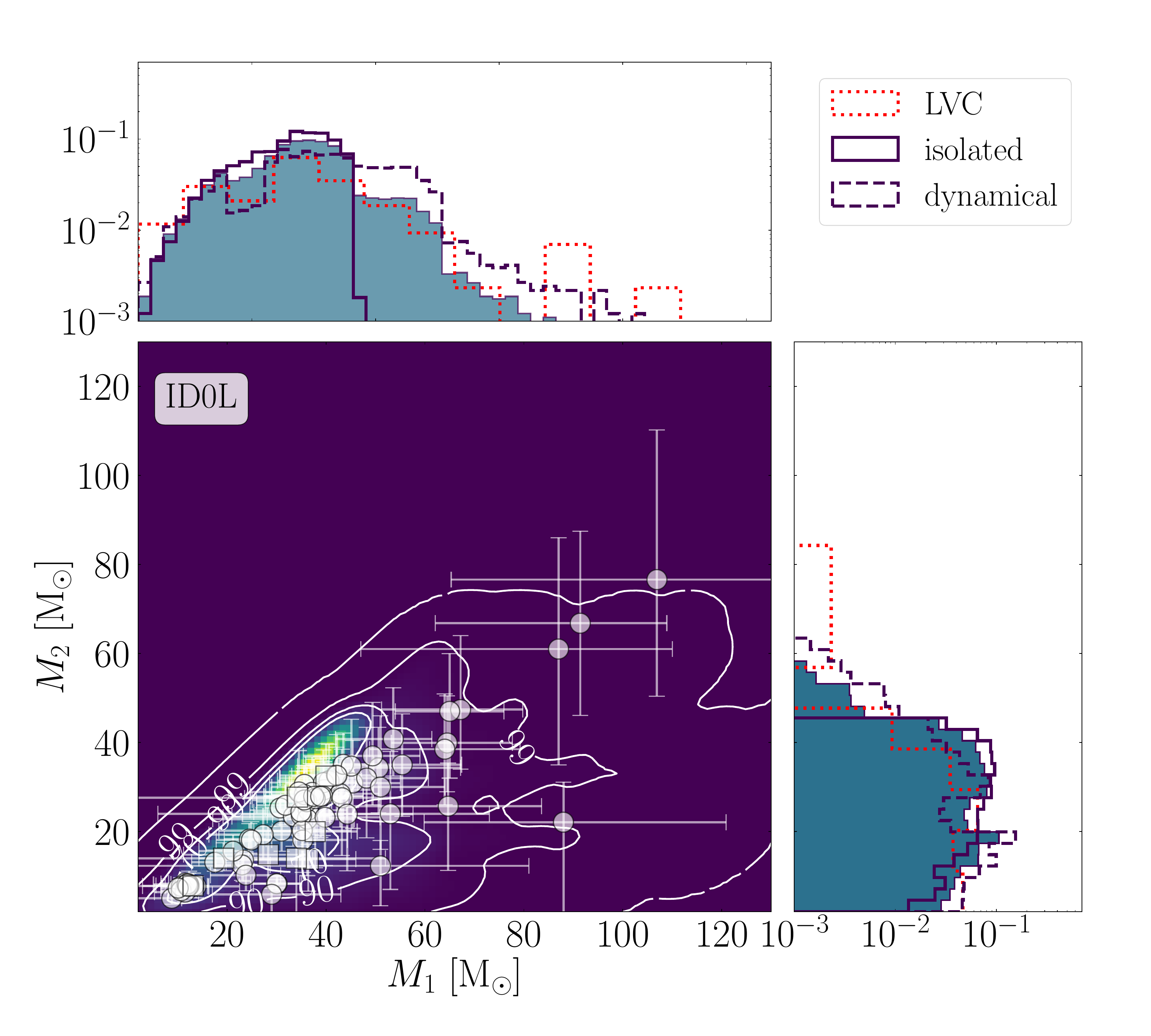}
\includegraphics[width=0.45\textwidth]{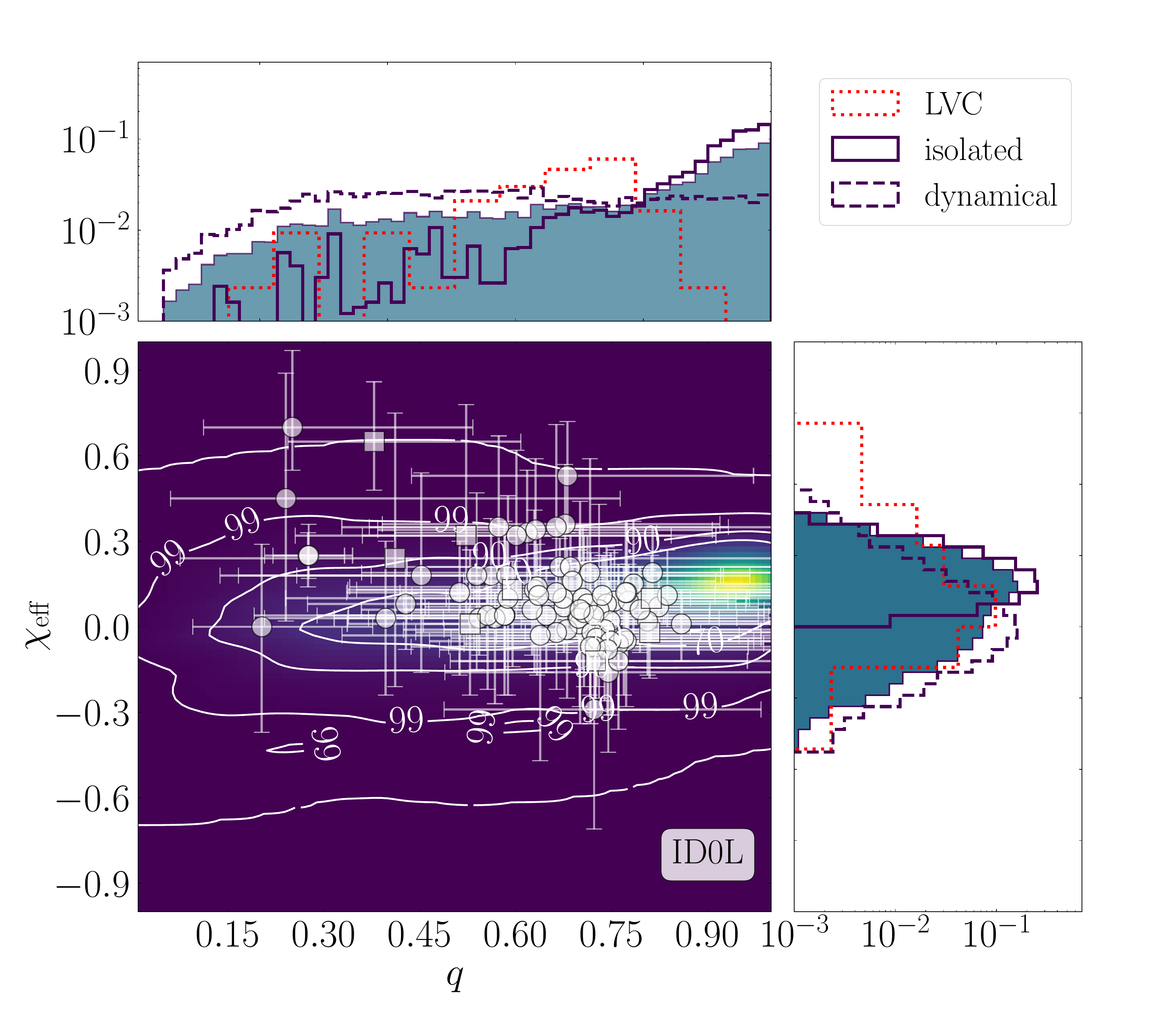}\\
\caption{Surface maps showing the combined distribution of component masses (left-hand panel) and of mass ratios and effective spin parameter (right-hand panel). In top (bottom) row panels, we draw BH spins from a Gaussian distribution centered on $\chi = 0.5(0.2)$. The surface maps are compared to GWTC-2.1 data \citep{gwtc2,gwtc21}. Contour lines encompass $70,\,{}90,\,{}99,\,{}100\%$ of the simulated BBH population. Marginal  histograms show the contribution of isolated (straight line steps) and dynamical (dashed steps) BBHs. Both panels show to the reference model.  
The BBH sample is weighted through the $VT-M_1$ and $VT-q$ relations to account for the main observational biases.}
\label{fig:comp}
\end{figure*}

In order to determine whether the BBH merger population in the reference model is representative of the GWTC-2 data, Figure \ref{fig:comp} compares mock data and observations in terms of combined distribution of component mass, mass-ratio, and effective spin parameter for both the low- and high-spin reference models. 

The combined mass distribution of BBH components lies in the same region as LVK detections in the $3-60\Ms$ mass range, with the contour plot enclosing all the mock binaries fully embracing the observed population of BBH mergers. 

Despite an apparent overlap between the distribution of different channels, it is possible to recognize in the parameter space some regions where one channel clearly dominates. 

For instance, isolated BBHs have higher mass-ratio and lower masses, on average, compared to dynamical BBHs. Given this, in the reference model we found a {\it sweet spot} in the component mass range $M_1 > 40\Ms$ and $M_2 < 30\Ms$ where $\simeq 98.3\%$ of mergers have a dynamical origin. 

Therefore, upon our main assumptions, mergers with masses in the aforementioned ranges could be characterised by a high probability to 
have formed in a star cluster.

A few sources appear to be outliers in our distribution. The heaviest source detected so far, GW190521, sits in the region of the $M_1-M_2$ plane containing only $1\%$ of our BBHs, populated by dynamical mergers that underwent multiple merger events. This suggests for GW190521 a dynamical origin triggered by a series of hierarchical mergers. Another interesting source is GW190517, a BBH merger with $M_1 = 36.4^{+11.8}_{-7.8} \Ms$ and $\chi_\eff = 0.53^{+0.20}_{-0.19}$. The large value of $\chi_\eff$ brings this source in the region dominated by isolated binaries in the high-spin model (ID0H), despite the observational uncertainties, whilst it lies in the region containing $\leq 1\%$ BBHs in the low-spin model (ID0L).

The choice of $n_\theta = 8$ in the reference model implies that $\sim 55\%$ of the isolated mergers have the spin-orbital angular momentum angles differing by less than $20\%$. As shown in the right panel of Figure \ref{fig:comp}, in the high-spin reference model (ID0H), the $n_\theta$ value adopted coupled with the overall high mass-ratio of isolated binaries, causes a clear overdensity in the $q-\chi_\eff$ plane around $q\simeq 0.9$ and $0.3 < \chi_\eff < 0.5$. Compared to the high-spin model, low spins lead to i) a richer population of mergers with $M_1>80-100\Ms$, ii) and a dearth of mergers with $|\chi_\eff| > 0.3$.

\begin{figure*}
\centering
\includegraphics[width=0.45\textwidth]{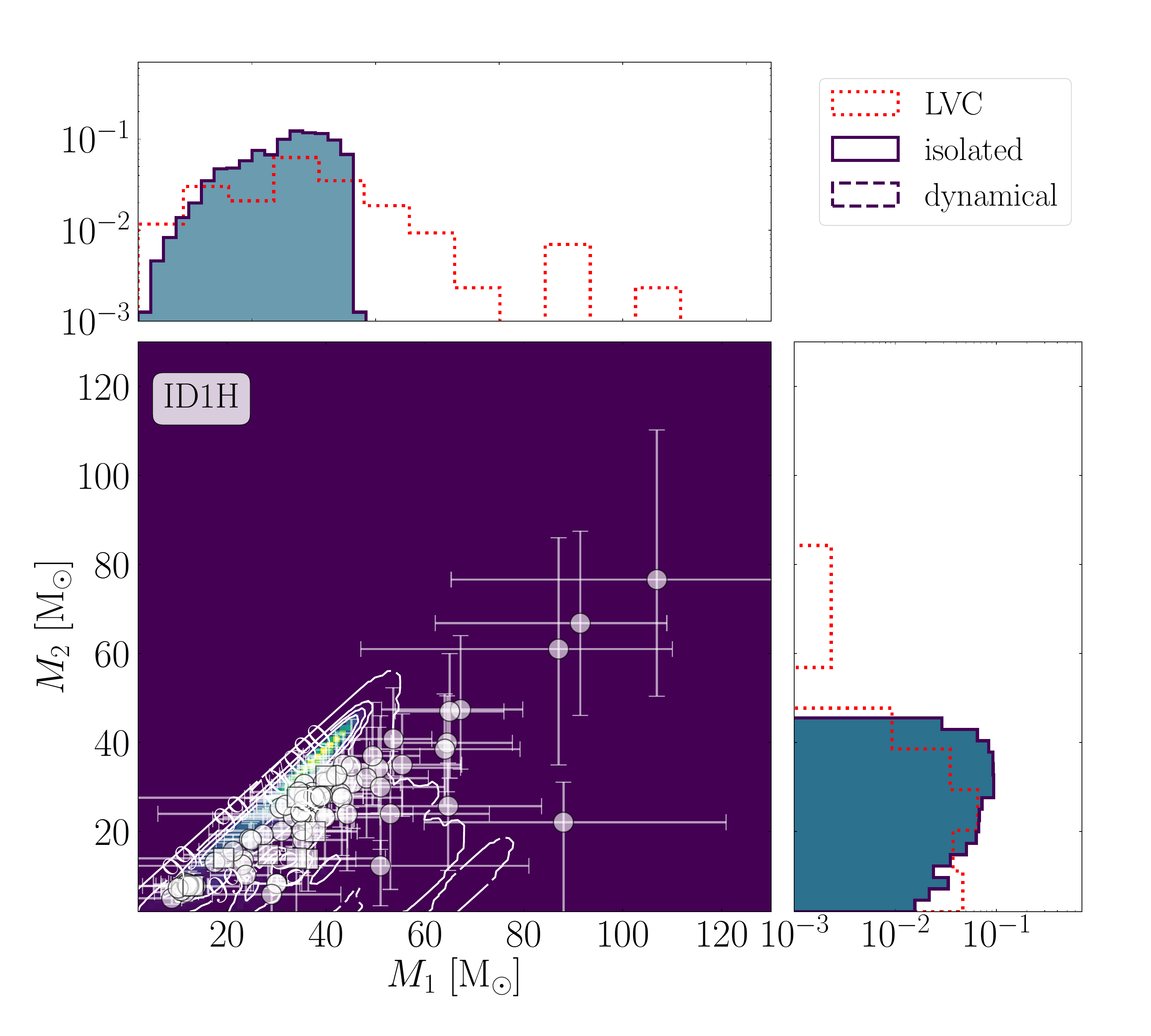}
\includegraphics[width=0.45\textwidth]{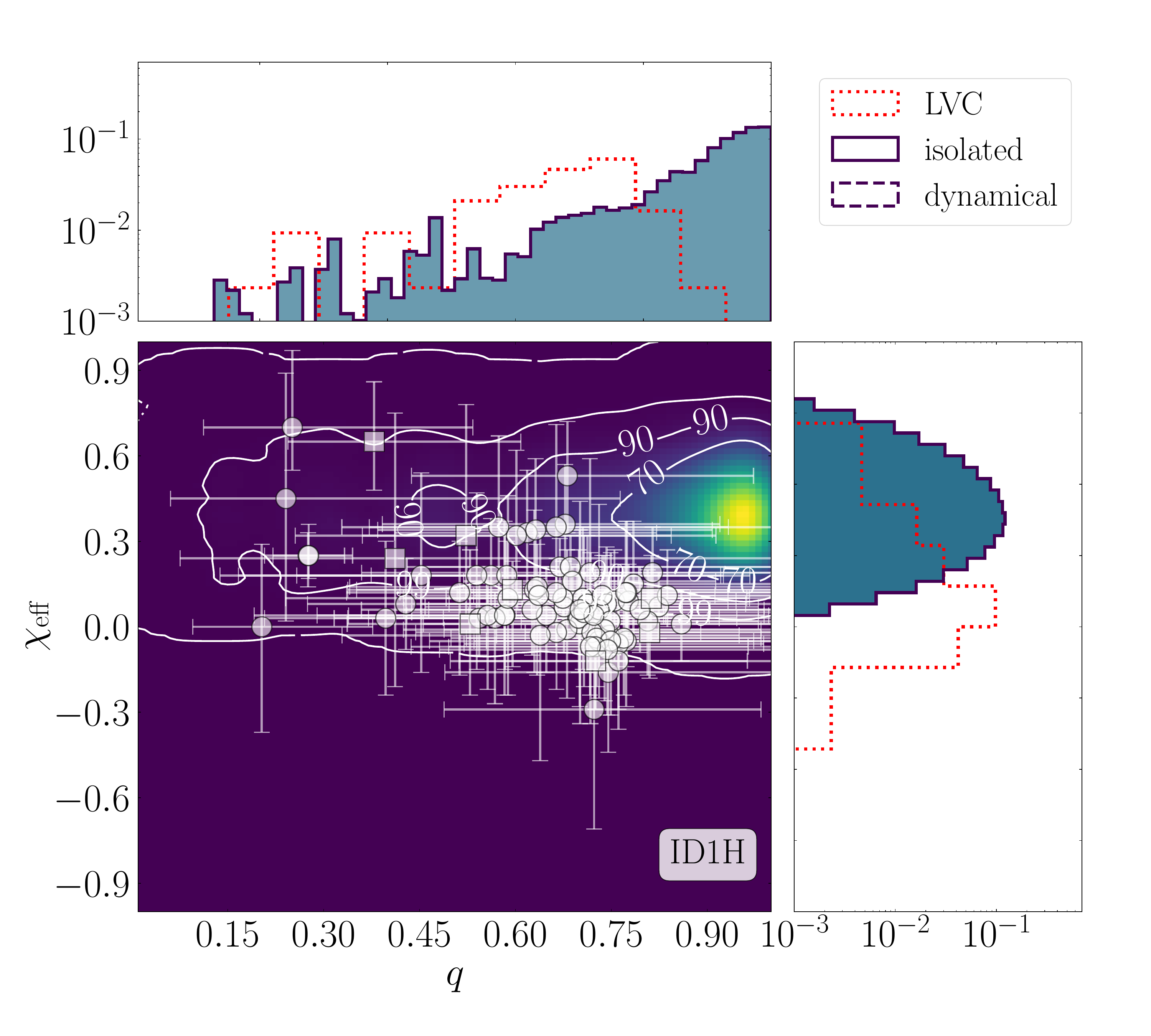}\\
\includegraphics[width=0.45\textwidth]{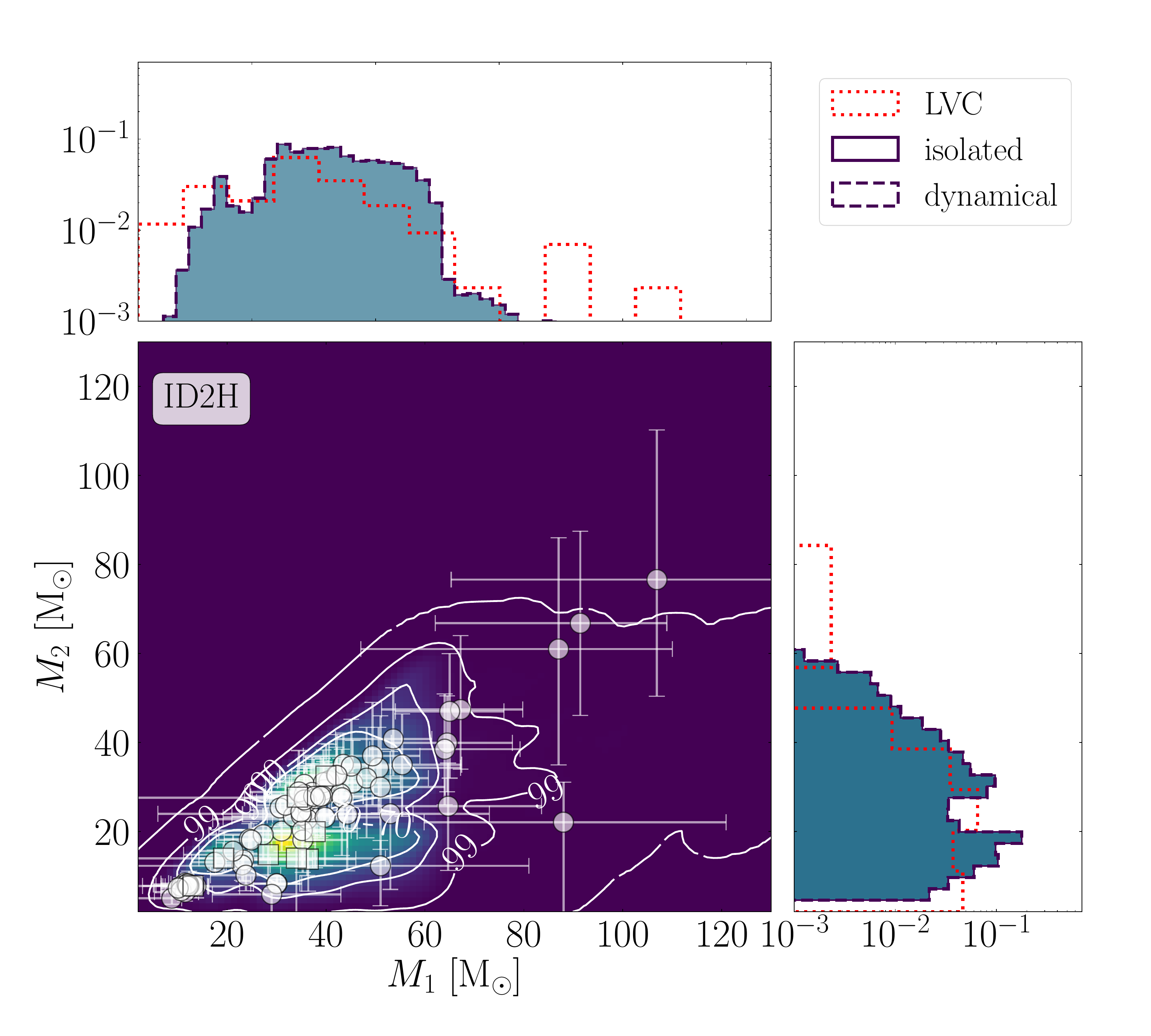}
\includegraphics[width=0.45\textwidth]{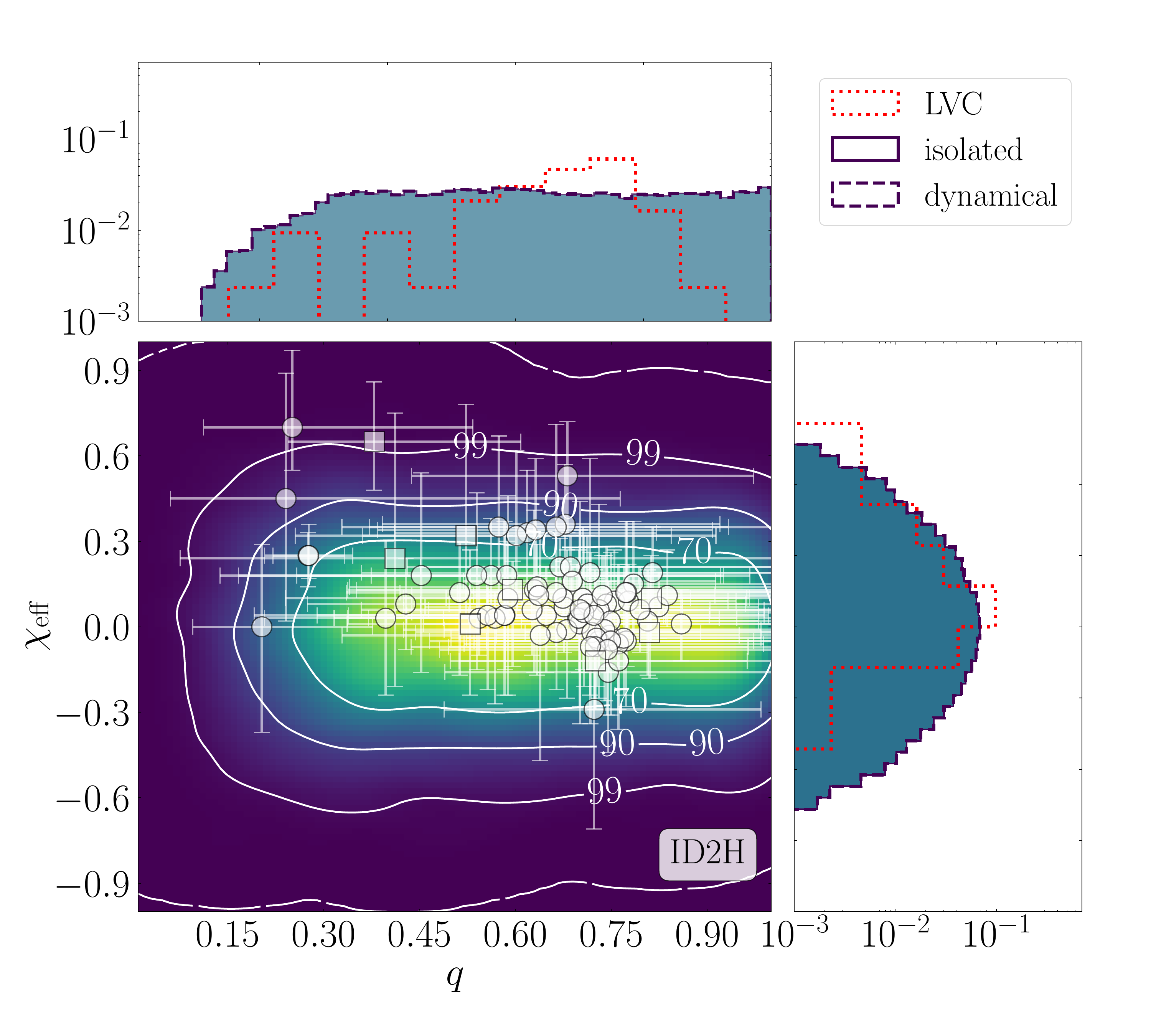}\\
\includegraphics[width=0.45\textwidth]{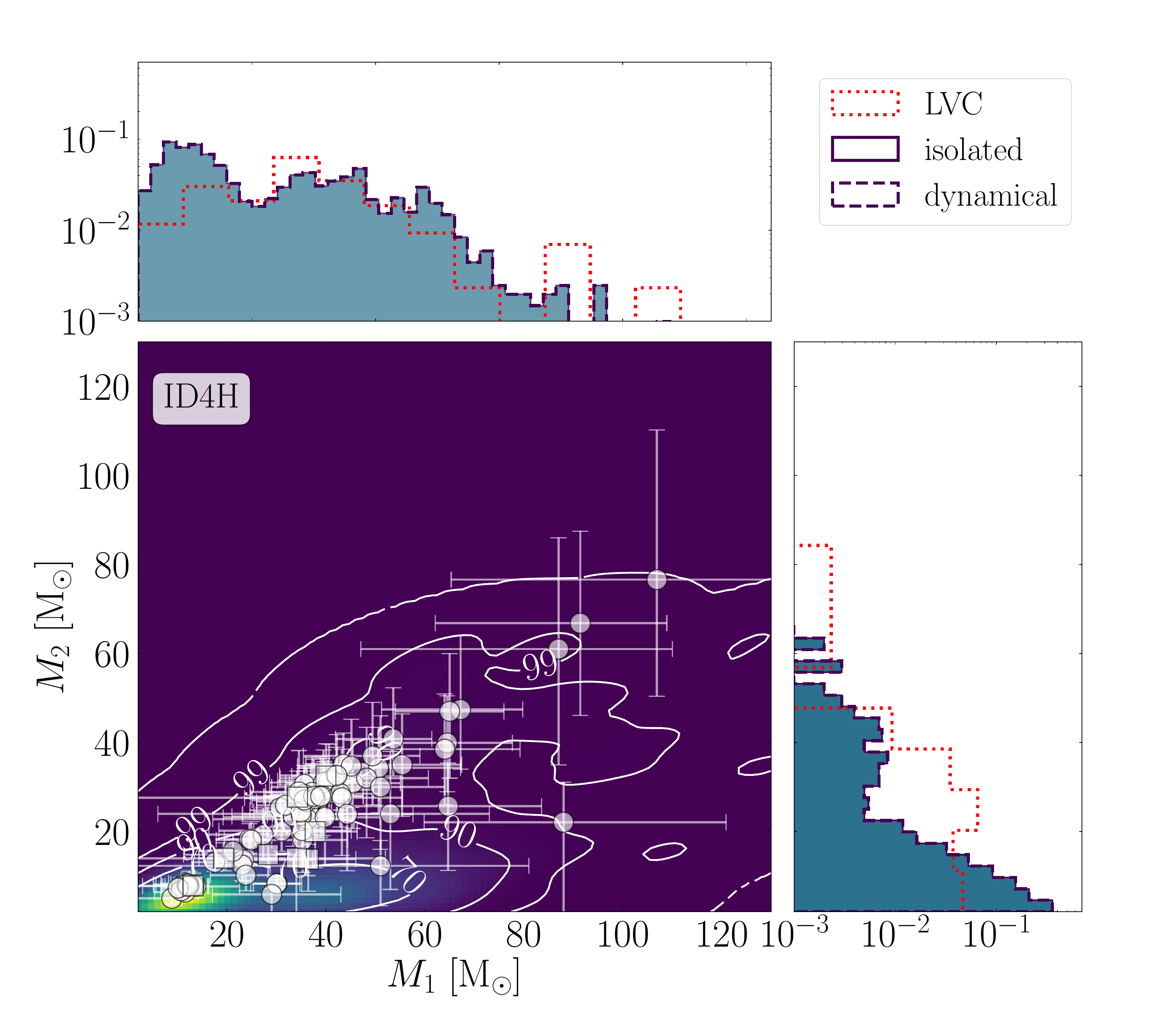}
\includegraphics[width=0.45\textwidth]{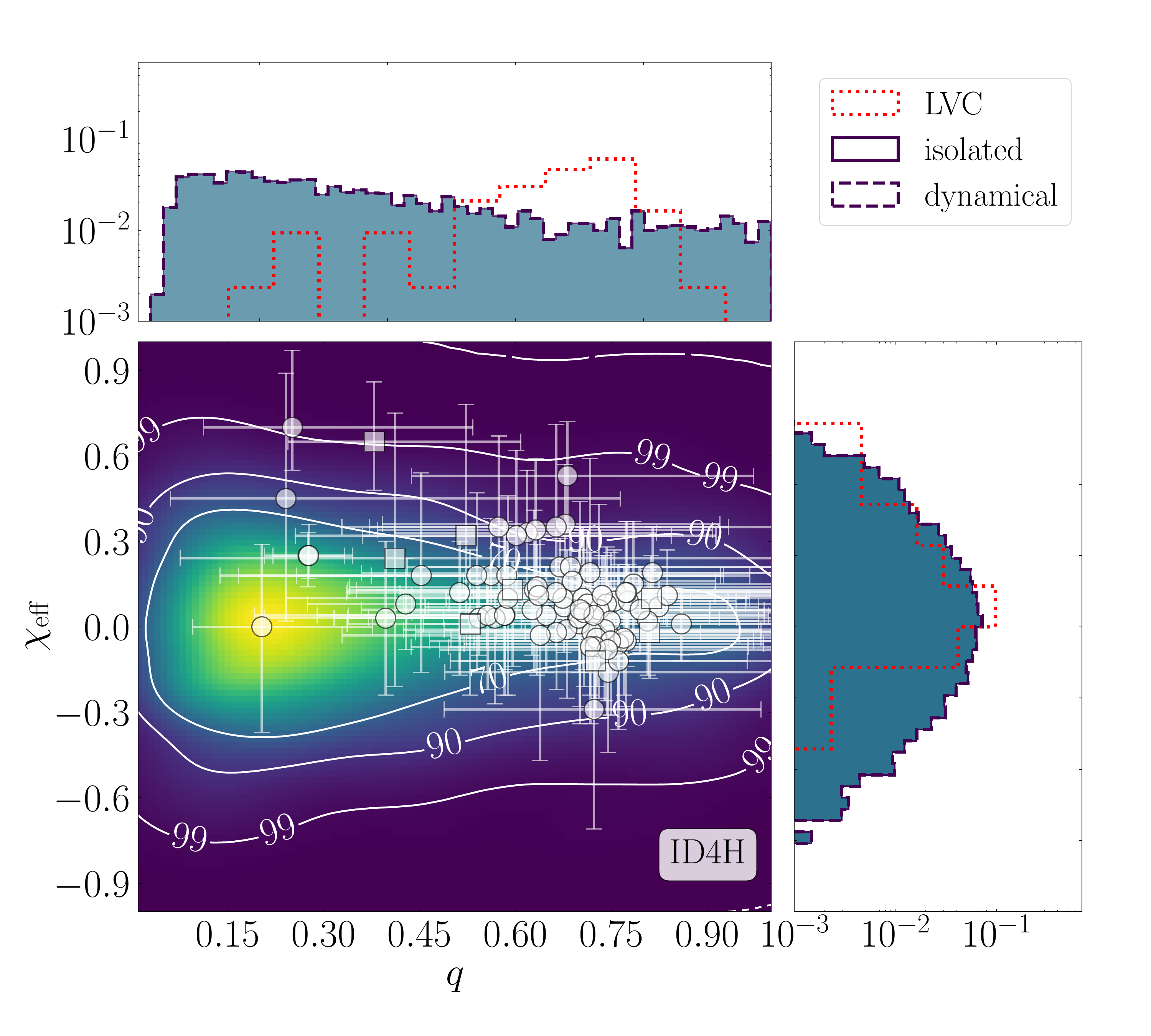}\\
\caption{
Same as Figure \ref{fig:comp}, but here we show a model that includes only isolated BBHs (top panels), or only dynamical BBHs with masses taken either from the SSBH  (central panels) or the MSBH mass spectrum (bottom panels). All models are characterised by natal spins following a Gaussian distribution peaked over $\chi = 0.5$.}
\label{fig:comp2}
\end{figure*}

In order to better highlight the properties of the two formation channels explored here, Figure \ref{fig:comp2} shows the $M_1 - M_2$ and $q-\chi_\eff$ planes for models in which the BBH population is either only isolated (panels in the upper row) or dynamical; in the latter case the BH masses are extracted either from the SSBH (panels in the central row) or the MSBH (panels in the lower row) mass spectra. For clarity's sake, here we refer to models with high-spins (all denoted with letter H).

In the case of isolated BBHs, which are shown in the top row panels of Figure \ref{fig:comp2}, the spin distribution peaking around $\chi_\eff = 0.5$ owes to a combination of factors: first, the choice $n_\theta = 8$ implies $\sim 55\%$ BBHs having $\theta_{1,2}$ that differ by less than $20\%$; second, isolated mergers have similar-mass BBHs on average; third, the choice of a Gaussian distribution peaked over $\chi = 0.5$ for BH natal spins. 

For nearly equal mass BBHs and spins $\chi_{1,2}\sim 0.5$, the condition $\chi_\eff > 0.25$ requires $\cos\theta_{1,2} > 0.5$. This condition is satisfied in the $P_{\chi_\eff > 0.25} = 5,~33,~58,~85\%$ for $n_\theta = 0,~2,~4,~8$. Decreasing the value of $n_\theta$ would bring the peak of the $\chi_\eff$ toward smaller values, but the clear tendency of isolated BBHs to feature mass ratio values $q>0.9$ would still cause a clear difference between observations and models, despite the large uncertainties associated with the observed mass ratios.
  
Focusing on BBHs with a mass ratio $q>0.9$, we find  that around $61\%$ of mergers in the reference model are isolated BBHs. Around $70\%$ of these isolated mergers have total masses $M = 50-90\Ms$, whilst $90\%$ of them have masses in the range $M=25-100\Ms$.

Figure \ref{fig:comp2} compares the total mass -- effective spin parameter distribution for isolated and dynamical mergers. Regarding dynamical BBHs, we find that a simple single mass spectrum (SSBH) seems to match the observed data in terms of component masses, mass ratio, and $\chi_\eff$. In the case of BBHs with component masses extracted from a mixed mass spectrum (MSBH), instead, we see that the $M_1-M_2$ distribution deviates significantly from the SSBH model, favouring the formation of small mass ratios ($q\lesssim 0.1-0.2$) and filling efficiently the region $M_1<40\Ms$ and $M_2<20\Ms$, which is poorly covered by the SSBH. The peculiar mass distribution obtained for dynamical binaries  owes to a combination of factors, among which the request that BHs have natal kicks smaller than the cluster escape velocity and the fact that, on average, heavier BHs are characterised by smaller kicks. This, in combination with the fact that the BH mass distribution in the MSBH configuration is peaked at lower values, i.e. $5-10\,{}\Ms$, compared to the SSBH case explains the difference between the two models.


As we will discuss in the next section, adopting a complex mass spectrum for BHs in dynamical mergers might be the key to understand the likely complex mass distribution of observed BBH mergers. 

\subsection{The primary mass distribution}

One of the main insights that can be inferred from the LVK database is the distribution of primary masses ($M_1$) in merging BBHs. As inferred from GWTC-2 data, the primary mass distribution is expected to be well described by a broken power-law characterised by a peak at masses $M_1\sim 40\Ms$ that truncates around $M_1\gtrsim 100\Ms$ \citep{gwtc2}. 
In our analysis, we reconstruct the overall population of merging BBHs first and then we derive the distribution of mergers sampled according to the observation selection criteria described in Section \ref{sec:model}. In the following, we will refer to the overall population of BH primaries as ``global primaries'' and to BHs sampled through the selection criteria as ``mock primaries''. Since the choice of BH natal spins does not critically impact  the $M_1$ distribution, in the following we show results for high-spin models only.

\begin{figure*}
\centering
\includegraphics[width=\columnwidth]{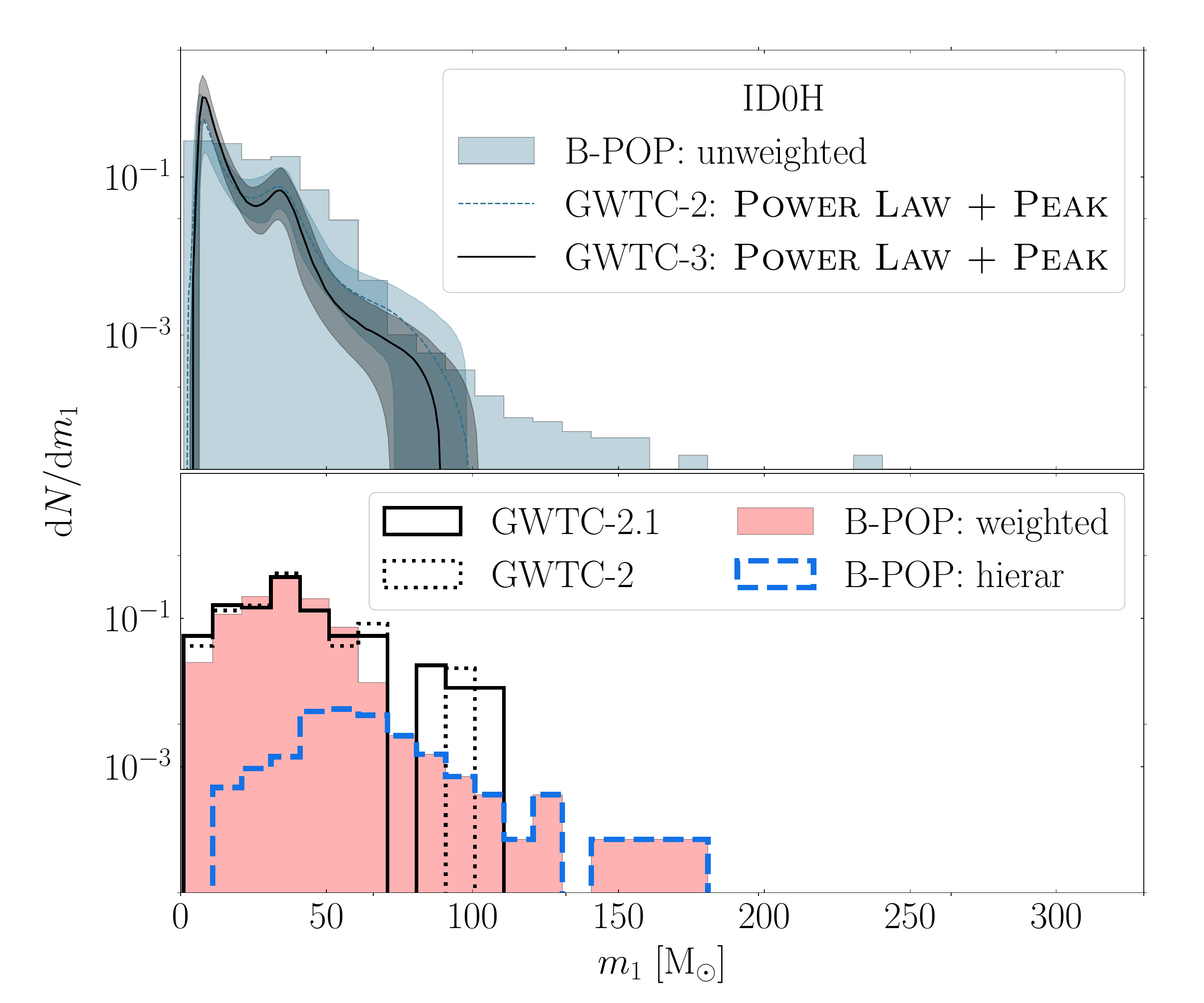}
\includegraphics[width=\columnwidth]{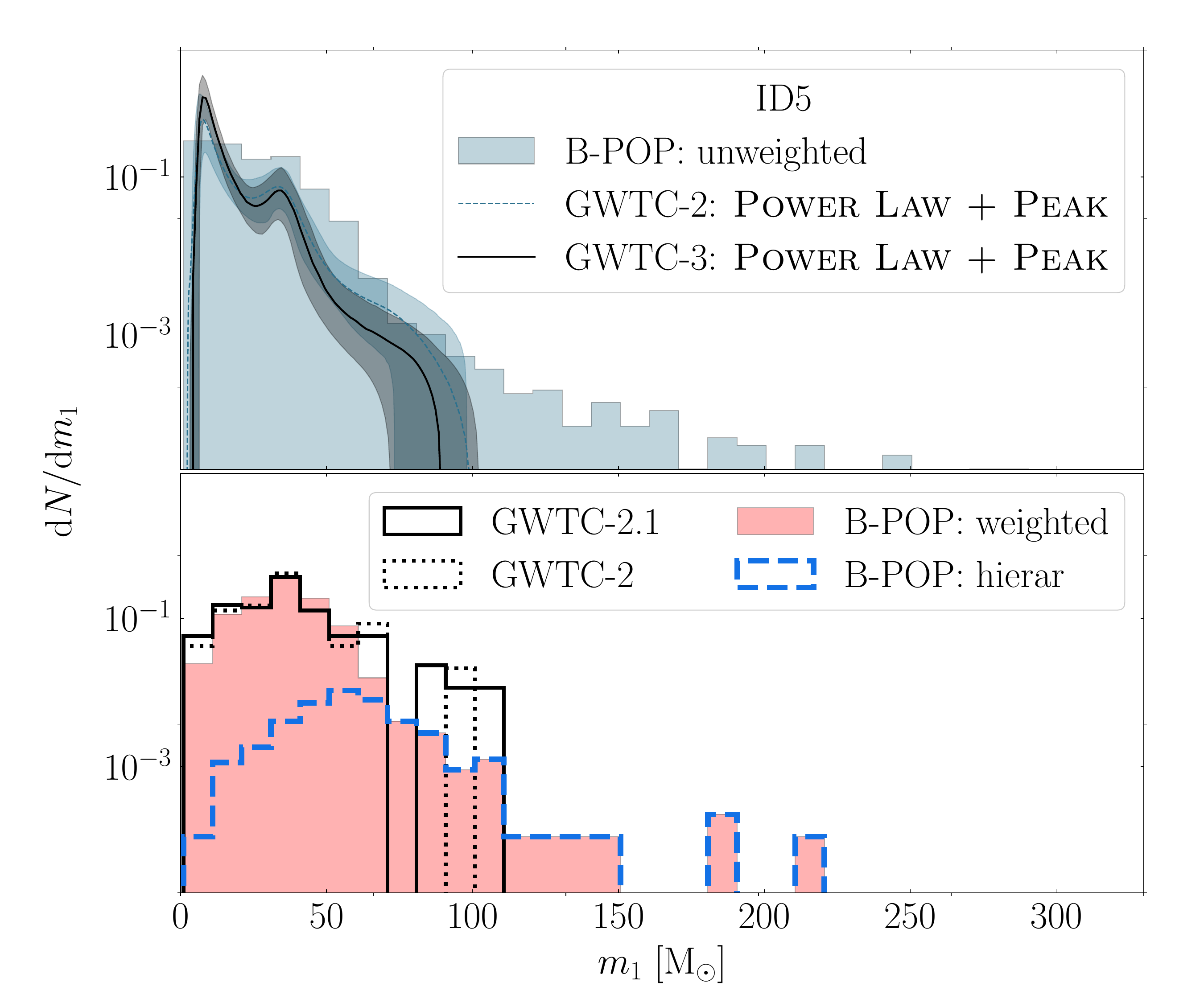}
\caption{Upper panels: primary mass ($M_1$) distribution of the overall  BBH merger population in the reference model (global primaries, blue filled step) compared to the {\sc power law + peak} model from \citet{gwtc2} (dashed line). Lower panels: same as above, but here we consider a BBH sub-population sampled through the $VT-M_1$ and $VT-q$ relation (mock primaries, red filled step) as compared to the median primary masses from GWTC-2.1 (open black steps) and GWTC-2 data (dotted black steps). Here, for each GW event we use only the median value without considering the uncertainties. The distribution for hierarchical mergers in the sample is highlighted (dashed grey open steps). The two panels show the reference model assuming for BH natal spin a Gaussian peaked on $\chi = 0.5$ (left) or a Maxwellian with dispersion 0.2 (right). }
\label{fig:lvc}
\end{figure*}

The upper panels of Figure \ref{fig:lvc} show  the mass distribution of global primaries. The reference model 
matches the {\sc power law + peak} model inferred by LVK at masses $<100\Ms$. Our model is characterised by a long tail extending beyond $300 \Ms$, mostly dominated by hierarchical merger products developed in dense clusters.  In the reference model with high spins (ID0H), mergers with $M_1 > 50\Ms$ constitute the $f_{\rm > 50} = 4\%$ of the overall BBH population, with $f_{\rm hier} \sim 18\%$ of them being hierarchical mergers. Similarly, the model with low spins (ID0L) is characterised by $f_{\rm > 50} = 4.4\%$ and $f_{\rm hier} \sim 33\%$.

The lower panels of Figure \ref{fig:lvc}, instead, compare our mock primaries with the median primary masses of GWTC-2 mergers \citep{gwtc2}. 
 For the reference model, we find that the observation biases lead to a primary mass distribution truncated at $M_1 < 150-170\Ms$. This happens because the highest primary masses are generally associated with lower mass ratios, leading to a lower detection probability given the $VT-q$ relation. 
Therefore, our analysis suggests that there is a population of unseen BHs with primary masses as high as $M_1 = 150-200 \Ms$ (around $0.47\%$ of all BBHs in the reference model) that escape LVK detection due to the $VT-M_1$ and $VT-q$ selection effects. Among mock sources, we find around 
$\sim 0.2\%$ mergers with a primary mass $>100\Ms$, and $(2-3)\%$ BBHs with a total mass $100 < M_{\rm bin}/\Ms < 170$.

\begin{figure}
\centering
\includegraphics[width=\columnwidth]{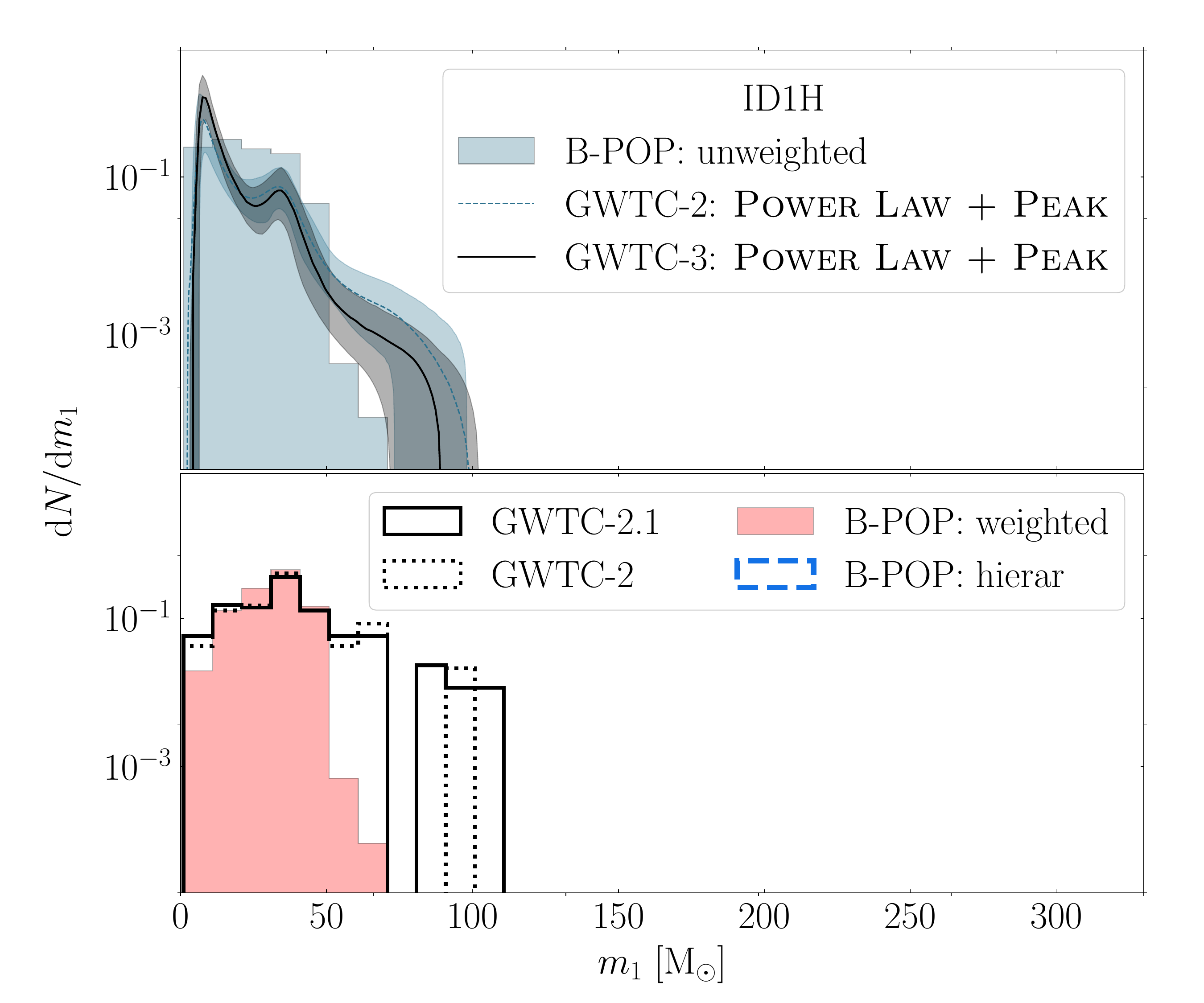}
\includegraphics[width=\columnwidth]{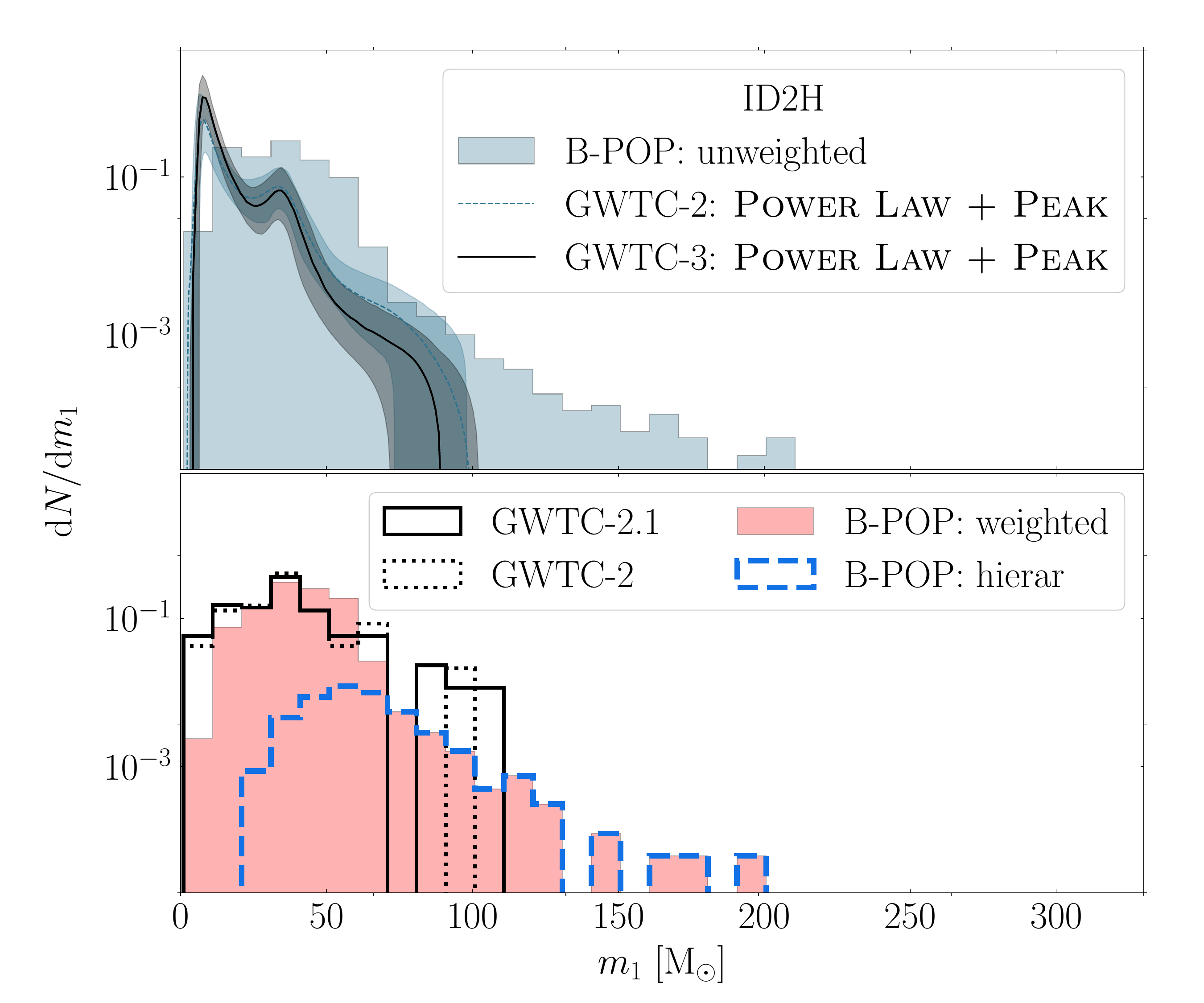}
\includegraphics[width=\columnwidth]{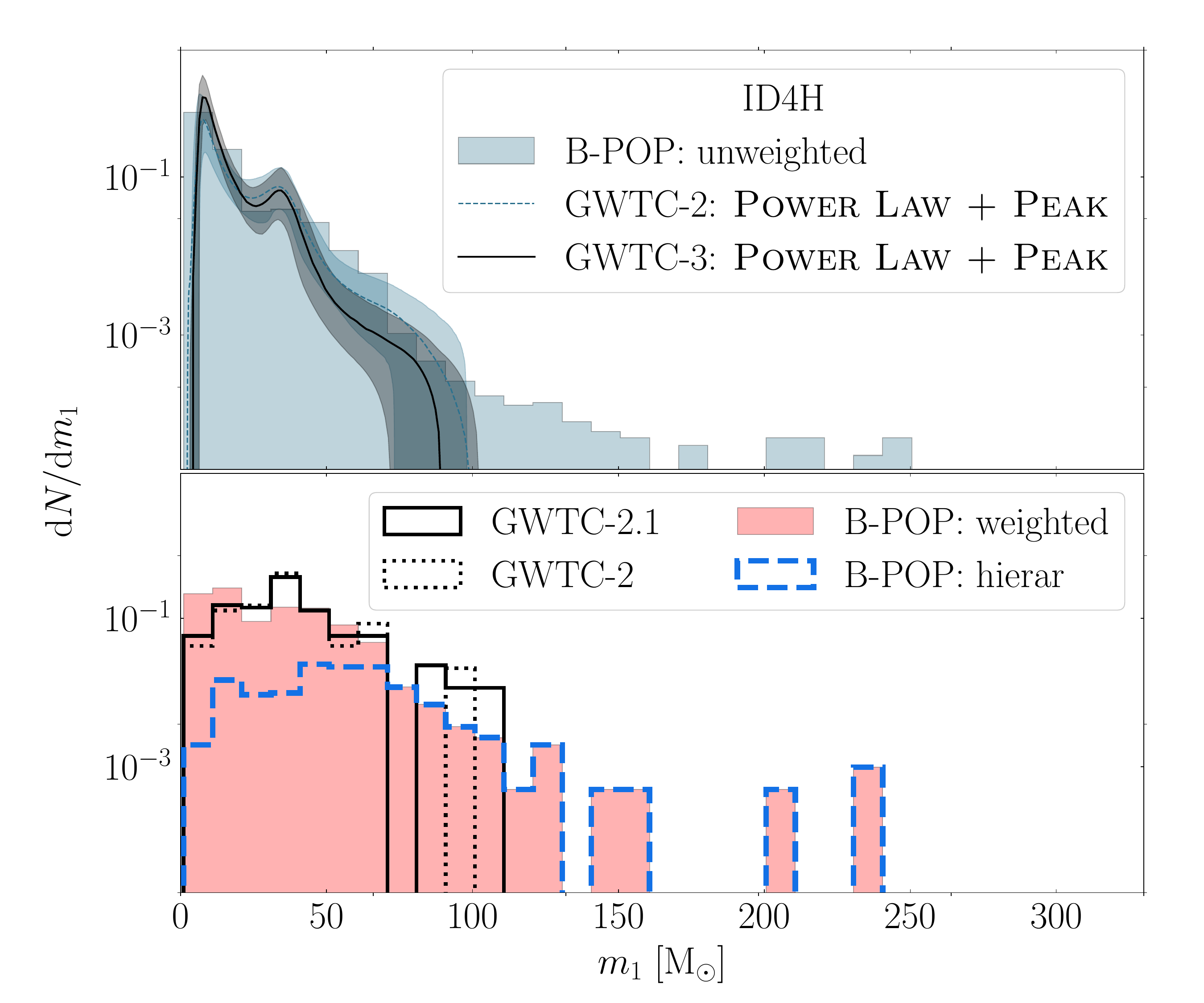}
\caption{Same as in Figure \ref{fig:lvc}, but here we show only isolated (upper panel) and dynamical mergers assuming for BH natal mass either the simple single mass spectrum (SSBH, central panel) or the mixed one (MSBH, lower panel).}
\label{fig:lvcb}
\end{figure}

Figures \ref{fig:lvc} and \ref{fig:lvcb} show the $M_1$ distribution for global and mock BBHs for other models. Adopting a Maxwellian distribution peaked over $\chi=0.2$ rather than a Gaussian peaked over $\chi = 0.5$ implies that merger remnant can get, statistically, lower kicks. This in turn can imply a larger fraction of hierarchical mergers. As a result, the $M_1$ distribution for the Maxwellian distribution case (ID 5) exhibits a slightly longer tail at $M_1 > 100\Ms$ that declines less sharply than the reference model ID0. This leads the percentage of primaries heavier than $100\Ms$ to $\simeq 0.6\%$, slightly larger than the reference model.

In the case of isolated mergers (ID1), the global distribution of the primary mass is sharply truncated at $M_1 \lesssim 50\Ms$. This is a clear consequence of 
binary evolution models adopted in \mobse{}. Nonetheless, our analysis suggests that sources with a primary mass $M_1 > 50\Ms$ are easy to explain with a dynamical origin\footnote{We note that $\sim 9$ sources in GWTC-2.1 have a median primary mass above $M_1 > 50\Ms$, and $5$ exceeds the this threshold at $90\%$ credibility level.}. 

In the "mock" sample, we find that around $2.4\%$ of BBHs are hierarchical mergers. Comparing them to the overall distribution of mock mergers, we find that hierarchical mergers dominate completely the range $M_1>60\Ms$, owing to the SSBH and MSBH mass spectra adopted.

\subsection{To spin or not to spin?}

As the number of detected BBHs increases, the constraints on the properties of BBH mergers become more robust. The current sample of detected BBH mergers suggests that merging BBHs are characterised by relatively low spins, with a possible peak of the distribution around $\chi_{1} \sim 0.2-0.3$ \citep{gw190521b,gwtc2}, and effective spin parameters narrowly distributed around zero \citep{gw190521b}, with a tail extending to positive values.

To quantify the impact of BH spins on our mock population, we create a series of variations of the reference model assuming that BBH mergers are characterised by either a Gaussian distribution centered on $\chi = 0.5$ or $0.2$ (ID0H/L), a Maxwellian with dispersion $0.2$ (ID5), or a fixed value of $\chi= 0.01$ (ID10). 
Additionally we vary the $n_\theta$ parameter, which regulates the amount of BBHs with aligned spins, setting it to $n_\theta = 8, ~4, ~2$ (models ID5, ID8H/L, and ID9H/L).
The high spin model matches the $|\chi_{\rm eff}| > 0.3$ range, which instead is poorly populated by low-spin models.
This has serveral implications for BH natal spins in single and binary systems. 

High-spin (e.g. ID0H) and Maxwellian spin models (e.g. ID5) are characterised by a wide distribution that extends beyond $|\chi_{\rm eff}| > 0.3-0.5$, whilst low- and non-spinning scenarios (e.g. ID0L or ID10) exhibit a narrower distribution peaking around 0.

The possible dearth of detected mergers with $\chi_\eff<0$ and the detection of sources having $\chi_\eff > 0.3$ could hint to differences in the distribution of natal spins for BHs in isolated and dynamical mergers, although the current observational uncertainties and the low statistics significantly affect the interpretation of observed sources. 

As recently suggested by \cite{fuller19}, efficient angular momentum transport driven by magnetic fields can lead to stellar BHs with natal spin as small as $\chi \sim 0.01$ for both single and binary stars, although in the latter case binary processes can spin-up the BH to large spin values. To test this idea, we build two further models, ID13 and 14, in which we assign to dynamical mergers a fixed spin $\chi=0.01$ following \cite{fuller19}, whilst we assign to isolated mergers a natal spin either drawn from a Maxwellian peaked over 0.2 (ID13) or from a Gaussian peaked over 0.5 (ID14), and we adopt $n_\theta = 8$ in both cases\footnote{This choice implies a $\sim 55\%$ probability to draw the angles between the spin directions and orbital angular momentum differing by less than $20\%$.}.

\begin{figure*}
    \centering
    \includegraphics[width=0.32\textwidth]{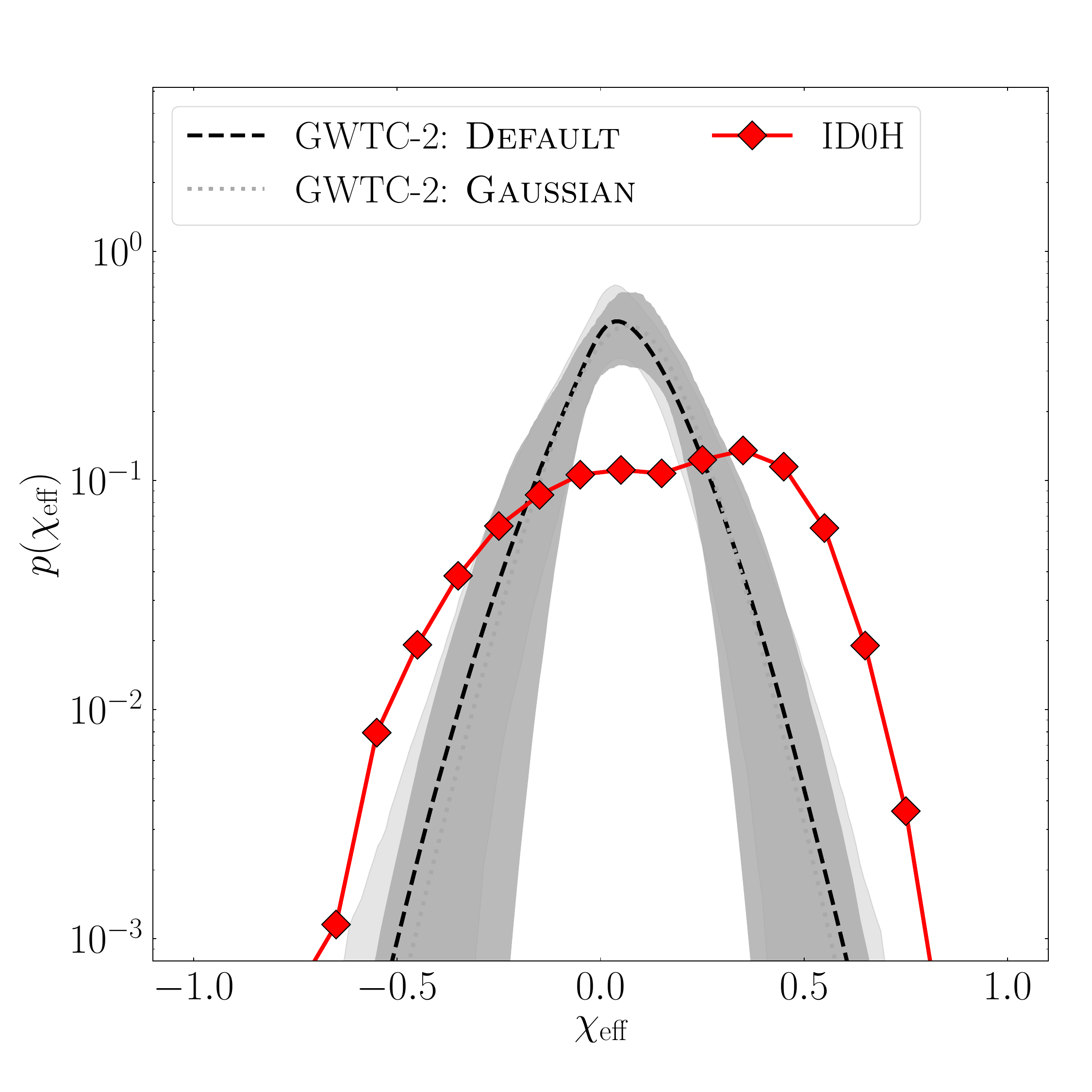}
    \includegraphics[width=0.32\textwidth]{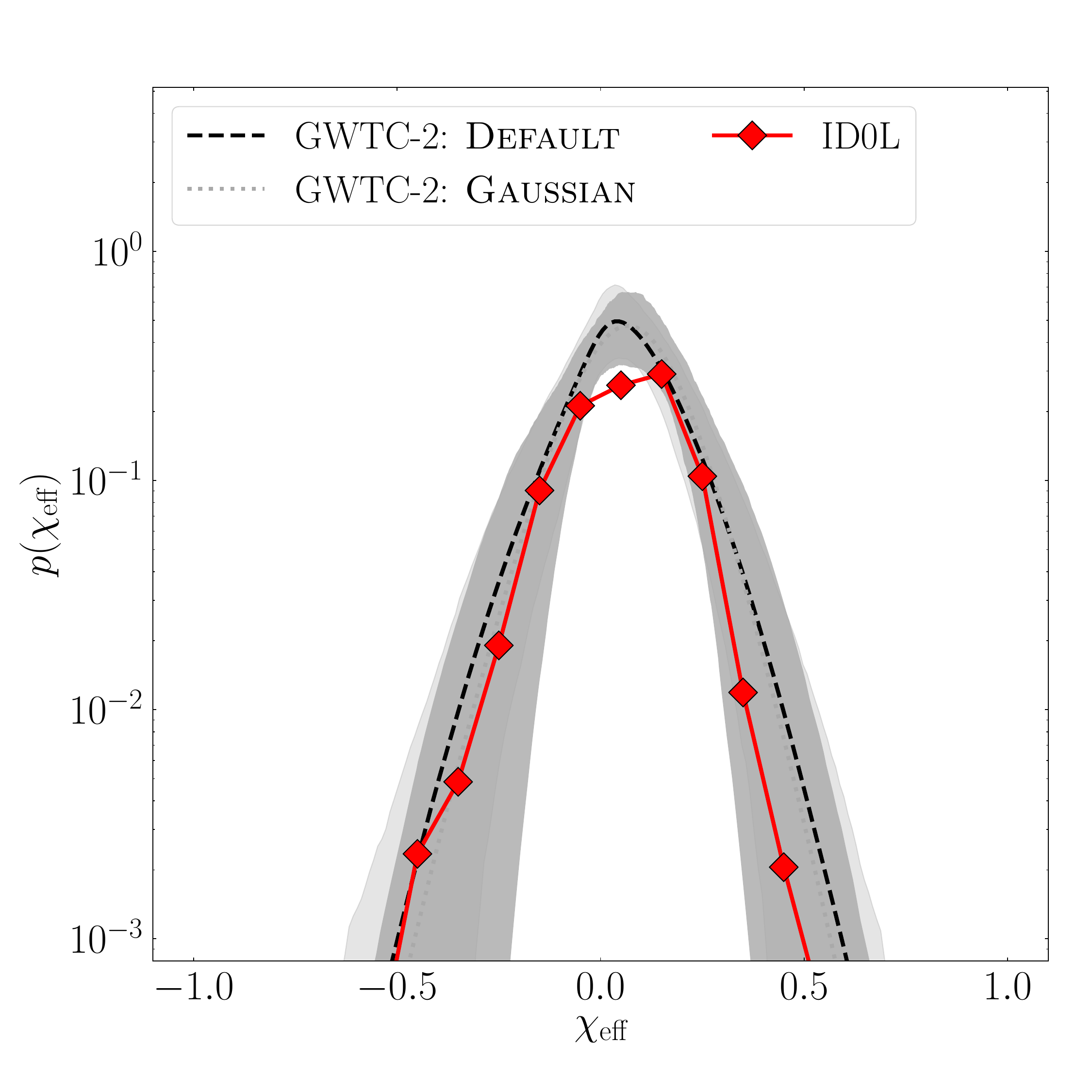}
    \includegraphics[width=0.32\textwidth]{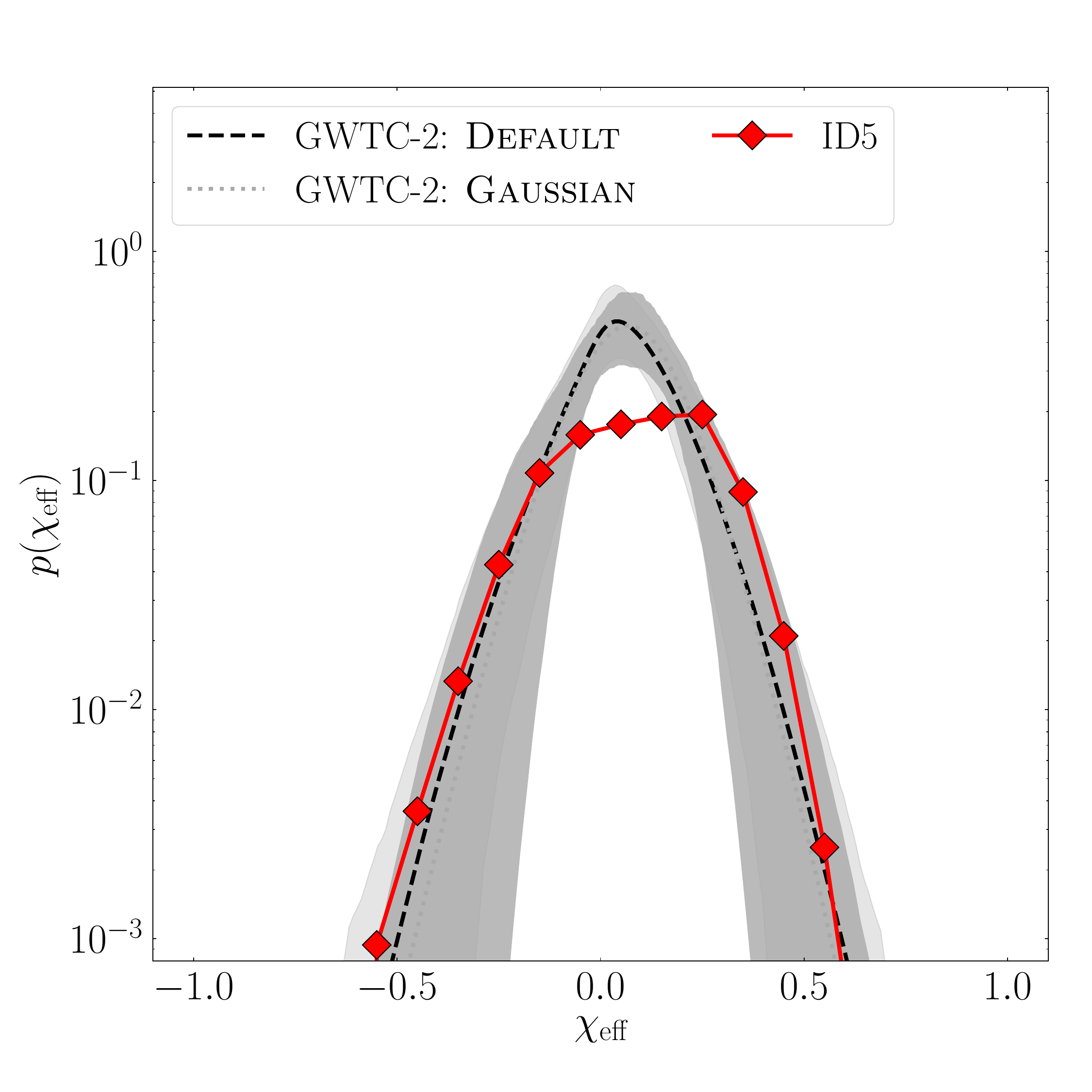}\\
    \includegraphics[width=0.32\textwidth]{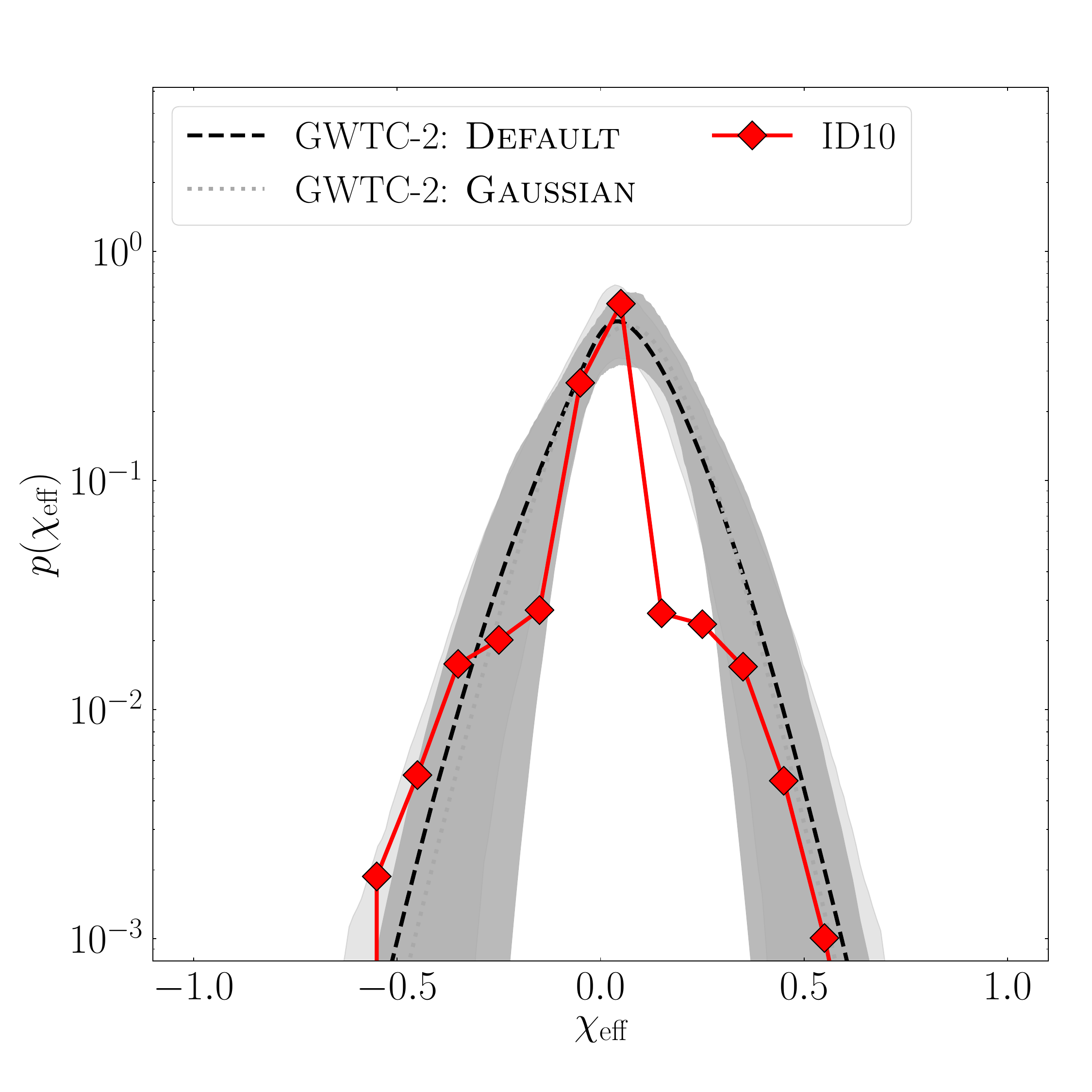}    
    \includegraphics[width=0.32\textwidth]{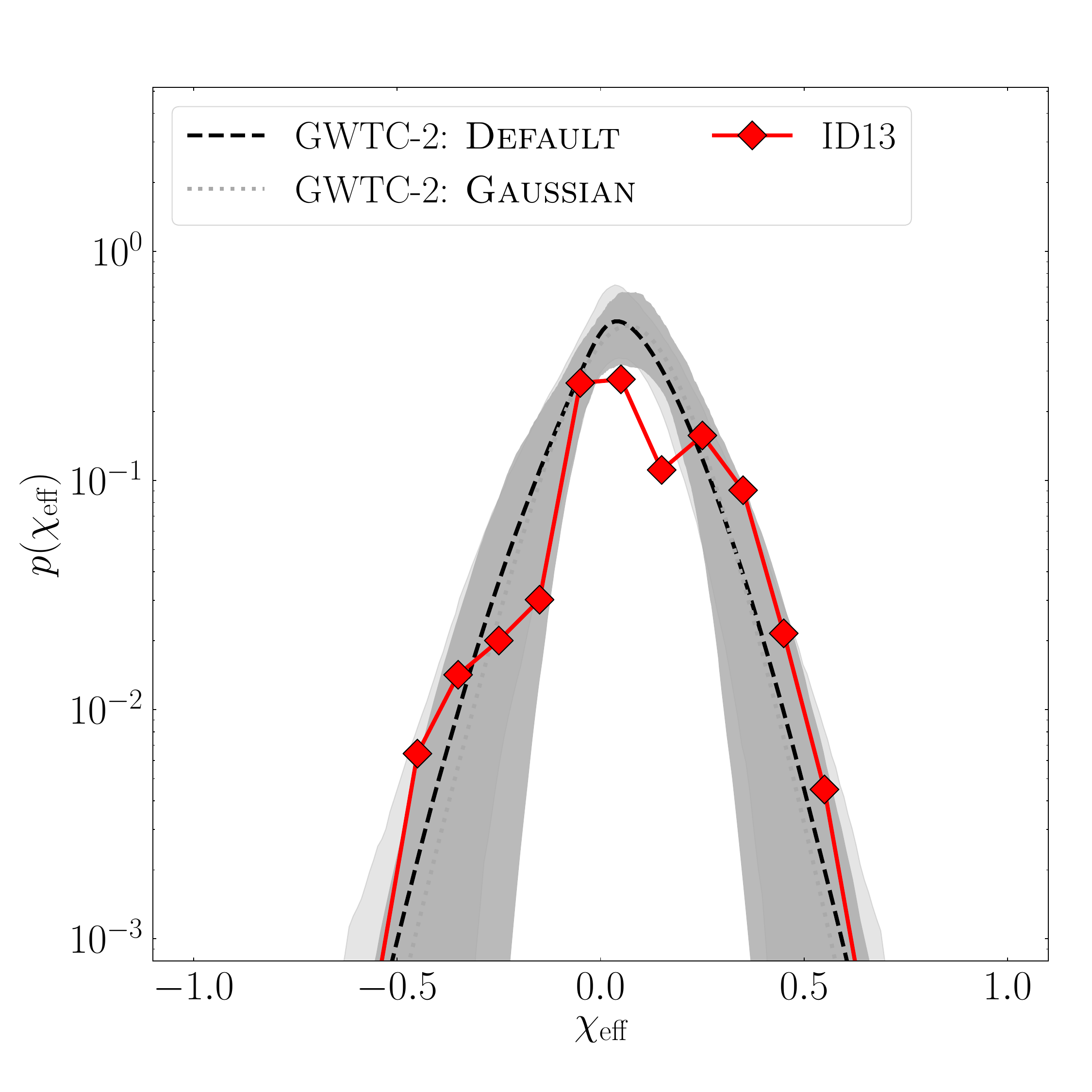}
    \includegraphics[width=0.32\textwidth]{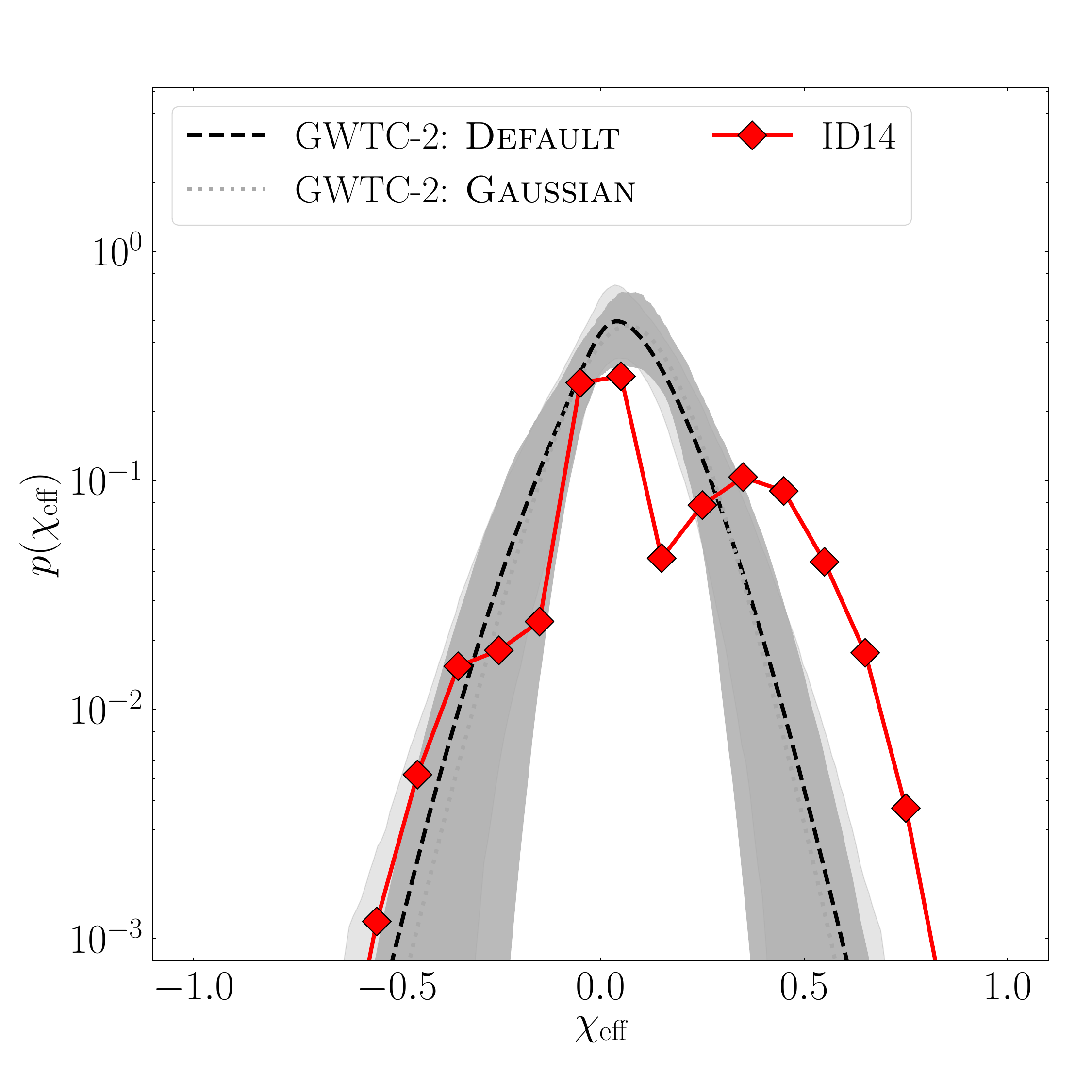}\\    
    \caption{Red solid lines with markers: distribution of $\chi_{\eff}$ in our reference model assuming: $n_\theta =8$ and a Gaussian BH natal spin distribution peaked over $\chi = 0.5,~0.2$ (models ID0H/L), a Maxwellian (model ID5), a mixed spin distribution in which single BHs have negligible spins and binary BH spins are taken from a Gaussian centered over $\chi = 0.5$ (model ID13) or from a Maxwellian with dispersion $0.2$ (ID14). Dashed black (Dotted gray) lines: posterior distribution of $\chi_\eff$  as derived from the  \textsc{Default} (\textsc{Gaussian}) LVK  models \citep{gwtc2}. The shadowed gray areas are the corresponding 90\% credible intervals.} 
    \label{fig:spins}
\end{figure*}

Figure \ref{fig:spins} shows the $\chi_{\rm eff}$ distribution for different models: different natal spin distributions in isolated and dynamical mergers clearly affect the overall $\chi_\eff$ distribution, possibly explaining both a dearth of mergers with $\chi_\eff < -0.3$ and a population of mergers with $\chi_\eff > 0.3$. These models suggest that BH merging in isolated or dynamical binaries might be characterised by different natal spin distributions, likely owing to the underlying different physical processes that contribute to the formation of mergers in each channel. In these regards, population synthesis tools like \bpop\ can readily serve as rapid and flexible parameter-space explorers, and can be exploited to compare models against the crescent number of observations. For instance, the detection of a few sources with negative $\chi_\eff$ could significantly help constraining the natal spin distribution of BHs in isolated and dynamical mergers.

\section{Discussion}
\label{sec:disc}

\subsection{Formation channels: the distribution of redshift and merging binary mass.}
 
\begin{figure}
    \centering
    \includegraphics[width=\columnwidth]{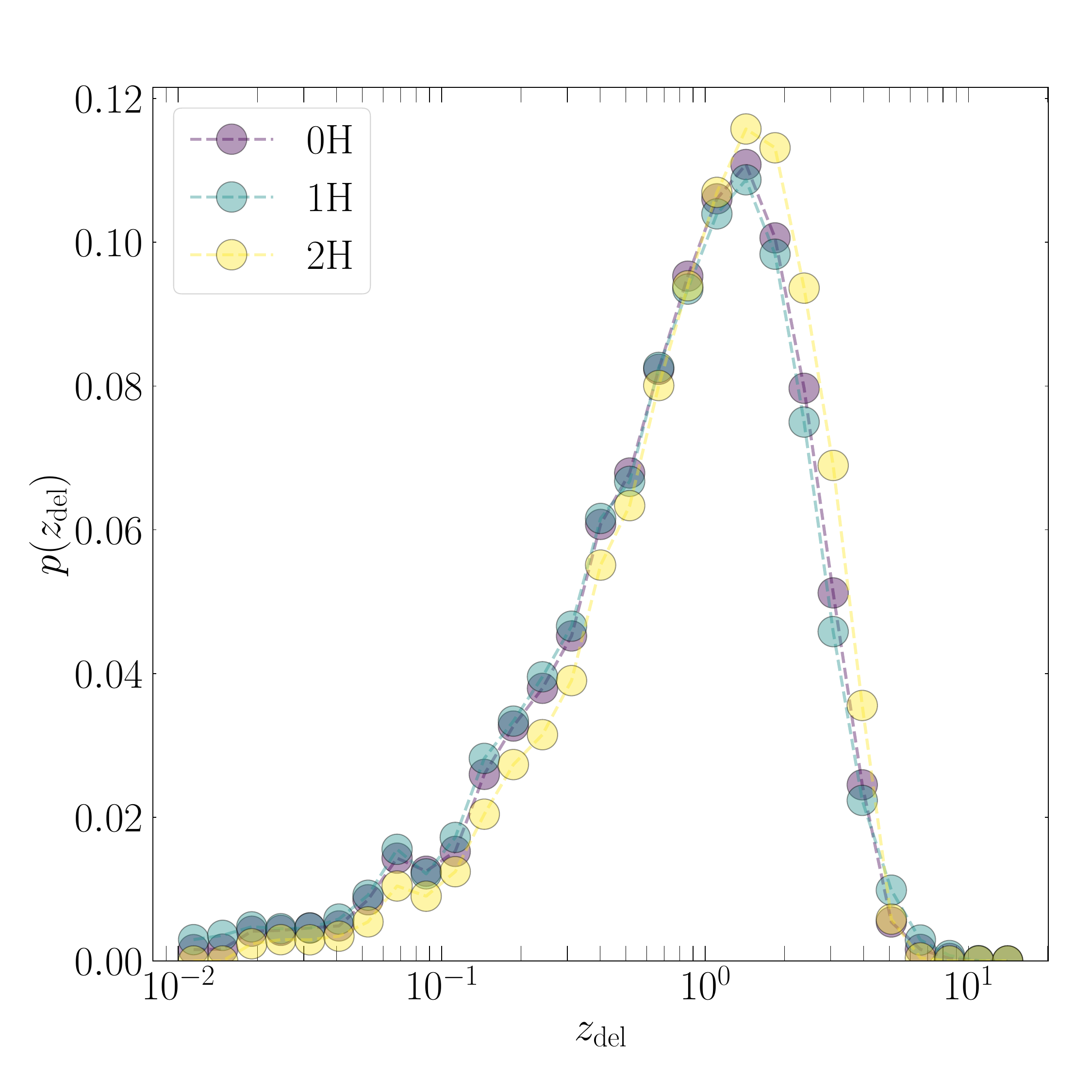}\\
    \includegraphics[width=\columnwidth]{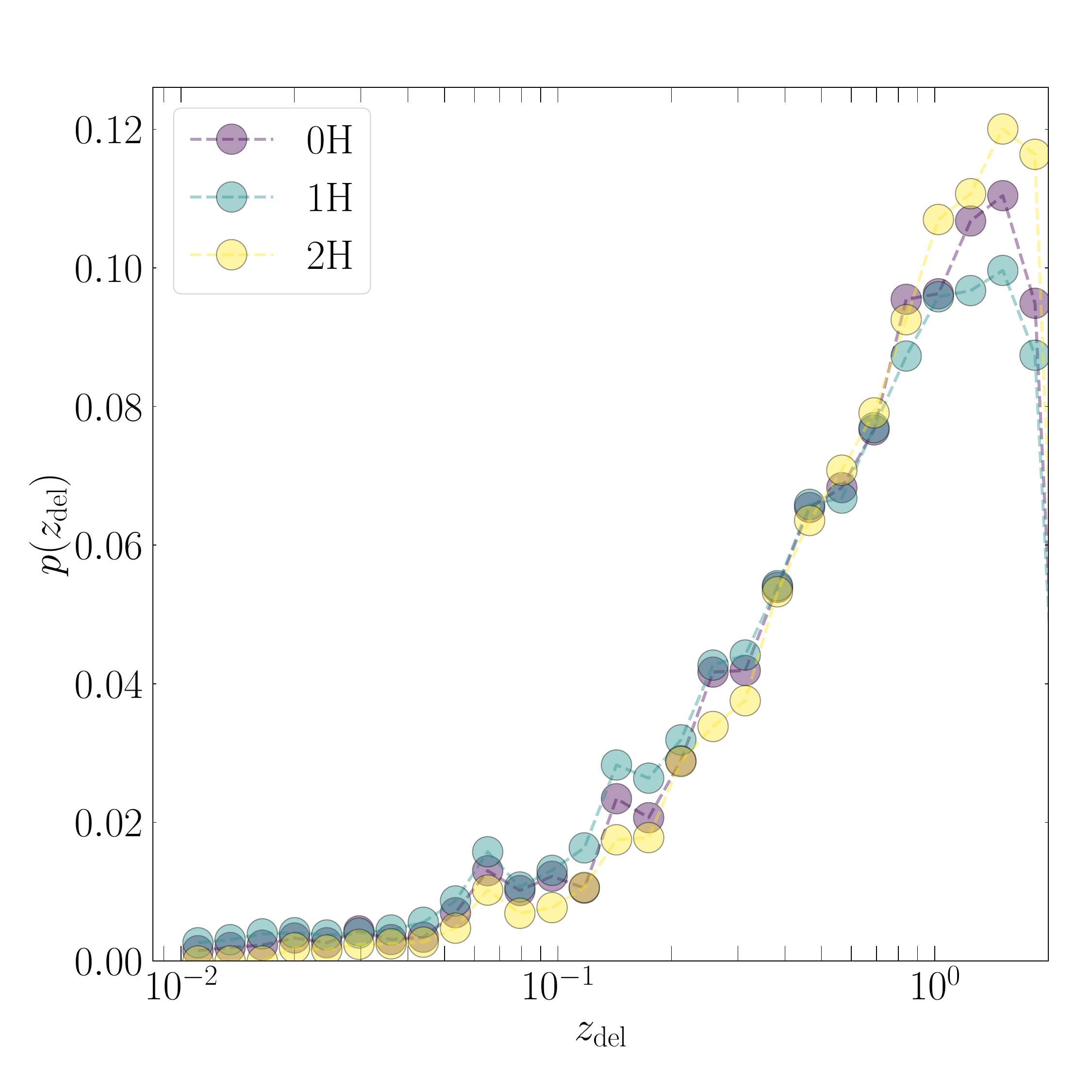}\\
    \caption{Redshift evolution for all (upper panel) and mock (lower panel) BBH mergers in the reference model (0H, purple), pure isolated channel (1H, blue), and pure dynamical  channel (2H, yellow).}
    \label{fig:z_distr}
\end{figure}

The redshift evolution of BBH mergers likely depends on many quantities, such as the cosmic star formation history, the adopted stellar evolution recipes, the star cluster properties. The top panel in Figure~\ref{fig:z_distr} shows the redshift distribution for all BBH mergers in the reference model (0H), the isolated model (1H), and the dynamical model assuming a SSBH mass spectrum (2H), while bottom panel shows the same quantity for the mock merger catalogue. The clear similarity among different models implies that isolated and dynamical mergers have a similar merger redshift distribution. 

The redshift distribution of isolated mergers is intrinsically due to the adopted stellar evolution model, and generally scales with the inverse of the delay time \citep[e.g.][]{dominik12}. This leads to a merger distribution that follows the same behaviour as the adopted star formation history but shifted at slightly lower redshift. 

\begin{figure}
    \centering
    \includegraphics[width=\columnwidth]{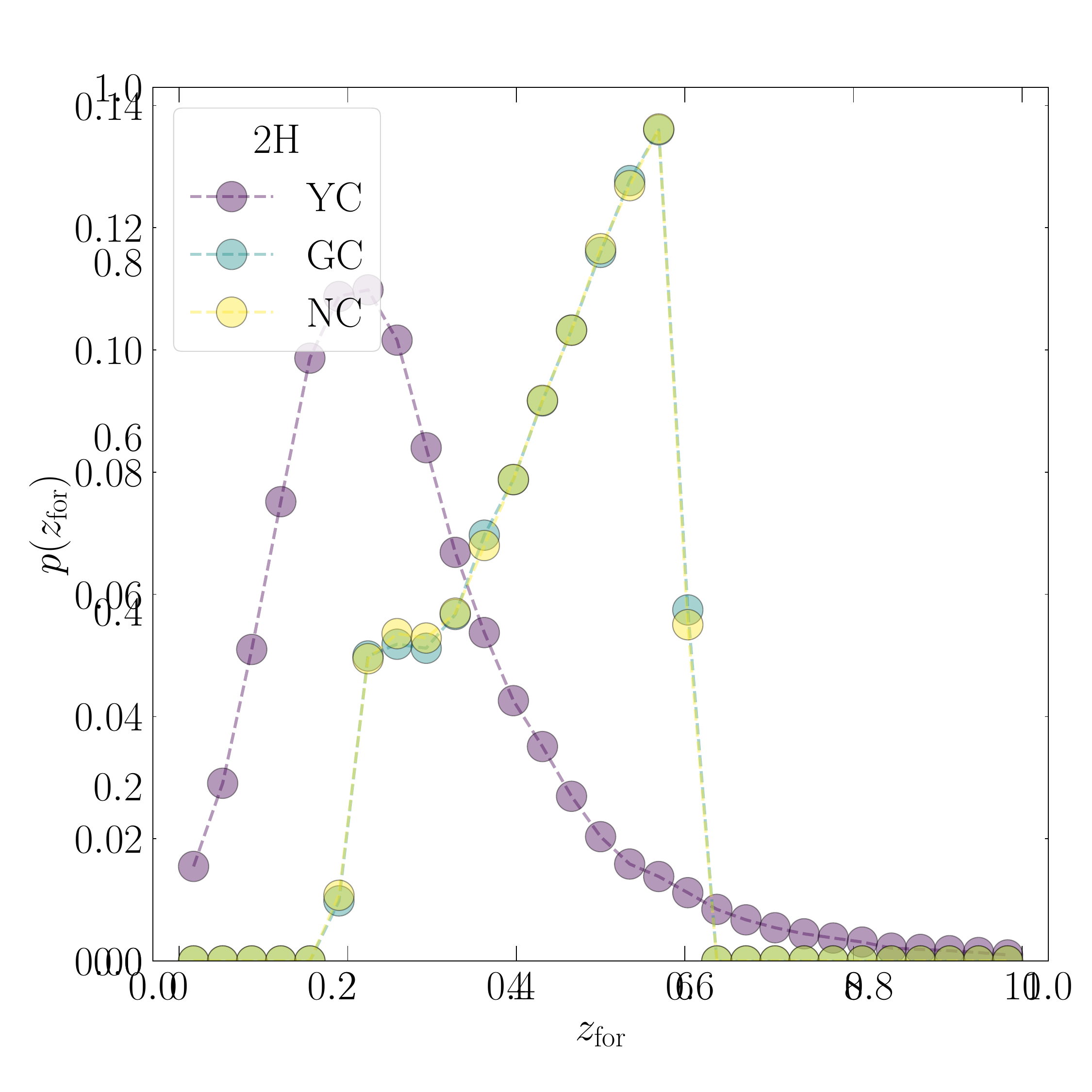}
    \includegraphics[width=\columnwidth]{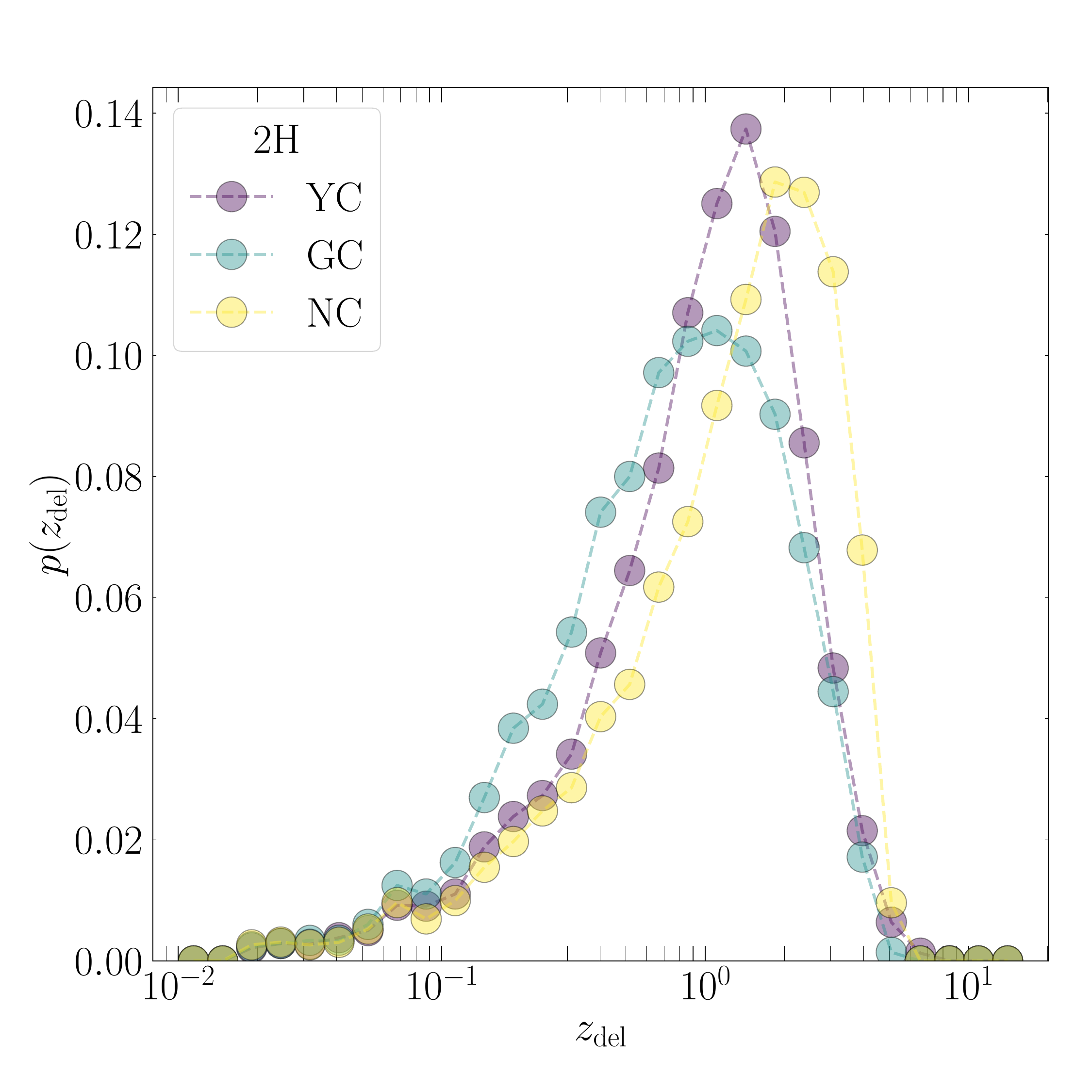}
    \caption{Redshift at formation (upper panel) and merger (lower panel) for all BBHs in YCs (purple), GCs (blue), and NCs(yellow) in the dynamical-only model (2H).}
    \label{fig:z_clus}
\end{figure}

The redshift distribution of dynamical mergers, instead, is determined by the timescales involved in the different dynamical processes that bring two unrelated BHs to pair and eventually merge \citep{mapelli22}. As shown by recent $N$-body simulations of star clusters, the delay time of dynamical mergers resembles that of isolated binaries \citep{2020ApJ...898..152S}, thus implying that, also for dynamical binaries, the resulting redshift distribution follows the star formation history shifted to lower redshift. To better investigate the role of dynamics in determining the overall merger rate of dynamical mergers, we show in Figure \ref{fig:z_clus} the distribution of the formation redshift of BBH merger progenitors (top panel) and the redshift at merger (bottom panel) for YCs, GCs, and NCs.

We see two important features. In the case of YCs, the distribution of formation redshift, with a peak at $z_{\rm for} = 2$ \citep{madau17}, is quite similar to the distribution of merger redshift, whose peak is shifted to $z_{\rm del} = 1-1.5$. In the case of GCs and NCs, we see that most of BBH merger progenitors form at relatively large redshifts. In spite of the same distribution of $z_{\rm for}$ for mergers in GCs and NCs, there are apparent differences in the distribution of $z_{\rm del}$, which is shifted toward larger values in the case of NCs. This happens because NCs, generally heavier and denser than other cluster types, are characterised by shorter delay times (given by the sum of the formation, pairing, hardening, and merger times).

Further quantities that can determine the redshift at merger are the mass and orbital properties of the stellar progenitors, the BBH orbital parameters at formation, the mass and size of the host cluster. For example, stellar evolution physics strongly affects the merging time of isolated binaries \citep[e.g.][]{giacobbo18a}, whilst cluster masses and sizes affect the time needed for two BHs to find each other and merge in dynamical environments. 

\begin{figure}
    \centering
    \includegraphics[width=\columnwidth]{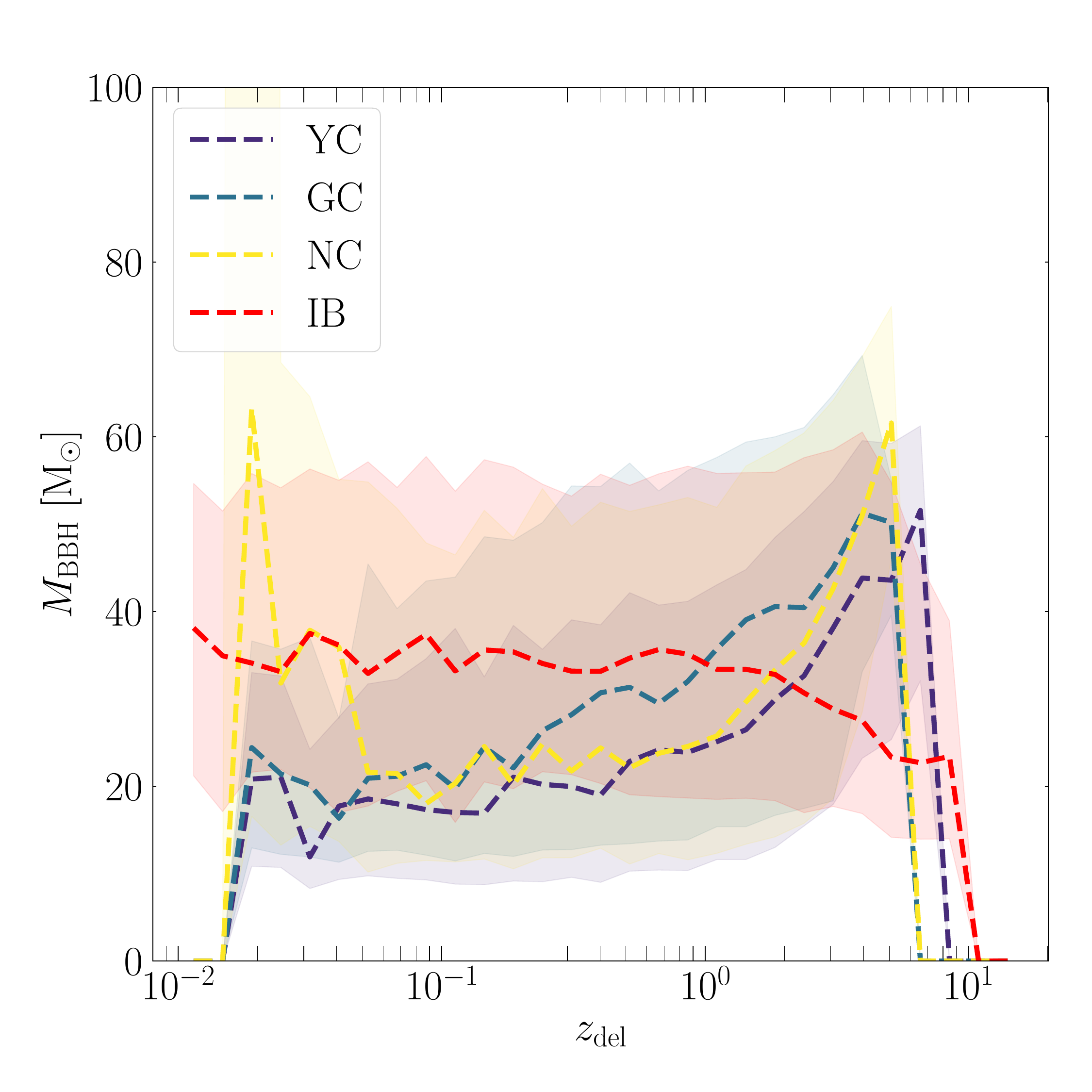} \\
    \caption{Top: Median value of the total BBH mass in YCs (purple), GCs (blue), NCs (yellow), and isolated binaries (red) in the reference model (0H). The shaded areas encompass the $60\%$ interval. }
    \label{fig:BBHmass}
\end{figure}

The top panel in Figure \ref{fig:BBHmass} shows the median BBH mass for all mergers in isolated binaries and different cluster types in our reference model (0H). In dynamical environments the median mass decreases at decreasing the redshift because the most massive BHs interacts on shorter timescales in star clusters \citep[e.g.][]{rodriguez16,askar17,2022ApJ...935..126B}. The evident increase in the median mass of BBHs in NCs at low redshift is driven by the formation of massive BHs via repeated mergers. Conversely, the median mass of isolated mergers increases at decreasing the redshift, attaining values around $M_{\rm BBH, IB} \simeq 40\,{}\Ms$. Also, our mergers have a maximum mass that is intrinsically set by the adopted stellar evolution, that is $M_{\rm BBH, max} \simeq 120\,{}\Ms$. 

\subsection{The impact of a mixed single BH mass spectrum on the population of BBH mergers}
As discussed in the previous sections, we can draw BH masses for dynamical mergers from three mass spectra: SSBH, MSBH, and HSBH. The first refers to the single BH mass spectrum output by \mobse, the second represents the population of BHs formed in binary systems modelled with \mobse\ that did not end their life in a compact binary merger, whilst the third represent the population of IMBH seeds possibly formed via stellar accretion processes and stellar collisions in dense clusters. 
In this section we discuss the different outcomes of the SSBH and MSBH mass spectra.

\begin{figure*}
    \centering
    \includegraphics[width=0.85\columnwidth]{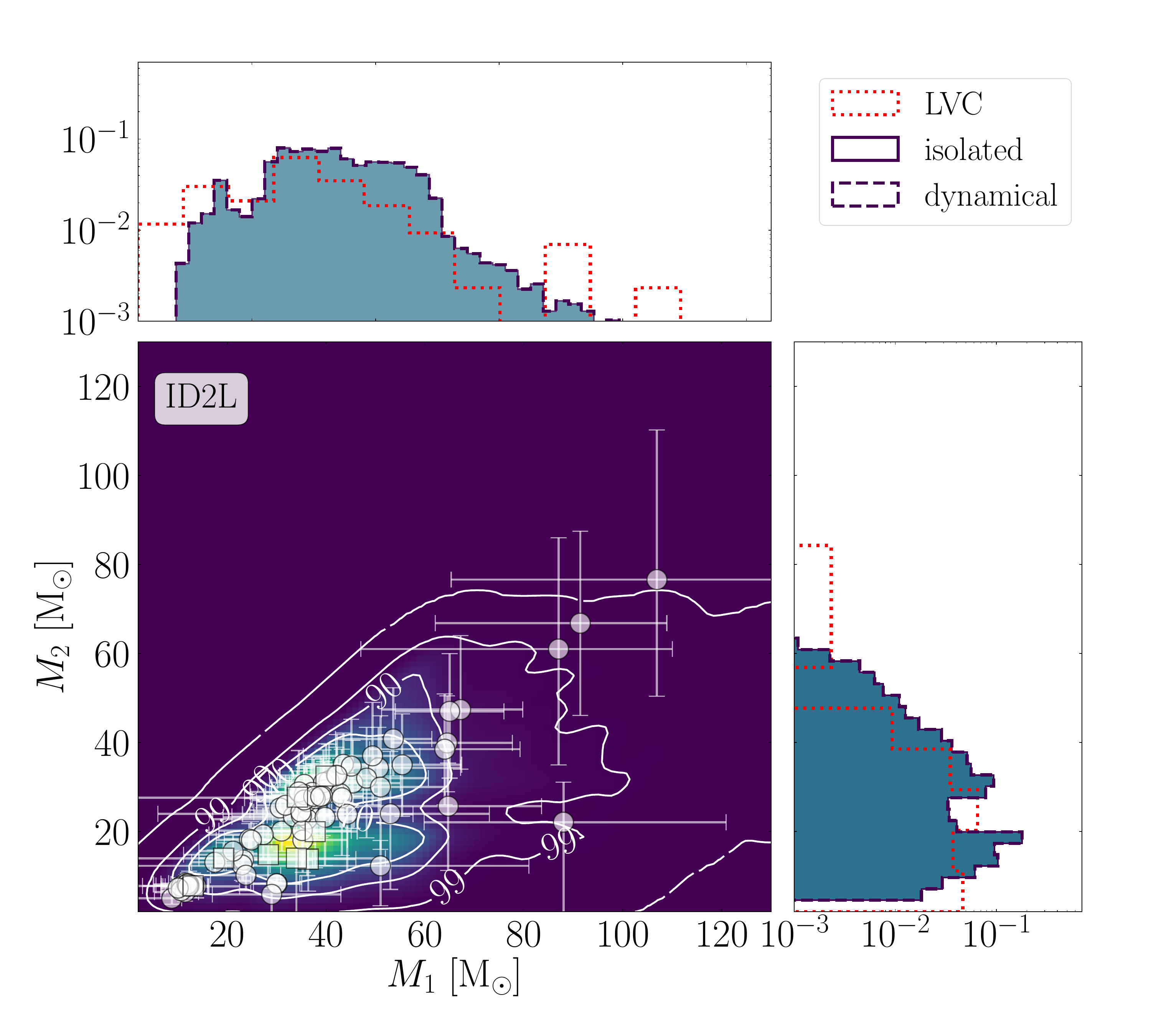}
    \includegraphics[width=0.85\columnwidth]{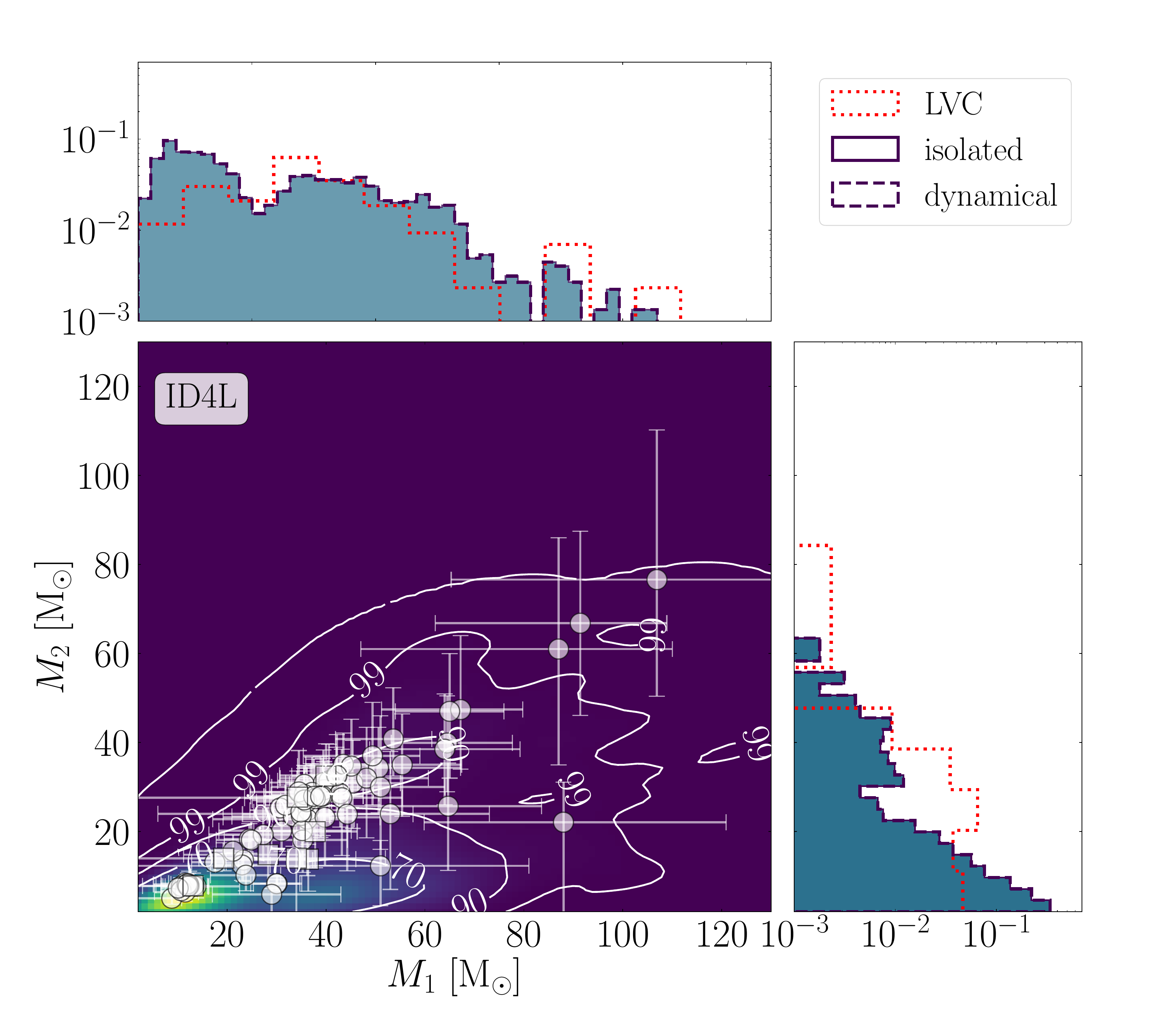}\\
    \includegraphics[width=0.85\columnwidth]{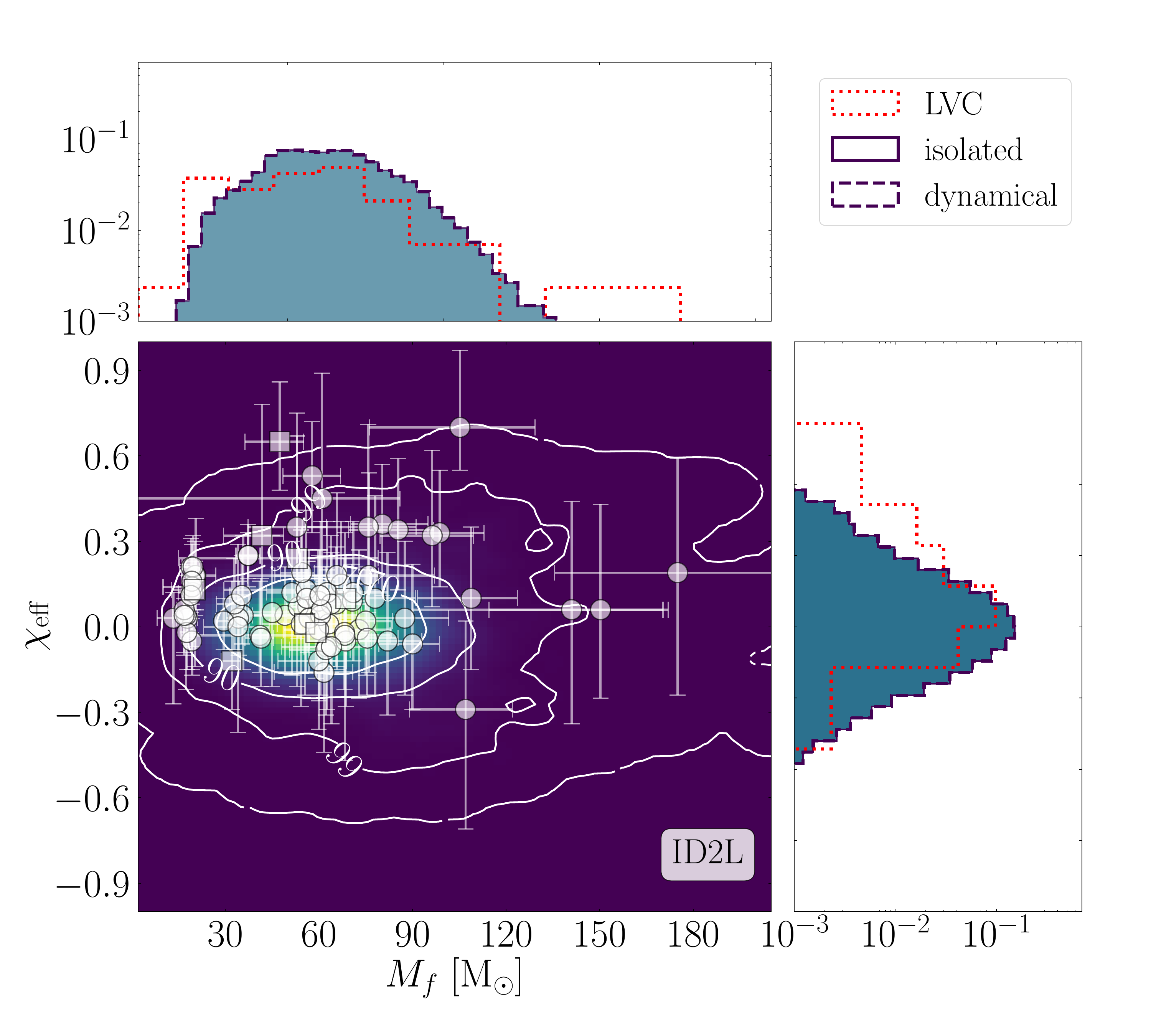}    
    \includegraphics[width=0.85\columnwidth]{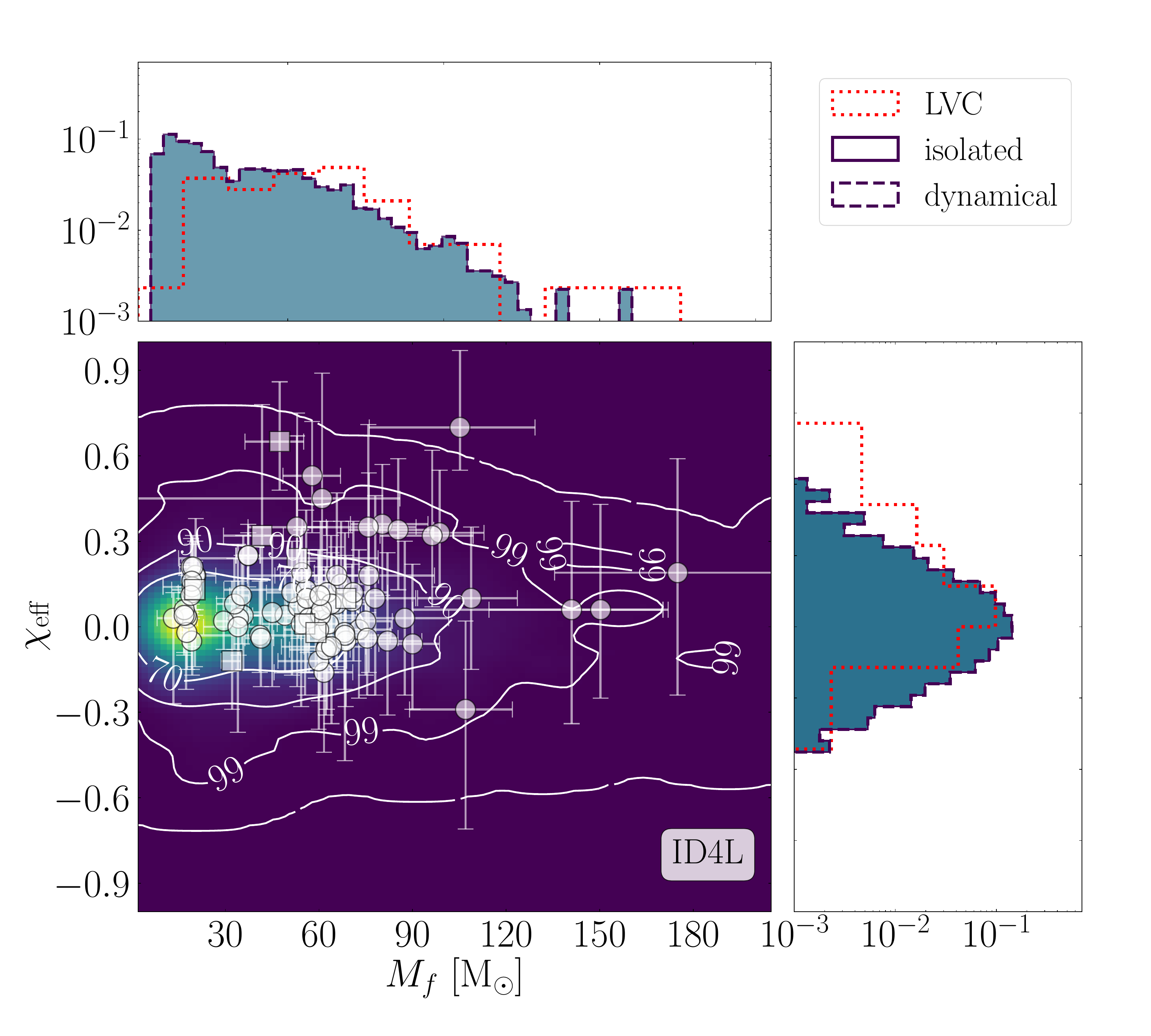}\\
    \includegraphics[width=0.85\columnwidth]{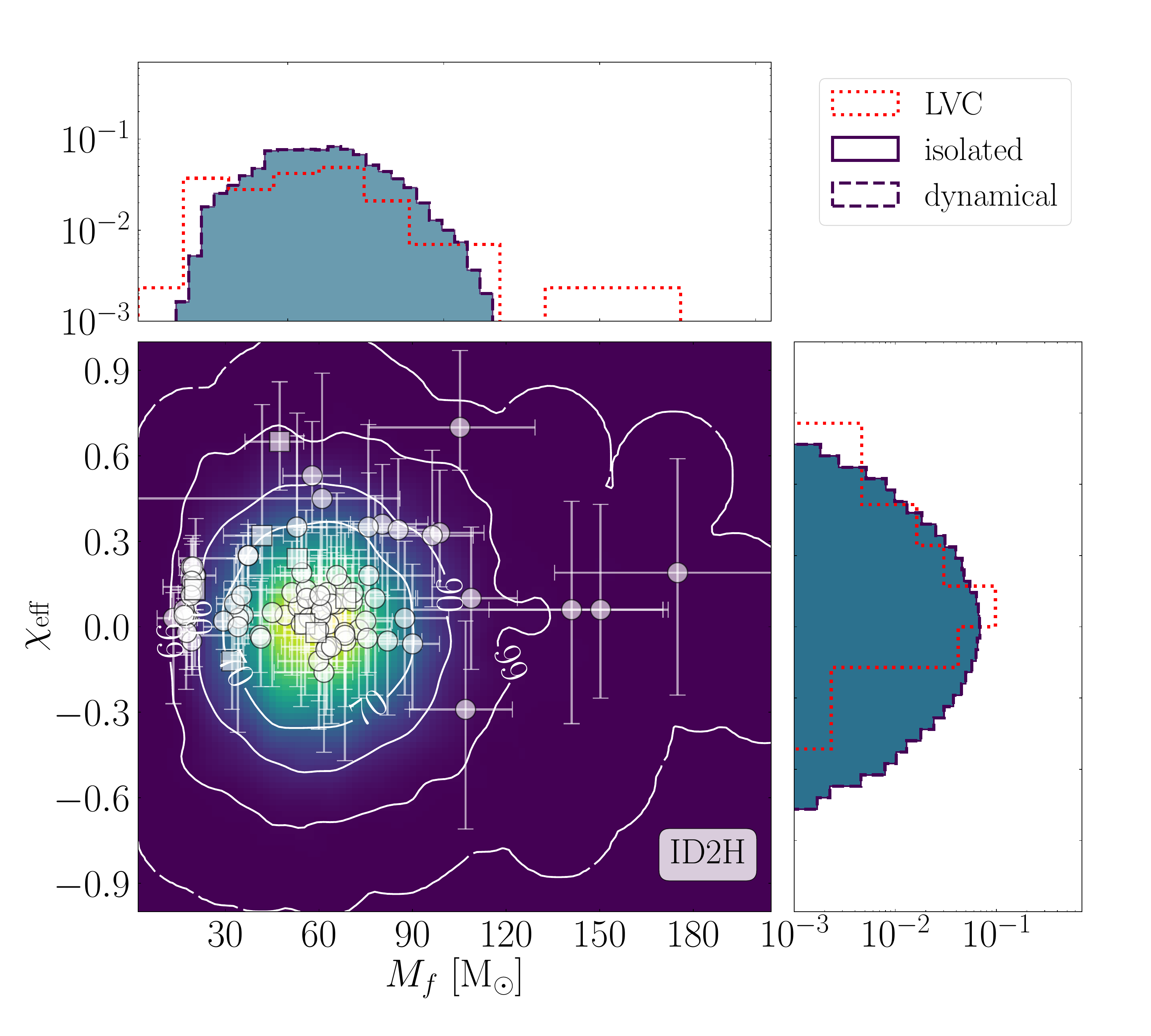}    
    \includegraphics[width=0.85\columnwidth]{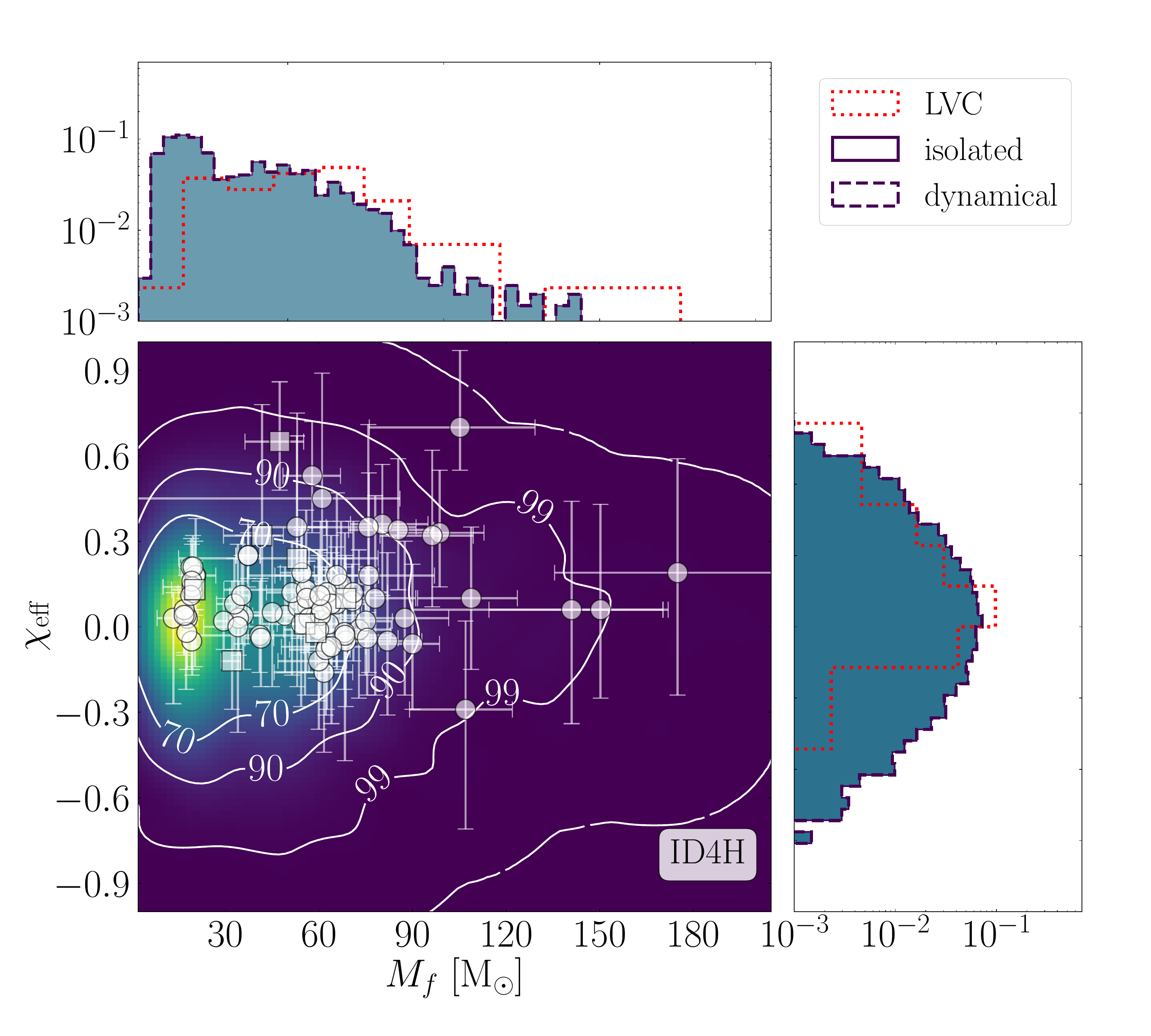}
    \caption{Component mass distribution (top panels) and remnant mass - effective spin parameter distribution (central and bottom panels) for models with only dynamical BBHs where the BH masses are extracted from either a single mass spectrum (SSBH, left panels) or from a mixed mass spectrum (MSBH, right panels). Top and central panels represent models in which BH spins are drawn from a Gaussian centered on $\chi = 0.2$, whilst bottom panels have BH spins drawn from a Gaussian centered on $\chi = 0.5$.}
    \label{fig:ssbhmsbh}
\end{figure*}

Figure \ref{fig:ssbhmsbh} shows the surface maps of component masses, remnant mass, and effective spin parameter for models ID2 and ID4, where we consider dynamical mergers only with masses taken from either SSBH or MSBH, respectively.

The ``standard'' single BH mass spectrum, SSBH, i.e. model ID2 with low (L) and high (H) spins, is characterised by a 2D distribution of component masses that encompasses the majority of detected BBH mergers, especially in the mass range $15 < M_1 /\Ms < 65$ and $M_2<45\Ms$. All detected mergers but one fall inside the region of the $M_f-\chi_\eff$ plane containing more the $99\%$ of mergers in model ID2L. Nearly half of detected mergers sit in the clear overdensity limited by $25< M_\rem/\Ms < 100$ and $|\chi_\eff|<0.25$. Conversely, BHs coming from the MSBH mass spectrum have more peculiar component mass distributions, which cover mostly the range $M_1 < 40\Ms$ and $M_2<20\Ms$. The $M_f - \chi_\eff$ distribution shows a clear overdensity that overlap quite well with 6 observed mergers which have a remnant mass $M_f\simeq 25\Ms$ and $|\chi_\eff|<0.25$. 

Interestingly, the mass ratio distribution of SSBH and MSBH are quite different, as shown in Figure \ref{fig:comp2}, with the former being characterised by $q>0.3$ and the latter showing a clear peak at smaller $q$ values, in the region $|\chi_\eff| < 0.25$ and $0.1<q<0.45$.  

Comparing low- and high-spin models in the bottom panels of Figure \ref{fig:comp2}, we see that whilst L models provide a better representation of the low-end of the $\chi_{\rm eff}$ distribution, H models are more suited to represent the high-end. 

\subsection{Hierarchical mergers: a route to form massive BH seeds in extremely dense environments}

When mergers take place in star clusters, sufficiently large escape velocities can favour the retention of the merger products and the development of multiple generation (hierarchical) mergers. 

\begin{figure}
    \centering
    \includegraphics[width=\columnwidth]{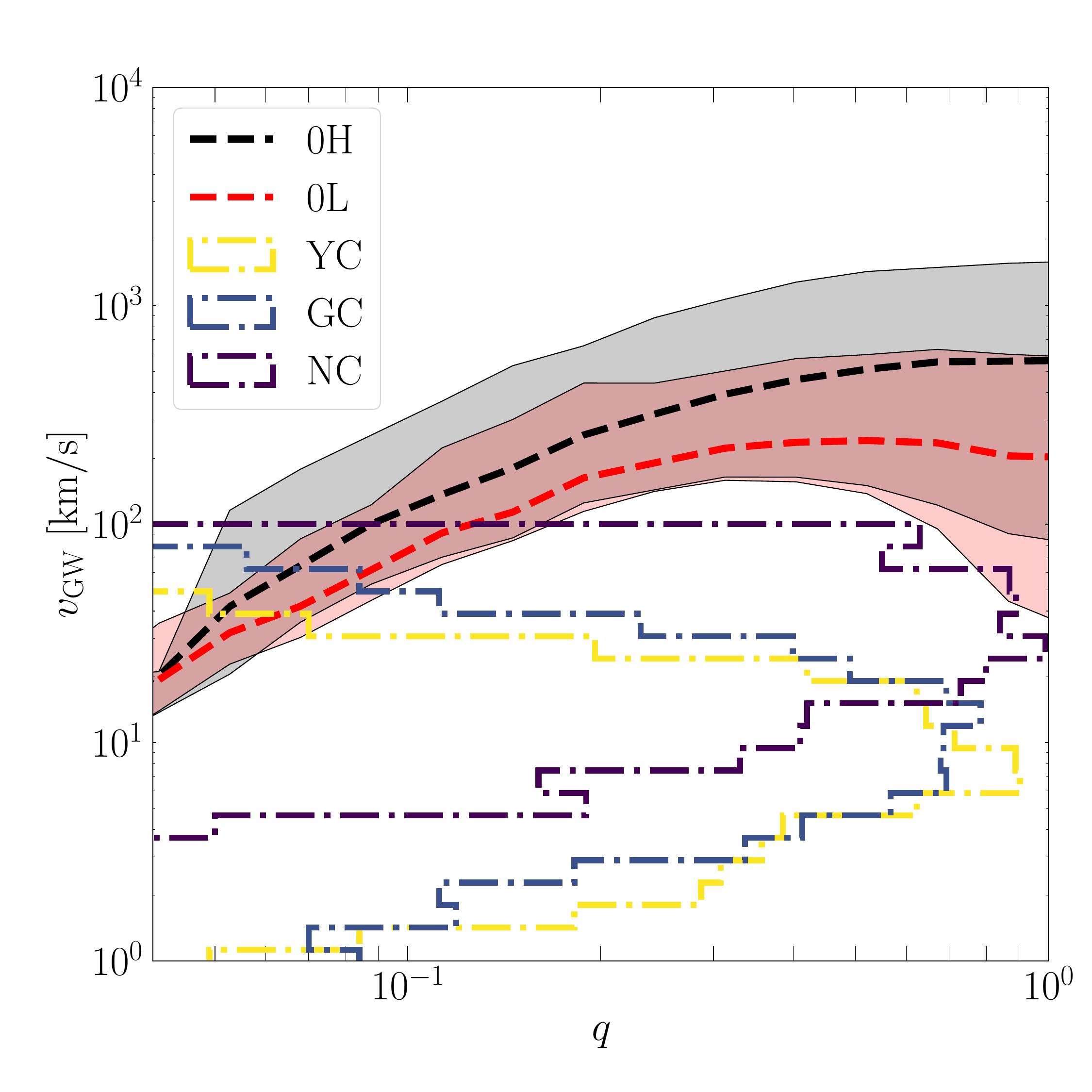}
    \caption{ Median value (dashed line) and $95$th percentile (shaded area) of the GW recoil kick for BBH mergers in high-spin (0H) and low-spin (0L) reference models, as a function of the mass ratio $q$. The typical distribution of escape velocities $v_{\rm esc}$ calculated at the half-mass radius are shown for YCs (yellow dotted histogram), GCs (blue dotted histogram), and NCs (dark purple dotted histogram). The $v_{\rm esc}$ histogram normalization is arbitrary for visibility's sake.}
    \label{fig:GWkicks}
\end{figure}

The retention probability, which represents the primary quantity affecting the development of multiple mergers, clearly depends on the environment in which the merger takes place. Typically, YCs have escape velocities in the $v_{\rm esc}= (1-10)$ km$/$s range, whilst NCs can be characterised by values as large as a few $10^2$ km$/$s \citep[e.g.,][]{antonini16}. The amplitude of the GW kick, $v_{\rm GW}$, depends on the merger properties, especially on the BH spin amplitude and binary mass ratio \citep[e.g.,][]{campanelli07,lousto08,lousto12}. To highlight the comparison among $v_{\rm esc}$ and $v_{\rm GW}$ in \bpop, Figure \ref{fig:GWkicks} shows  the median value and $5-95\%$ quartile for $v_{\rm GW}$ in both the low- and high-spin reference models (0L and 0H), compared to the typical distribution of escape velocities in YCs, GCs, and NCs. This Figure shows that loose environments like YCs or sparse GCs are unlikely to support multiple mergers, and suggest that retained mergers in such environments are likely the byproduct of highly asymmetric BBHs with mass ratios $q \lesssim 0.1$. 
As expected, we find that the vast majority of multiple generation mergers occur in dense GCs and NCs.

\begin{figure}
    \centering
    \includegraphics[width=\columnwidth]{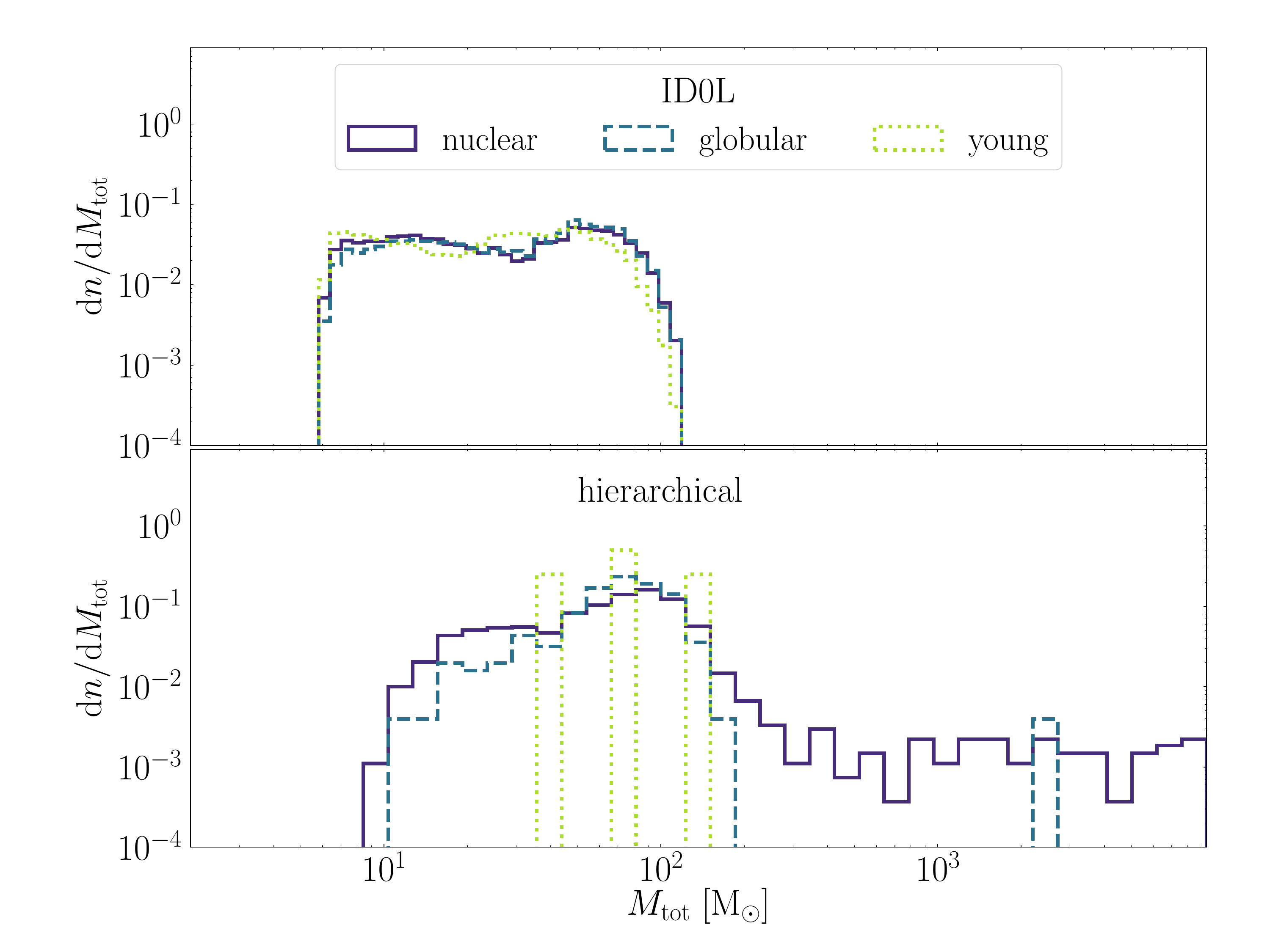}
    \caption{Mass distribution of first generation (top panel) and multiple generation (bottom panel) dynamical mergers for different cluster types and assuming the reference model with low spins (ID0L). The histograms refer to the overall population of mergers. We do not show the population of BBH remnants with mass $>10^4\Ms$.}
    \label{fig:recBHsp}
\end{figure}

Figure \ref{fig:recBHsp} shows the mass distribution of single and recycled mergers in YCs, GCs, and NCs for the low-spin reference model (ID0L). The mass spectrum of first-generation mergers is characterised by a well defined distribution that poorly depends on the cluster type, showing two clear peaks at $15\Ms$ and $\sim 60\Ms$. The mass distribution of hierarchical mergers, instead, depends crucially on the environment, with clusters characterised by higher densities and masses favoring the formation of heavier BH remnants, on average. YSCs host a handful of repeated mergers with total mass $\sim 100\Ms$. GCs exhibit a peak around $70\Ms$ with a large dispersion in the range $40-200\Ms$ and a sharp truncation at masses $>200\Ms$. For NCs, instead, the mass distribution peaks at $100\Ms$ and slowly decreases down to $500\Ms$. 

In the heaviest and densest NCs (${\rm Log} M/\Ms>7.5$), dynamical interactions can give rise to ``cascades'' of BH mergers, leading to massive BHs with mass as large as $10^{4-6}\Ms$. However, given the rarity of such dense and massive star clusters, these ``oversized'' BHs are expected to be extremely rare \citep[see also][]{mapelli21,fragione20}. As a consequence, the mass function of hierarchical mergers in NCs displays a clear rise beyond $M_{\rm tot}>10^3\Ms$. Clusters with mass $0.5 < M_c/10^7\Ms < 3.5$ and half-mass radius $0.2 < r_h/{\rm pc} < 1$ have sufficiently high density ($>10^7\Ms~{\rm pc}^{-3}$) and velocity dispersion  ($400 < v_{\rm esc}/{\rm km ~s}^{-1} < 2000$) to harbour a {\it merger avalanche} that builds up BHs $\gg (0.5-1)\times 10^4\Ms$ over timescales $<5-10$ Gyr. In our reference model, we find the formation of such massive BHs in  $\sim{12}\%$ NCs. We leave the investigation of such massive seeds to a follow-up study.

\begin{figure*}
    \centering
    \includegraphics[width=0.65\columnwidth]{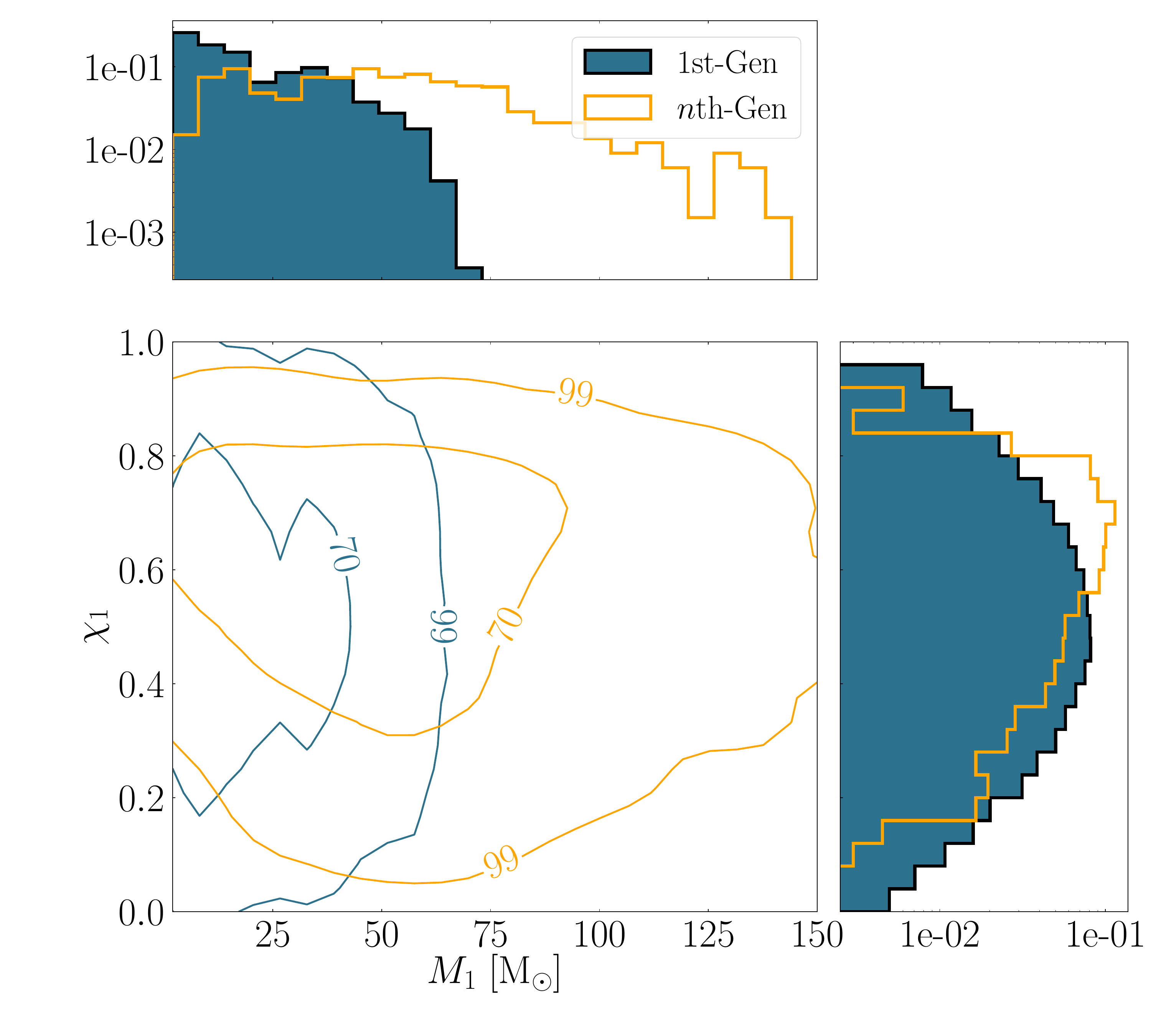}
    \includegraphics[width=0.65\columnwidth]{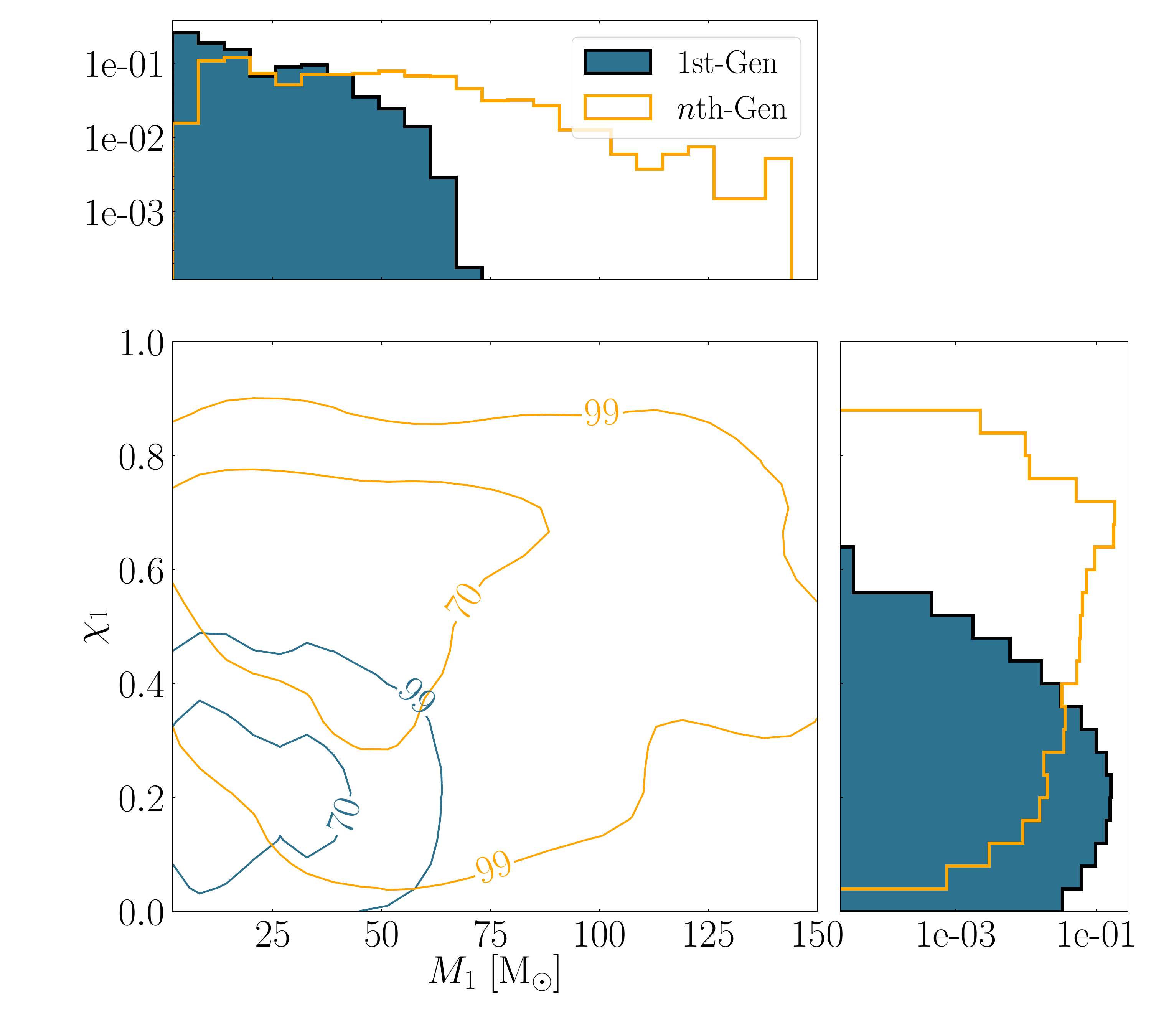}
    \includegraphics[width=0.65\columnwidth]{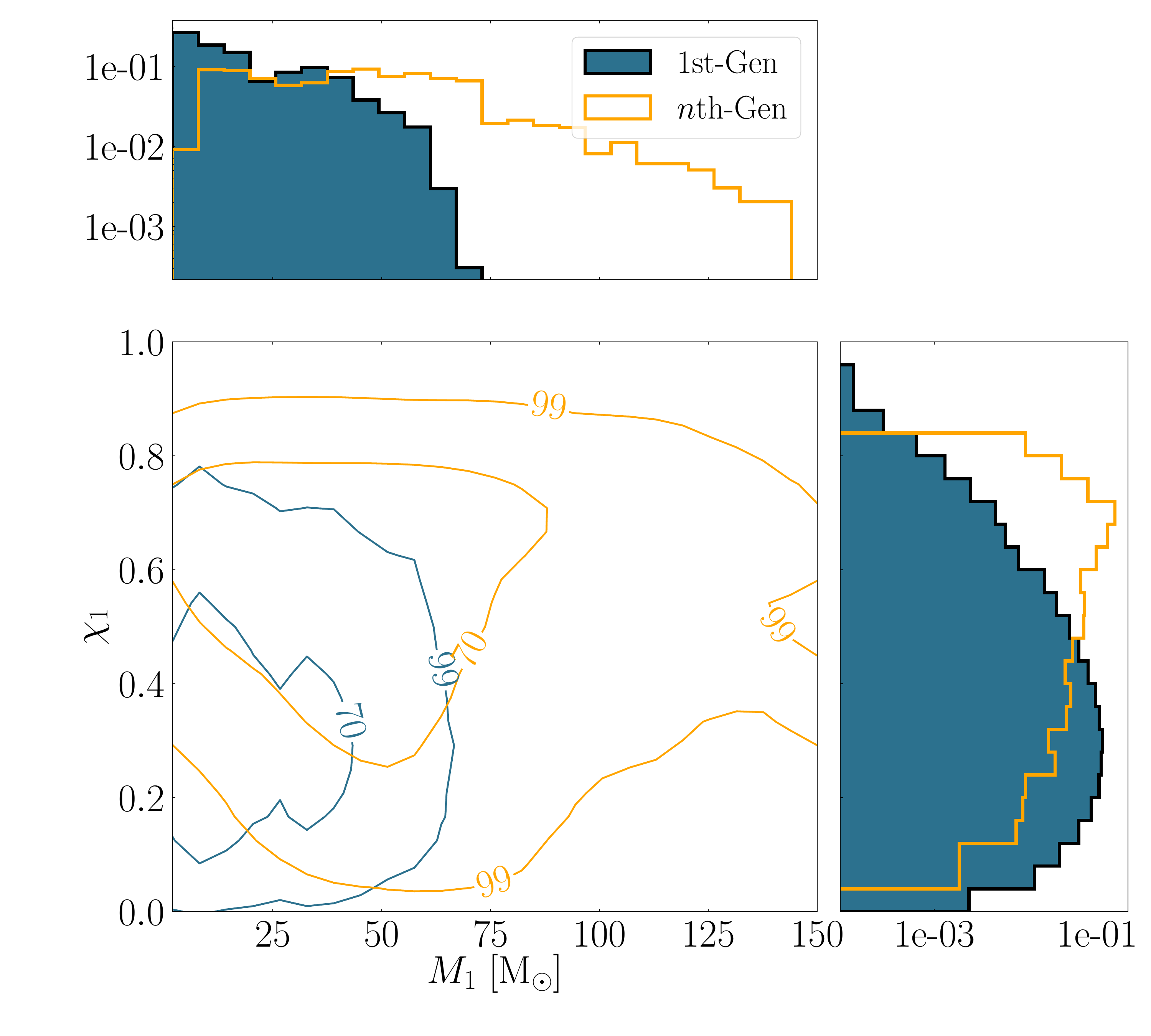}
    \caption{Primary spin and mass surface map distribution for first generation (blue contours and filled steps) and multiple generation BBHs (orange contours and open steps), assuming a BBH population equally contributed by isolated and dynamical BBHs and adopting  a Gaussian distribution peaked on either $\chi_1 = 0.5$ (left panel, ID0H) or $\chi_1 = 0.2$ (central panel, ID0L), and a Maxwellian with dispersion $\sigma_{s_1}=0.2$ (right panel, ID5) for BH natal spins.}
    \label{fig:hier}
\end{figure*}

Figure \ref{fig:hier} compares the combined distribution of primary mass and spin for single and repeated mergers in the case of spins drawn from a Gaussian centered on $\chi_\eff = 0.5$ or $0.2$ (ID0H/L) or a Maxwellian (ID5) distribution.
The primary mass distribution for repeated mergers seems to poorly depend on the spin distribution, being nearly flat in the $M_1=(25-75)\,{}\Ms$ mass range. However, the natal spin distribution affects evidently the primary spin of repeated mergers, with the low-spin model  (ID0L) being characterised by a narrower $\chi_\eff$ distribution compared to the corresponding high-spin model (ID0H). This happens because lower spins imply smaller GW recoil and, thus, a larger probability for hierarchical mergers.

In the high-spin model ID0H,  $\sim{23}\%$ of repeated mergers have a primary $M_1>50\Ms$ and $\chi_1 > 0.6$, while only $2\%$ of first-generation mergers have such high mass and spins. Changing the value of $n_\theta$, thus the amount of nearly aligned isolated mergers, does not affect appreciably the distribution. The percentage of high mass and spin hierarchical mergers remain almost the same adopting a low-spin model, but the amount of single generation mergers with such properties drops to $0.4\%$. Therefore, in the framework of low natal spins for merging BHs, the detection of mergers with $\chi_\eff \gtrsim 0.6$ and masses $M_1 > 40\Ms$ could represent a strong indication of a dynamical origin \cite[but see also][]{gerosa21}.
 In the semi-plane $45 < M_1/\Ms < 85$ and $\chi_1 > 0.3$, hierarchical mergers are the $14\%$, $61\%$, and $22\%$  of the population in the high-spin reference model (ID0H),   in the ID0L model and  in the Maxwellian spins model (ID5), respectively.

\subsection{The role of massive IMBH seeds}
The initial evolutionary phases of dense clusters can favour the growth of an IMBH seed with a mass in the range $100-500\Ms$ \citep{zwart02,giersz15,mapelli16,dicarlo19,dicarlo21,rizzuto21,gonzalez21,arca21b}. If retained in the parent cluster, these seeds can capture a BH companion \citep{dicarlo21,rizzuto21} and undergo coalescence \citep[e.g.][]{arca20c,rizzuto21,arca21b}. 

To explore the impact of such IMBH seeds onto the population of BBH mergers, we explore two further models, assuming that a certain fraction of dynamical BBHs have a primary with mass falling in the range $100-500\Ms$, i.e. in the IMBH mass range. 
For these IMBH seeds, we adopt a power-law mass spectrum with slope $-2$. This choice implicitly assumes that the IMBH mass scales linearly with the cluster mass, as happens for supermassive BHs and galactic nuclei, and that the overall cluster mass function follows a power-law with slope $-2$ \citep{lada03}. Nonetheless, we note that the IMBH mass spectrum is highly uncertain, owing to the dearth of thorough studies about the formation of these objects in clusters in a wide mass range.  

In model ID11, we assume that $40\%$ of dynamical BBH mergers have masses taken from the ``standard'' BH mass spectrum (SSBH), $40\%$ from the mixed BH mass spectrum (MSBH), and $20\%$ are comprised of IMBH seeds. In model ID12, we assume that $85\%$ of BHs have masses taken from SSBH, $5\%$ from MSBH, and the remaining $10\%$ is composed of IMBH seeds. We do not find appreciable differences between models with high or low BH natal spins.

\begin{figure*}
    \centering
    \includegraphics[width=\columnwidth]{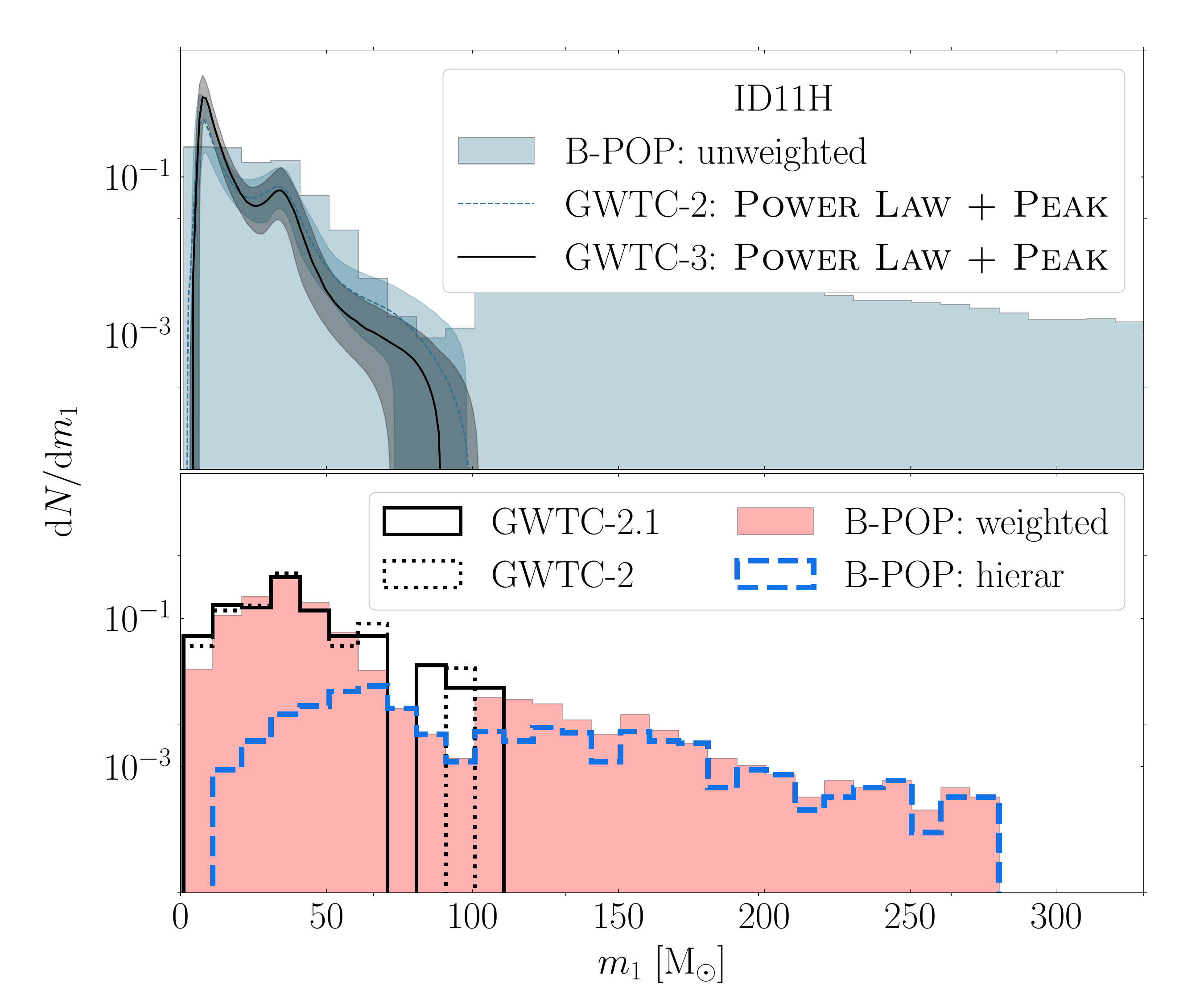}
    \includegraphics[width=\columnwidth]{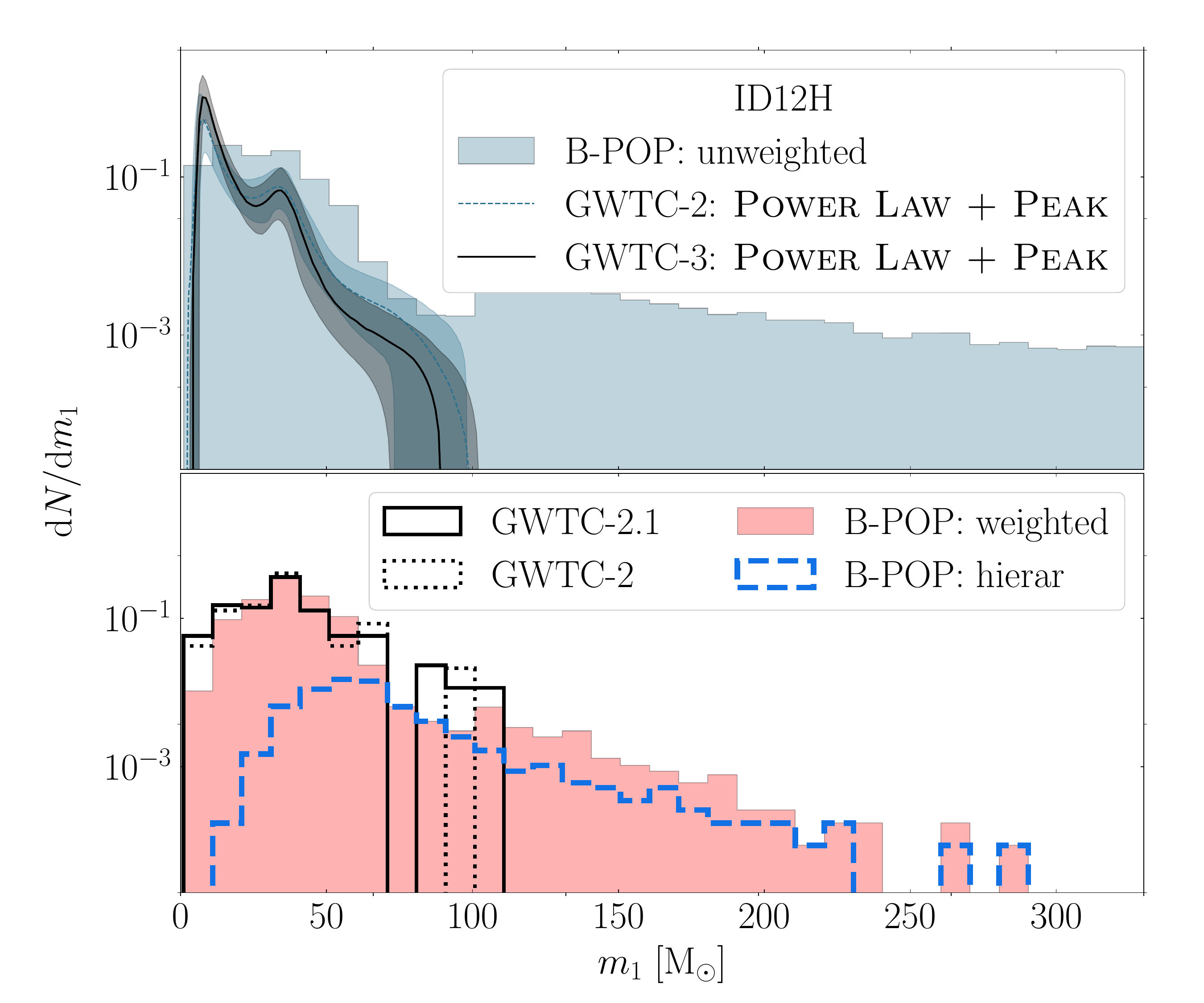}\\
    \caption{Same as in Figure \ref{fig:lvc} but for high-mass seed models with high spins (ID 11H and 12H). In model ID 11H, $40(40)\%$ of dynamical mergers have masses taken from SSBH (MSBH) and the remaining masses are taken from the IMBH seed spectrum adopted. In model ID 12H, instead, we assume that $85\%$ of mergers have masses taken from SSBH, $5\%$ from MSBH, and $10\%$ from the IMBH seed mass distribution.}
    \label{fig:lvcseed}
\end{figure*}

Figure \ref{fig:lvcseed} shows the primary mass distribution for global and mock BBHs in models ID11 and ID12. The impact of IMBH seeds is apparent from this figure, namely the high-end of the $M_1$ distribution is densely populated by these objects. When the observation selection criteria are applied, the mock $M_1$ distribution, which is shown in the lower panel of Figure \ref{fig:lvcseed}, is still characterised by a long tail extending beyond $100-300\Ms$, which contains $\gtrsim 10\%$ of the mock BBH population. Comparing the result of this model with the reference one, we see that increasing the amount of detected GW sources is crucial to constrain the formation of IMBH seeds in dense star clusters.

\begin{figure}
    \centering
    \includegraphics[width=\columnwidth]{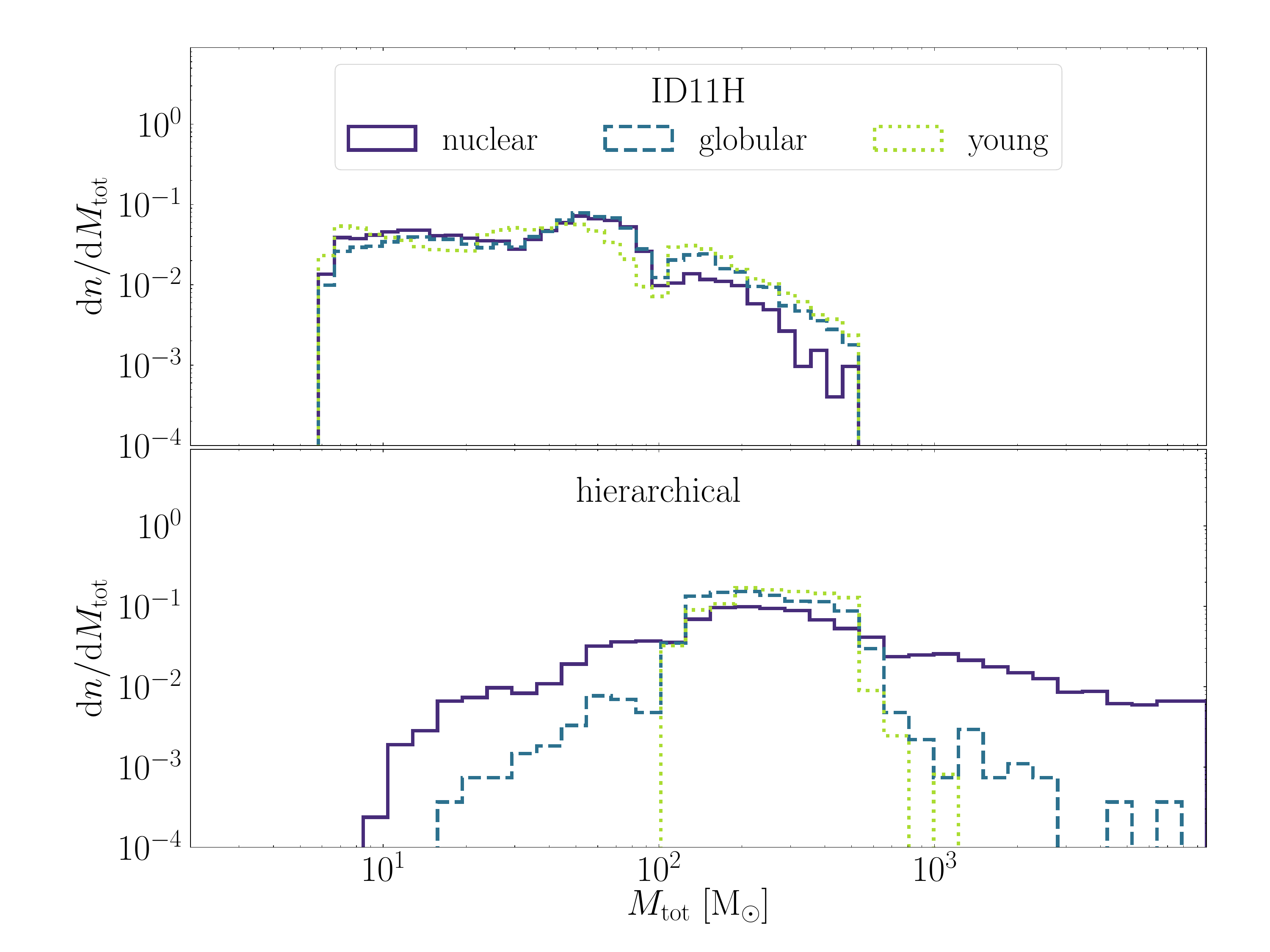}
    \includegraphics[width=\columnwidth]{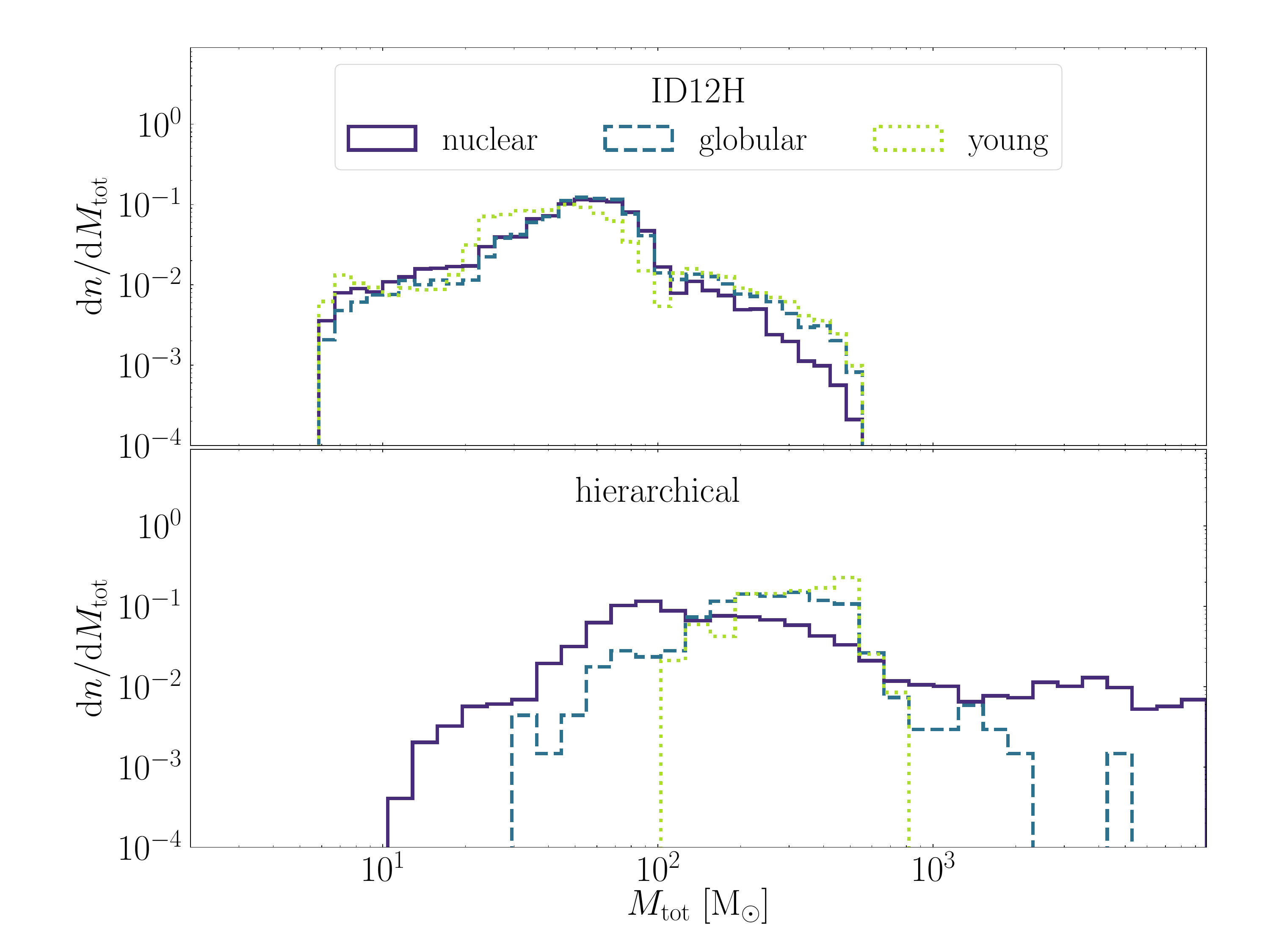}
    \caption{Same as in Figure \ref{fig:recBHsp}, but for models ID11H and 12H.}
    \label{fig:repHigh}
\end{figure}

Figure \ref{fig:repHigh} compares the mass spectrum of first-generation and repeated mergers in the case of model ID11H(12H), i.e. $\sim 20(10)\%$ of mergers involving IMBH seeds and $40(5)\%$ of BH masses taken from the MSBH spectrum. A substantial population of heavy seeds can significantly impact the mass of hierarchical mergers in all cluster types. The percentage of hierarchical mergers in the overall BBH merger population is $18.7-23.7\%$ for models 11H and 11L, and $5.8-9.7\%$ in models ID12H and 12L. The larger amount of hierarchical mergers in models denoted with L owes to the fact that lower spins lead generally to lower GW recoils, whilst the larger amount of mergers in models denoted with number 11 owes to the larger amount of IMBH seeds allowed in the overall population.

In YCs, the presence of IMBH seeds triggers the formation of a population of hierarchical mergers with masses in the range $100-900\Ms$. Most of them are mergers with $M_1\gg M_2$, whose remnants might receive kicks sufficiently small to be retained inside the parent cluster. The mass distributions of first-generation and repeated mergers in GCs are similar, although the latter is shifted by $0.7$ dex toward larger values. NCs, instead, are characterised by a population of mergers with masses $>500\Ms$, with a small sub-population of mergers ($\sim 0.052\%$) reaching masses $M_{\rm tot} > 10^4\Ms$ over a $5-10$ Gyr timescale.

 Considering the overall population of mock BBHs, we find around $2.7-7.5\%$ of mergers with $M_1 > 100\Ms$ and $6.6-11.0\%$ with $M_1+M_2 > 100\Ms$ in the high-mass seed models (ID11H/L and ID12H/L). For comparison, in the low- and high-spin reference models (ID0H/L) we found $3.7-5\%$ of mergers with $M_1+M_2 > 100\Ms$. Note that, in general, models with high-spins are characterised by a lower percentage of high-mass mergers, owing to the lower probability for hierarchical mergers to occur.
Interestingly, GWTC-2 contains $4$ mergers out of 47 detections having a median mass above this threshold and only $1$ exceeding $M_1+M_2 > 100\Ms$ at $90\%$ credible level, corresponding to the $2.1-8.5\%$ of the sample. Increasing the amount of GW sources in this mass range will thus help unveiling the impact of IMBH seeds onto the population of merging BBHs, at least at relatively low redshift.

\subsection{The impact of cluster evolution on the formation of massive BBHs. }
\label{sec:GCevo}

Following the cluster evolutionary scheme depicted in Section \ref{dynBBH}, in models 14, 15, and 16 we explored  the role of cluster mass-loss and expansion on the properties of dynamical mergers.

To simplify the picture and focus on the effect of cluster evolution, in these models we consider only GCs with masses in the range $M_{\rm cl} = (10^4-5\times 10^6)\Ms$ that formed at redshift $z=4$ with a metallicity $Z = 0.01$ Z$_\odot$. 
Thus, we focus only on dynamical mergers ($f_{\rm dyn} = 1$) formed in GCs ($f_{\rm GC}=1$).

Mass-loss and cluster expansion cause a decrease in the velocity dispersion, escape velocity, and density, thus inevitably affecting the dynamical timescales of BBH formation, hardening, and merger. 

\begin{figure}
    \centering
    \includegraphics[width=\columnwidth]{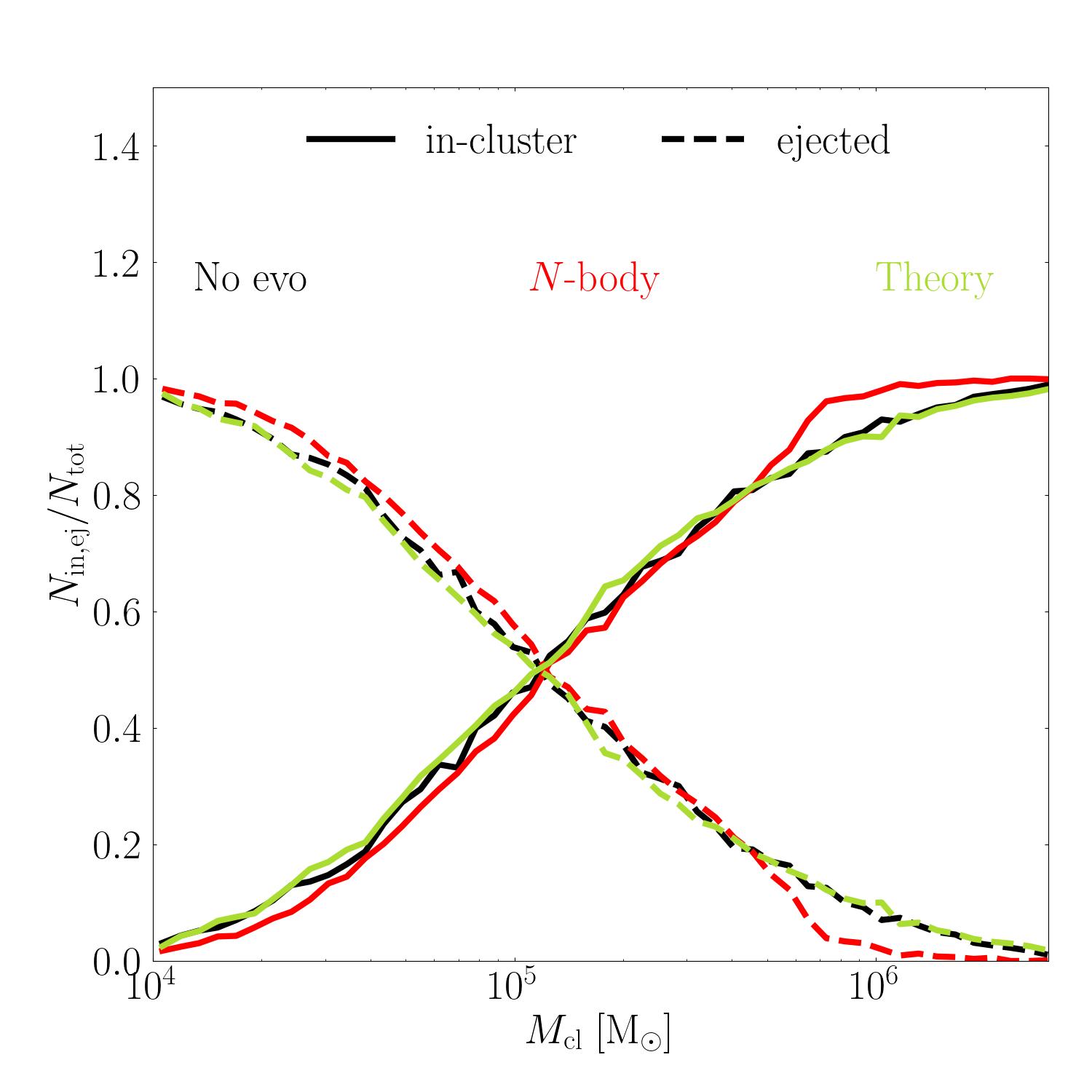}\\
    \includegraphics[width=\columnwidth]{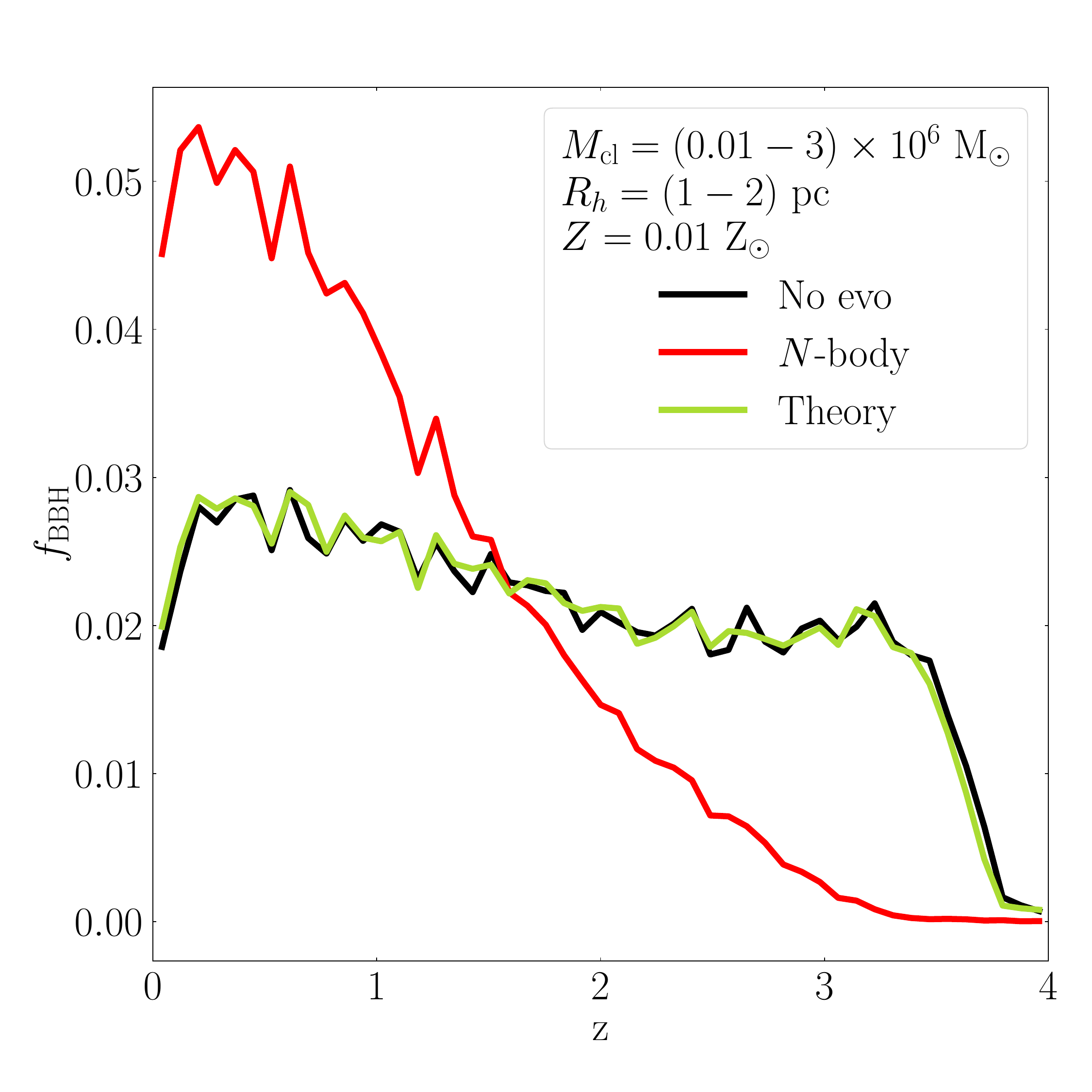}
    \caption{Top panel: Fraction of BBH mergers occurring inside the parent cluster (solid lines) or after ejection (dashed lines) assuming that the cluster does not evolve (black), or undergoes evolution according to $N$-body (red) or theoretical (green) prescriptions. Bottom panel: Fraction of mergers as a function of redshift for model 14 (no cluster evolution, black line), 15 ($N$-body cluster evolution, red line), and 16 (theoretical cluster evolution, green line).}
    \label{fig:ret}
\end{figure}

One possible effect driven by the cluster evolution could be a variation in the mutual fraction of mergers occurring inside the host cluster or after ejection. Note that in \bpop, mergers are labelled as "in-cluster" or "ejected" depending on the $a_{\rm ej}/a_{\rm gw}$ ratio (see Equations \ref{eq:ejorin} and \ref{eq:ejorin2}). The top panel in Figure \ref{fig:ret} shows the fraction of in-cluster and ejected mergers for evolving and non-evolving clusters, highlighting that the effect of cluster evolution is rather minimal on the relative amount of in-cluster/ejected mergers. This counter-intuitive result owes to the fact that the adopted evolutionary prescriptions cause a variation of the $a_{\rm ej}/a_{\rm gw}$ ratio by less than a factor $1.7-2$ over 100 relaxation times, thus making the cluster evolution irrelevant in determining the ejection of a merging BBH. 

There is a cluster mass threshold $M_\mathrm{cl}\simeq 10^5\Ms$ below(above) which the population of BBHs is dominated by ejected(in-cluster) mergers. Our semi-analytic predictions match well the outcomes of recent Monte Carlo models of star clusters \citep[see e.g.][]{rodriguez18}, highlighting the importance of the cluster potential well in favouring in-cluster mergers.  

The cluster evolution could affect also the delay time, or merger redshift, of dynamical binaries. A comparison among the merger redshift distribution for non-evolving and evolving cluster models is shown in the bottom panel of Figure \ref{fig:ret}. The model with a cluster evolution tailored to represent $N$-body simulations differs substantially from the other models, showing a clear shift of the distribution toward lower redshifts. In contrast, the difference is negligible in the cluster evolution model based on theoretical arguments. This striking difference owes to the fact that, over 10 relaxation times, the $N$-body model predicts a mass-loss of $p_{\rm loss}\sim 80\%$, much less than the theoretical model, for which $p_{\rm loss}< 20\%$.

\begin{figure}
    \centering
    \includegraphics[width=\columnwidth]{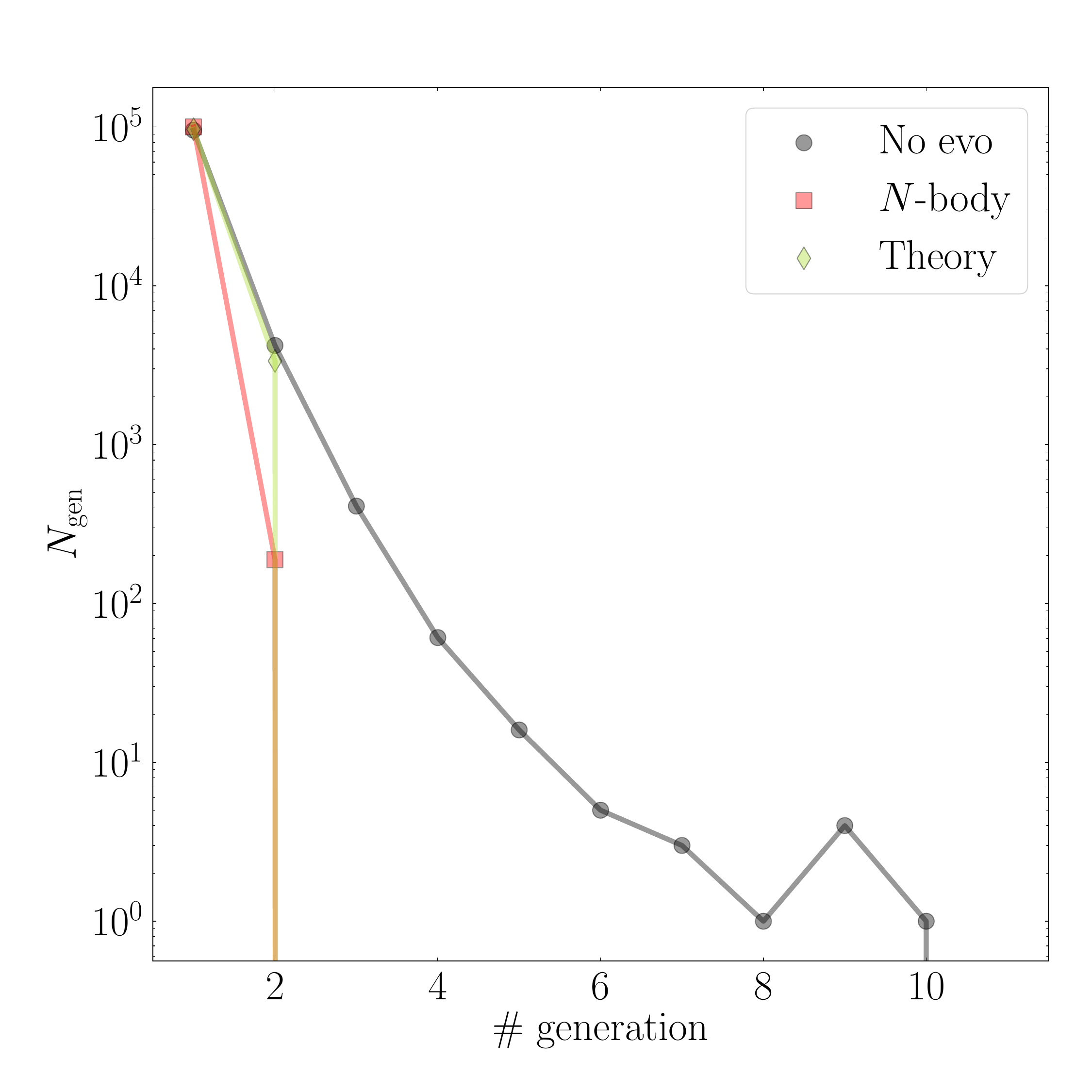}\\
    \caption{Number of mergers as a function of the merger generation for non-evolving (black points), $N$-body (red squares), and theoretical (green diamond) cluster models.}
    \label{fig:ret2}
\end{figure}

A further effect that the cluster evolution can have on the formation of merging BBHs is the development of multiple-generation mergers. In the non-evolving model, we find that $0.5\%$ of mergers are $2$nd--$10$th generation, while evolving clusters do not produce mergers beyond the $2$nd generation. Figure \ref{fig:ret2} shows how many mergers we get per merger generation, highlighting the importance of cluster evolution in determining a clear cut in the merging probability beyond the 3rd generation. 
A side effect of the cluster evolution is a reduction of the maximum BBH merger mass that the clusters can produce. In these test models, evolving clusters produce only 1st and 2nd generation merger remnants with masses $M_{\rm rem}<150\,{}\Ms$, while non-evolving clusters can produce remnants as massive as $M_{\rm rem} \simeq 250\,{}\Ms$. 
 
Therefore, the main effect of cluster evolution seems to be a drastic reduction of the probability for long hierarchical merging chains to occur and thus a possible effect on the high-end tail of the merging remnant mass distribution. The overall properties of dynamical mergers remain more or less the same, unless the cluster mass-loss is extremely fast, as it is shown in the comparison between models 15 and 16 in Figure \ref{fig:ret}. Nonetheless, the cluster evolution should be mitigated in NCs, where long merger chains are more likely to occur, because they are well embodied in their host galaxy centre. Sitting to the bottom of the potential well, and surrounded by the galactic bulge, NCs are less likely to undergo strong mass loss or expansion, thus in these environments it should be still possible to maintain a small population of mergers with large masses.

\section{Summary}
\label{sec:end}
We presented \bpop, a semi-analytic tool that enables population synthesis for BBH mergers taking place either in isolated binaries or in young, globular, and nuclear clusters. 
In its current version, \bpop\  exploits a library of single and binary BHs modelled with \mobse, though it can be easily fed with other stellar evolution libraries. The code implements a semi-analytic technique to model the dynamical formation of BBHs in star clusters, and a flexible interface to set a wide variety of parameters, like BH natal spins for isolated and dynamical mergers, the amount of IMBH seeds possibly forming in clusters, the relative amount of mergers occurring in different environments, and the star formation history of galaxies and clusters. Additionally, \bpop\  includes observation selection criteria that filter the modelled population of BBHs and returns a sub-sample of mergers as might be seen with second-generation ground-based GW detectors. 
 
To summarize, \bpop -- whose embryo was conceived and implemented in our previous papers \citep{arca19b, arca20} -- is a highly flexible, semi-analytic tool that enables the user to model the cosmic formation of both isolated and dynamical mergers. Its flexibility lies in the possibility to easily change the stellar evolution recipes, the observational selection criteria, the metallicity distribution, the star formation history, and the fractional amount of mergers forming through one channel or another. 
A further forthcoming upgrade of \bpop\ will permit to interface it with other semi-analytic codes tailored to model star cluster evolution in their host galaxies \citep[see, e.g.,][]{arca14,belczynski18,leveque22a, leveque22b}, thus enabling the simultaneous simulation of both BH dynamics in star clusters and star cluster dynamics in galaxies.

These properties place \bpop\ in between population synthesis codes targeting isolated BBHs \citep[e.g.][]{belckzinski10, dominik12, belczynski16, giacobbo18a, giacobbo18b, neijssel19, broekgaarden22, riley22, fragos22} and other recent semi-analytic tools that model dynamical mergers in star clusters. The latter codes share some common traits with \bpop. Some of them, like {\sc fastcluster} \citep{mapelli21, mapelli21b,mapelli22} model the formation of dynamical mergers in a given star cluster type with a given metallicity. Other semi-analytic codes, like cBHBd \citep{antonini19, antonini20, antonini22}, focus on the co-evolution of a single star cluster and its whole population of BBHs, thus representing a more self-consistent simulation of star cluster evolution and BBH formation at the expense of the possibility to model simultaneously the formation of dynamical mergers in many cluster types with different formation times and metallicities. Other semi-analytic tools focus on quiescent and active galactic nuclei \citep{antonini16b, arca20b, tagawa21a}. Finally, some codes incapsulate BBH formation processes, either semi-analytical or from numerical simulations, into a larger tool that models also the co-evolution of star clusters and their host galaxy \citep{gnedin14,arca14}, thus providing a picture on the connection between merging compact objects, their parent clusters, and the host galaxy \citep[e.g.][]{fragione18,belczynski18,leveque22a,leveque22b}.

All the aforementioned semi-analytic codes model either dynamical or isolated mergers only.

\bpop, instead, creates a synthetic Universe where both dynamical and isolated processes proceed simultaneously, taking into account the contribution of all those parameters (e.g. metallicity distribution, cosmic star formation rate, merger efficiency, cluster properties) that strongly affect the formation and merger of BBHs across space and time. 

In the following, we summarize our main findings:
\begin{itemize}
    \item assuming that isolated and dynamical binaries are equally distributed, we find that observation selection criteria lead to a ``mock'' population of mergers composed of $\sim 34\%\,{}(66\%)$ of isolated(dynamical) BBHs;
    \item we find that the isolated channel likely dominates the BBH population at low redshift, whilst dynamically assembled BBHs are dominant at $z>1$;
    \item in our reference model, the primary mass distribution of mock mergers matches GW observations; 
    \item the reference model produces a sub-population of mergers with masses heavier than $100-200\Ms$, whose detection might be hindered by observation selection criteria;
    \item assuming that BHs in isolated binaries have relatively high spins and that single BHs have low or even negligible spins leads to an effective spin distribution of mock mergers characterised by a dearth of mergers with $\chi_{\rm eff} < -0.3$ and a sub-population of mergers with $\chi_{\rm eff} > 0.3$;
    \item in our reference model, around $4.6-7.9\%$ of mock mergers are the byproduct of multiple (hierarchical) mergers, mostly developing in GCs and NCs, with total masses extending beyond $10^3\Ms$. In a small fraction of cases ($\sim 0.03-0.06\%$), hierarchical merger remnants can reach a mass $>10^4\Ms$;
    \item depending on BH natal spins, hierarchical mergers are $20-60\%$ of all the detectable BBHs with $\chi_1 > 0.3$ and primary mass $45 < M_1 /\Ms <85$;        
	\item the inclusion of the cluster dynamical evolution, driven by mass loss and expansion, does not have a dramatical impact on the total number of mergers, but strongly affect the development of mergers beyond the third generation, hampering {\it de facto} the growth of very large ($>10^3\Ms$) IMBHs;
    \item we explore the impact of IMBH seeds formed out of stellar collisions on the overall BBH population. If we assume that $10-20\%$ of all BBH mergers involve an IMBH seed formed via stellar collisions, around $2.7-7.5\%$ of mock mergers have $M_1 > 100 \Ms$.
\end{itemize}

\section*{Acknowledgements}
MAS acknowledges funding from the European Union’s Horizon 2020 research and innovation programme under the Marie Skłodowska-Curie grant agreement No. 101025436 (project GRACE-BH, PI: Manuel Arca Sedda).

MAS acknowledges support from the Alexander von Humboldt Foundation and the Federal Ministry for Education and Research for the research project "The evolution of black holes from stellar to galactic scales" and the Volkswagen Foundation Trilateral Partnership through project No. I/97778 ``Dynamical Mechanisms of Accretion in Galactic Nuclei''. 

MM acknowledges financial support from the European Research Council for the ERC Consolidator grant DEMOBLACK under contract no. 770017, and from 
the Italian Ministry of University and Research for the PRIN grant METE under contract no. 2020KB33TP.   

This work benefited from support by the International Space Science Institute (ISSI), Bern, Switzerland, through its International Team programme ref. no. 393 The Evolution of Rich Stellar Populations \& BH Binaries (2017-18), by the Deutsche Forschungsgemeinschaft (DFG, German Research Foundation) -- Project-ID 138713538 -- SFB 881 ``The Milky Way System''), and by the COST Action CA16104 ``GWverse''.

\section*{Data Availability}

The data and the code associated with the present study are available upon reasonable request to the corresponding author.


\bibliographystyle{mnras}
\bibliography{biblio} 


\appendix

\bsp	
\label{lastpage}
\end{document}